\pdfoutput=1
\documentclass[10pt,twocolumn,letterpaper]{article}

\usepackage[pagenumbers]{cvpr}

\definecolor{dark_green}{rgb}{0, 0.5, 0}

\usepackage{soul}
\usepackage[normalem]{ulem}
\setuldepth{foobar}

\usepackage{multirow}

\newcommand{\lucent}{\mbox{JADE}\xspace}

\newcommand{\solvetime}{0.19--0.23\,s}

\usepackage{comment}

\renewcommand{\topfraction}{0.9}
\renewcommand{\bottomfraction}{0.8}
\renewcommand{\textfraction}{0.07}
\renewcommand{\floatpagefraction}{0.75}
\renewcommand{\dbltopfraction}{0.9}
\renewcommand{\dblfloatpagefraction}{0.75}

\renewcommand{\paragraph}[1]{\vspace{.25em}\noindent\textbf{#1}}

\newcommand{\cmark}{{\color{ForestGreen}\checkmark}}

\newcommand{\xmark}{{\color[RGB]{204,0,0}$\times$}}

\usepackage{placeins}

\newcommand{\TabObjects}{
\begin{table}[t]
\caption{\textbf{Relighting comparison.} Mean over objects, on black; the synthetic objects also hold out cameras.
``Net.'': weights of the learned networks~(per-element parameters are of similar size for \lucent and SSD-GS).
``best'': number of objects with the best PSNR.
Per-object results are in the supplementary material.}
\label{tab:objects}
\centering\footnotesize
\setlength{\tabcolsep}{3pt}
\begin{tabular}{@{}l r ccc c@{}}
\toprule
Method & Net. & PSNR$\uparrow$ & SSIM$\uparrow$ & LPIPS$\downarrow$ & best \\
\midrule
\multicolumn{6}{@{}l}{\textit{(a) Real captures, held-out lights~(10 objects)}} \\
SSS-GS~\cite{dihlmann2024sssgs} & 7.1k & 33.79 & .633 & \textbf{.019} & 2 \\
SSD-GS~\cite{zheng2026ssdgs}    & 193k & 33.95 & \textbf{.692} & .032 & 1 \\
\lucent                         & 151k & \textbf{35.68} & .685 & .028 & \textbf{7} \\
\cmidrule(l){1-6}
\multicolumn{6}{@{}l}{\quad learned part removed~(retrained; \lucent: physics only)} \\
SSS-GS  & -- & 28.59 & .442 & .050 & \\
SSD-GS  & -- & \textbf{32.80} & \textbf{.658} & \textbf{.036} & \\
\lucent & -- & 30.11 & .499 & .054 & \\
\midrule
\multicolumn{6}{@{}l}{\textit{(b) Synthetic, held-out cameras and lights~(4 objects)}} \\
SSS-GS  & 7.1k & 36.95 & .872 & \textbf{.017} & 0 \\
SSD-GS  & 193k & \textbf{39.09} & \textbf{.922} & .021 & \textbf{3} \\
\lucent & 151k & 38.51 & .888 & .028 & 1 \\
\cmidrule(l){1-6}
\multicolumn{6}{@{}l}{\quad learned part removed} \\
SSS-GS  & -- & 29.72 & .723 & .058 & \\
SSD-GS  & -- & \textbf{32.51} & \textbf{.825} & \textbf{.039} & \\
\lucent & -- & 30.89 & .779 & .057 & \\
\bottomrule
\end{tabular}
\end{table}
}

\newcommand{\TabObjectsFull}{
\begin{table*}[!tbp]
\caption{\textbf{Per-object relighting under held-out lights, with and without the learned part}~(summarized in \cref{tab:objects}), on black: whole-image PSNR~(dB), SSIM over the object~(ground-truth mask eroded by $3$\,px) and LPIPS~(AlexNet, whole image); $256$ views per real object~($32$ cameras $\times$ $8$ held-out lights), $32$ per synthetic object~(one held-out light per held-out camera); every method trained on the same images.
\lucent ``full'' adds the $150.8$k-weight object residual~(\cref{sec:method:residual}, size selected on validation lights in \cref{tab:residual_sizes}) to the physics-only model; switching it off returns that model exactly.
``w/o network(s)'': the baseline retrained from scratch without its learned appearance networks~(\cref{sec:exp:share}).
Bold: best per metric among the full models and among the models without learned part.
``SSD-GS, refin.\ on'': SSD-GS retrained on the real captures with its pose and light refinement on, in its release's protocol~(\cref{sec:supp:calib}); it is left out of the bold marks.
The residual sees up to $64$ of the $57$--$84$ training lights per camera on the real captures and $56$ of $112$ on the synthetic objects.
On the joyful yell, five cameras that the dataset mis-registers~(by up to a head width) are left out of our training; every method is scored on the same views.
On the hair clip, the residual fit floors its per-image loss weights at $10\%$ of the physics error, since the default weighting diverged.}
\label{tab:objects_full}
\centering\scriptsize
\setlength{\tabcolsep}{1.55pt}
\resizebox{\textwidth}{!}{%
\begin{tabular}{@{}l ccc ccc ccc ccc ccc ccc ccc@{}}
\toprule
 & \multicolumn{12}{c}{Full model} & \multicolumn{9}{c}{Without the learned part} \\
\cmidrule(lr){2-13}\cmidrule(l){14-22}
 & \multicolumn{3}{c}{\lucent} & \multicolumn{3}{c}{SSS-GS~\cite{dihlmann2024sssgs}} & \multicolumn{3}{c}{SSD-GS~\cite{zheng2026ssdgs}} & \multicolumn{3}{c}{SSD-GS, refin.\ on} & \multicolumn{3}{c}{\lucent physics only} & \multicolumn{3}{c}{SSS-GS w/o network} & \multicolumn{3}{c}{SSD-GS w/o networks} \\
\cmidrule(lr){2-4}\cmidrule(lr){5-7}\cmidrule(lr){8-10}\cmidrule(lr){11-13}\cmidrule(lr){14-16}\cmidrule(lr){17-19}\cmidrule(l){20-22}
Object & PSNR$\uparrow$ & SSIM$\uparrow$ & LPIPS$\downarrow$ & PSNR$\uparrow$ & SSIM$\uparrow$ & LPIPS$\downarrow$ & PSNR$\uparrow$ & SSIM$\uparrow$ & LPIPS$\downarrow$ & PSNR$\uparrow$ & SSIM$\uparrow$ & LPIPS$\downarrow$ & PSNR$\uparrow$ & SSIM$\uparrow$ & LPIPS$\downarrow$ & PSNR$\uparrow$ & SSIM$\uparrow$ & LPIPS$\downarrow$ & PSNR$\uparrow$ & SSIM$\uparrow$ & LPIPS$\downarrow$ \\
\midrule
red car             & \textbf{35.93} & \textbf{.725} & .037 & 32.10 & .641 & \textbf{.028} & 32.69 & .704 & .050 & 32.00 & .683 & .057 & 30.42 & .583 & .070 & 28.57 & .533 & .064 & \textbf{31.85} & \textbf{.692} & \textbf{.052} \\
hair clip           & \textbf{38.48} & .618 & .011 & 36.98 & .545 & \textbf{.009} & 38.00 & \textbf{.637} & .012 & 37.38 & .606 & .015 & 33.98 & .332 & .025 & 33.89 & .407 & \textbf{.017} & \textbf{36.52} & \textbf{.571} & \textbf{.017} \\
blue bottle         & 33.70 & \textbf{.769} & .016 & \textbf{34.43} & .752 & \textbf{.009} & 33.51 & .768 & .017 & 34.22 & .781 & .016 & 26.55 & .615 & .043 & 22.46 & .314 & .068 & \textbf{33.10} & \textbf{.765} & \textbf{.018} \\
clear crystal       & 34.91 & .564 & .030 & 34.62 & .583 & \textbf{.016} & \textbf{35.36} & \textbf{.640} & .027 & 35.43 & .639 & .027 & 31.07 & .439 & .042 & 29.74 & .378 & .033 & \textbf{34.16} & \textbf{.611} & \textbf{.029} \\
marmalade  & \textbf{36.85} & \textbf{.709} & .027 & 33.32 & .575 & \textbf{.021} & 33.38 & .657 & .039 & 32.86 & .637 & .043 & 29.98 & .451 & .054 & 32.07 & .560 & \textbf{.032} & \textbf{32.17} & \textbf{.621} & .043 \\
small liqueur       & \textbf{34.82} & \textbf{.681} & .026 & 32.09 & .606 & \textbf{.020} & 32.80 & .672 & .030 & 33.05 & .674 & .029 & 28.71 & .406 & .064 & 27.18 & .374 & .063 & \textbf{32.27} & \textbf{.655} & \textbf{.032} \\
nail polish rem.    & \textbf{40.20} & .767 & .016 & 36.51 & .782 & \textbf{.009} & 36.98 & \textbf{.807} & .015 & 37.42 & .811 & .013 & 32.30 & .640 & .039 & 25.65 & .291 & .057 & \textbf{36.25} & \textbf{.799} & \textbf{.015} \\
massage ball~(red)  & \textbf{39.48} & \textbf{.728} & \textbf{.016} & 36.62 & .596 & .017 & 36.31 & .670 & .030 & 36.17 & .666 & .030 & 34.68 & .566 & .036 & 33.06 & .503 & \textbf{.033} & \textbf{35.55} & \textbf{.642} & \textbf{.033} \\
joyful yell         & \textbf{33.42} & .713 & .043 & 31.71 & .694 & \textbf{.029} & 31.33 & \textbf{.730} & .049 & 31.35 & .722 & .050 & 27.87 & .547 & .082 & 27.06 & .550 & .076 & \textbf{28.56} & \textbf{.659} & \textbf{.058} \\
physalis            & 29.02 & .575 & .059 & \textbf{29.52} & .560 & \textbf{.034} & 29.13 & \textbf{.634} & .051 & 29.30 & .627 & .055 & 25.49 & .415 & .083 & 26.18 & .508 & \textbf{.059} & \textbf{27.61} & \textbf{.566} & .060 \\
\midrule
soap~(synth.)       & \textbf{42.63} & .948 & .018 & 37.01 & .921 & .012 & 42.28 & \textbf{.972} & \textbf{.007} & -- & -- & -- & 31.40 & .874 & .062 & 28.22 & .783 & .081 & \textbf{35.27} & \textbf{.911} & \textbf{.030} \\
candle~(synth.)     & 40.33 & .912 & .015 & 40.19 & .886 & \textbf{.008} & \textbf{41.95} & \textbf{.933} & \textbf{.008} & -- & -- & -- & 30.76 & .737 & .043 & 29.47 & .687 & .044 & \textbf{33.32} & \textbf{.821} & \textbf{.023} \\
dragon~(synth.)     & 35.09 & .830 & .043 & 34.56 & .804 & \textbf{.034} & \textbf{35.59} & \textbf{.877} & .043 & -- & -- & -- & 31.31 & .760 & .065 & 29.47 & .646 & .065 & \textbf{31.84} & \textbf{.777} & \textbf{.061} \\
statue~(synth.)     & 35.99 & .863 & .035 & 36.04 & .876 & \textbf{.016} & \textbf{36.52} & \textbf{.904} & .024 & -- & -- & -- & 30.07 & .745 & .057 & \textbf{31.72} & .777 & \textbf{.041} & 29.59 & \textbf{.792} & \textbf{.041} \\
\midrule
mean, real~(10)     & \textbf{35.68} & .685 & .028 & 33.79 & .633 & \textbf{.019} & 33.95 & \textbf{.692} & .032 & 33.92 & .685 & .033 & 30.11 & .499 & .054 & 28.59 & .442 & .050 & \textbf{32.80} & \textbf{.658} & \textbf{.036} \\
mean, synthetic~(4) & 38.51 & .888 & .028 & 36.95 & .872 & \textbf{.017} & \textbf{39.09} & \textbf{.922} & .021 & -- & -- & -- & 30.89 & .779 & .057 & 29.72 & .723 & .058 & \textbf{32.51} & \textbf{.825} & \textbf{.039} \\
\bottomrule
\end{tabular}}
\end{table*}
}

\newcommand{\FigQual}{
\begin{figure*}[t]
  \centering
  \includegraphics[width=\textwidth,trim=0 188.40 0 0,clip]{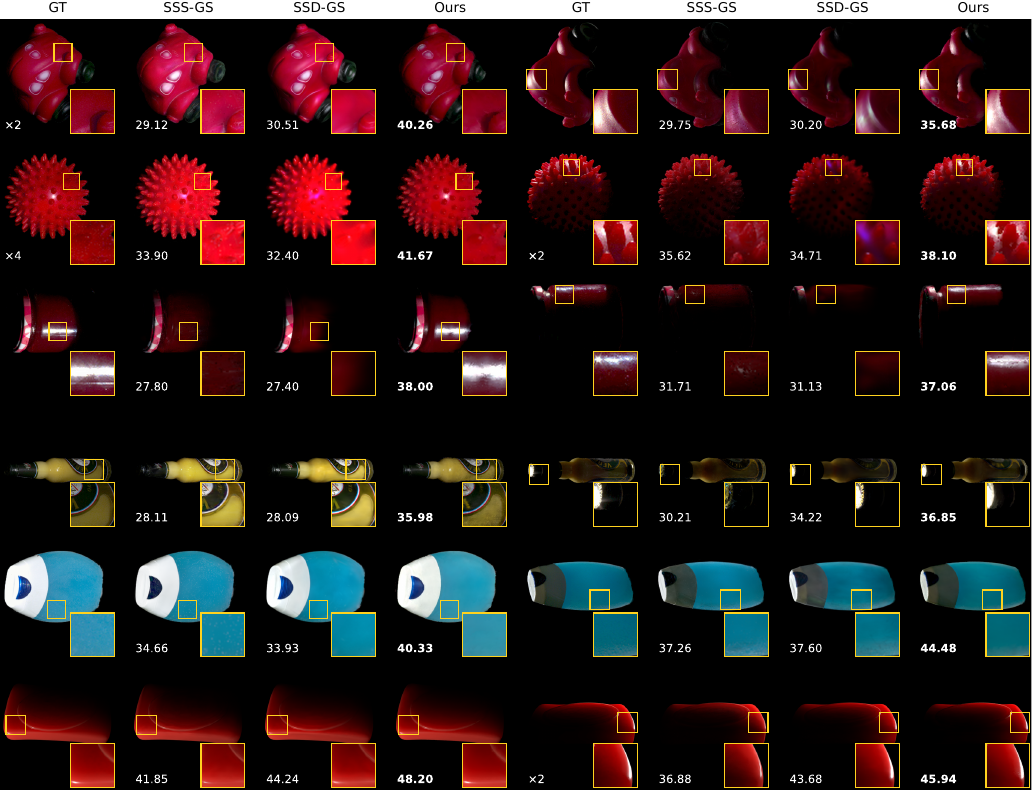}
  \caption{\textbf{Qualitative comparison under held-out lights.} Car, massage ball and marmelade, each in one front-lit~(left) and one back-lit~(right) view.
  The views are selected among those where \lucent leads both baselines; \cref{tab:objects} scores all views, and the supplementary material shows more objects.
  ``Ours'' is the full model, and insets magnify the boxed region.
  PSNR~(dB) on black, best in bold.
  Dark views are brightened by the factor shown on the GT, after computing PSNR.}

  \label{fig:qual}
\end{figure*}
}

\newcommand{\FigDecomposition}{
\begin{figure*}[t]
  \centering
  \includegraphics[width=\textwidth,trim=0 111 0 0,clip]{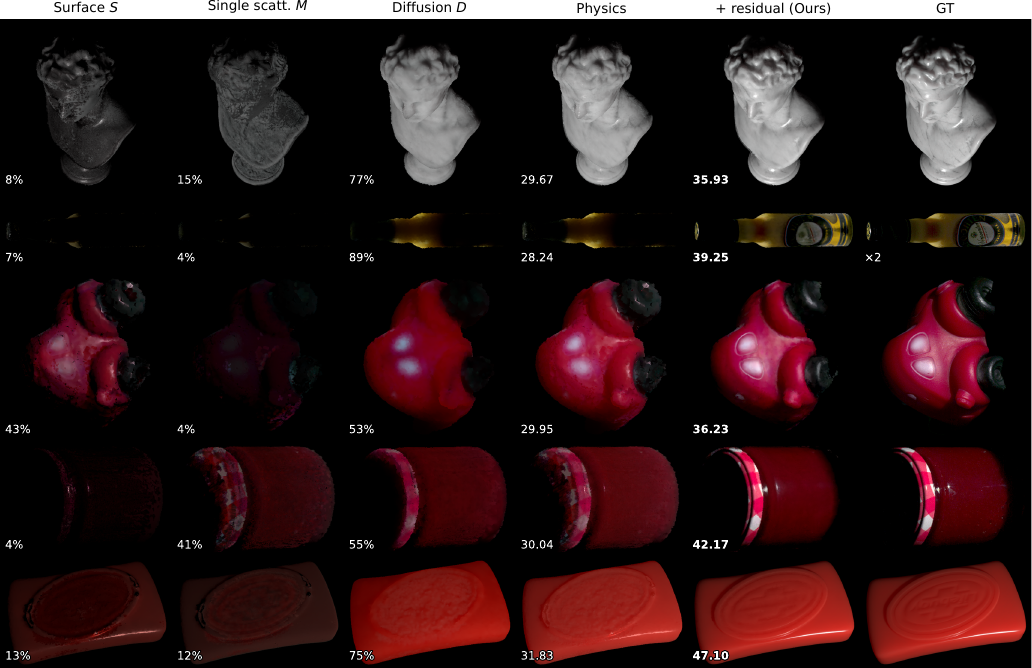}
  \caption{\textbf{What each part of \lucent renders}, on selected held-out views: the synthetic statue and the real liqueur and car~(more rows in the supplementary material).
  The physics is the sum of the surface $S$, single scattering $M$ and diffusion $D$; percentages give each term's share of its light.
  The number on $D$ is the PSNR~(dB) of the physics $S{+}M{+}D$, and the number on Ours that of the full model.
  Rows share one tone mapping~(liqueur $2\times$).
  On the statue the residual mainly corrects brightness and shading; on the real objects it also adds what one homogeneous medium cannot hold, such as the liqueur's label and the car's windows and wheels.}
  \label{fig:decomposition}
\end{figure*}
}

\newcommand{\FigLearned}{
\begin{figure}[t]
  \centering
  \includegraphics[width=0.94\linewidth]{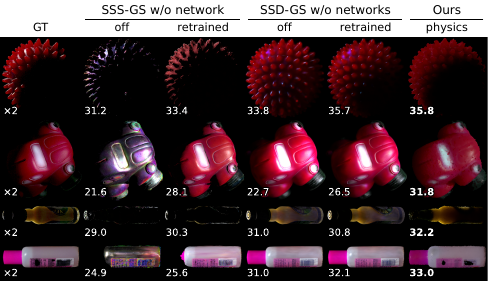}
  \caption{\textbf{What each method's learned part carries}, under held-out lights: baselines with their networks off or retrained without them, \lucent with its residual off.
  PSNR~(dB), best in bold; $\times k$: display gain.
  Without its network, SSS-GS loses the object's color and SSD-GS turns over-bright; \lucent keeps shape, color and transmitted light.}
  \label{fig:learned_share}
\end{figure}
}

\newcommand{\FigEditing}{
\begin{figure*}[t]
  \centering
  \includegraphics[width=\linewidth]{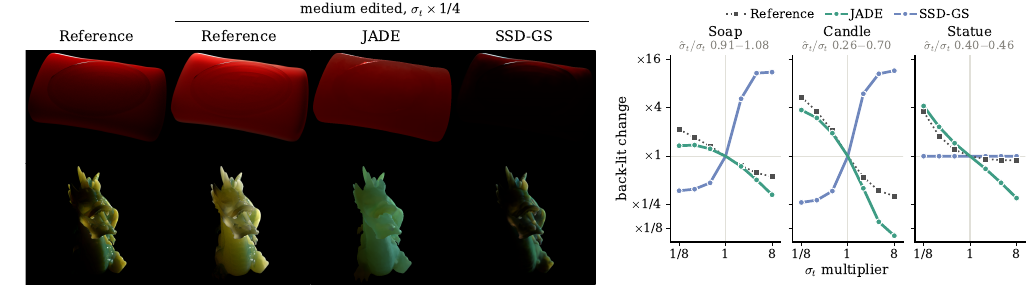}
  \caption{\textbf{Editing the medium.}
  Only \lucent has a medium to edit; SSS-GS and SSD-GS bake the subsurface light into networks and per-Gaussian parameters.
  Left: back-lit views at $\boldsymbol{\sigma}_t\times\frac14$, without retraining~(\lucent: physics only; SSD-GS: the $\sigma_s'$ and $\sigma_a$ predicted by its network, scaled alike).
  The dragon reference is rendered in a re-created scene, since the dataset has no Blender file for it.
  Right: back-lit change against the $\boldsymbol{\sigma}_t$ multiplier on the synthetic objects; gray: recovered over true $\boldsymbol{\sigma}_t$.
  \lucent moves with the Blender reference, with an overshoot on the statue; SSD-GS moves the opposite way or not at all.
  Surface and interior edits are in the supplementary material.}
  \label{fig:editing}
\end{figure*}
}

\newcommand{\TabSigma}{
\begin{table}[t]\centering\footnotesize\setlength{\tabcolsep}{5pt}
\caption{\textbf{Recovered extinction} against the true one~(Blender, per scene unit, RGB).}
\label{tab:sigmagt_full}

\begin{tabular}{l ccc ccc}
\toprule
 & \multicolumn{3}{c}{true $\boldsymbol{\sigma}_t$} & \multicolumn{3}{c}{recovered / true} \\
\midrule
Soap   & 12.6 & 62.8 & 62.8 & 0.91 & 0.97 & 1.08 \\
Candle & 35.9 & 120  & 359  & 0.70 & 0.51 & 0.26 \\
Statue & 126  & 157  & 157  & 0.46 & 0.40 & 0.40 \\
\bottomrule
\end{tabular}
\end{table}
}

\newcommand{\TabPriors}{
\begin{table}[!tbp]
\caption{\textbf{Monocular priors of the geometry stage}, candle, $11$ held-out views, geometry stage without $\mathcal{L}_{\text{sil}}$.}
\label{tab:priors}
\centering\small
\setlength{\tabcolsep}{3pt}
\resizebox{\linewidth}{!}{
\begin{tabular}{@{}l cccc@{}}
\toprule
 & Normal MAE$\downarrow$ & PSNR$\uparrow$ & SSIM$\uparrow$ & LPIPS$\downarrow$ \\
\midrule
SDF                                     & 21.40 & 17.71 & 0.759 & 0.258 \\
\quad + normal prior                    & 16.29 & 19.18 & 0.790 & 0.206 \\
\quad + depth prior                     & 20.42 & 18.91 & 0.747 & 0.213 \\
\quad + normal + depth prior~(\lucent)  & \textbf{14.56} & \textbf{21.41} & \textbf{0.806} & \textbf{0.164} \\
\bottomrule
\end{tabular}}
\end{table}
}

\newcommand{\TabAblation}{
\begin{table}[t]
\caption{\textbf{Ablation of the transport}: diffusion against more scattering orders, physics only, same frozen geometry and surface~(10k iterations).
Object PSNR on $16$ front-, $4$ side- and $12$ back-lit test views; s/iter on one A100.
Best in bold, second underlined.}
\label{tab:ablation}
\centering\small
\setlength{\tabcolsep}{3pt}
\newlength{\ablW}\setlength{\ablW}{1.42\linewidth}
\resizebox{\linewidth}{!}{
\begin{tabular}{@{}l@{}}
\begin{tabular*}{\ablW}{@{\extracolsep{\fill}}l cccccc@{}}
\toprule
\multicolumn{7}{@{}l}{\textit{Light transport}: synthetic candle, physics only, object PSNR~(dB)} \\
 & Front-lit$\uparrow$ & Back-lit$\uparrow$ & All$\uparrow$ & SSIM$\uparrow$ & LPIPS$\downarrow$ & s/iter \\
\midrule
Single scatter only~($K{=}1$)           & 19.40 & 18.59 & 18.88 & 0.618 & 0.063 & 0.62 \\
$K{=}2$ octaves                         & \underline{19.55} & \underline{20.85} & \underline{20.03} & \underline{0.671} & \underline{0.054} & 0.63 \\
$K{=}4$ octaves                         & 17.92 & 18.73 & 18.12 & 0.466 & 0.074 & 0.56 \\
$K{=}8$ octaves                         & 16.98 & 18.16 & 17.38 & 0.440 & 0.082 & 0.56 \\
$K{=}1$ + diffusion~(\lucent)           & \textbf{20.06} & \textbf{21.97} & \textbf{20.72} & \textbf{0.718} & \textbf{0.046} & 0.92 \\
\bottomrule
\end{tabular*}
\end{tabular}}
\end{table}
}

\newcommand{\TabCapabilities}{
\begin{table}[t]
\caption{\textbf{Capabilities} of relightable translucent reconstruction methods.
\cmark: yes; \xmark: no; --: not applicable~(no volumetric transport); grid: an auxiliary voxel grid; dipole: an analytic dipole profile instead of transport through a volume.
$^\dagger$Evaluated per Gaussian, with scattering and absorption predicted by a network from the light and view directions.
$^*$Concurrent work.}
\label{tab:capabilities}
\centering\small
\setlength{\tabcolsep}{3pt}
\resizebox{\linewidth}{!}{
\begin{tabular}{l ccccc}
\toprule
 & \shortstack{Closed\\surface} & \shortstack{Interior\\volume} & \shortstack{Simulated\\transport} & \shortstack{Medium indep.\\of light, view} & \shortstack{No auxiliary\\grid} \\
\midrule
SSS-GS~\cite{dihlmann2024sssgs}      & \xmark & \xmark & \xmark & \xmark & -- \\
SSD-GS~\cite{zheng2026ssdgs}        & \xmark & \xmark & dipole$^\dagger$ & \xmark & -- \\
GS$^3$~\cite{bi2024gs3}              & \xmark & \xmark & \xmark & \xmark & -- \\
VolS-GS~\cite{ahn2026volsgs}$^*$     & \xmark & grid & \cmark & \xmark & \xmark \\
Deng~\etal~\cite{deng2022translucent} & \cmark & \xmark & dipole & \cmark & -- \\
\textbf{\lucent}                     & \cmark & \cmark & \cmark & \cmark & \cmark \\
\bottomrule
\end{tabular}}
\end{table}
}

\newcommand{\FigEditCandle}{
\begin{figure}[t]
\centering
\setlength{\tabcolsep}{0.5pt}\renewcommand{\arraystretch}{0.4}\scriptsize
\begin{tabular}{@{}c@{\hspace{2pt}}cccc@{}}
 & recovered & $\boldsymbol{\sigma}_t\times2$ & $\boldsymbol{\sigma}_t\times\frac14$ & \shortstack{$\boldsymbol{\sigma}_t$ per\\channel} \\
\rotatebox{90}{\makebox[0.395\linewidth][c]{front light}} & \includegraphics[width=0.205\linewidth]{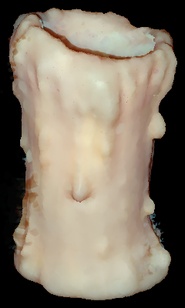} & \includegraphics[width=0.205\linewidth]{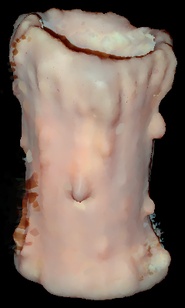} & \includegraphics[width=0.205\linewidth]{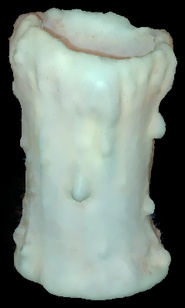} & \includegraphics[width=0.205\linewidth]{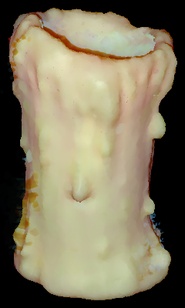} \\
\rotatebox{90}{\makebox[0.395\linewidth][c]{back light}} & \includegraphics[width=0.205\linewidth]{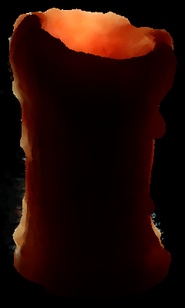} & \includegraphics[width=0.205\linewidth]{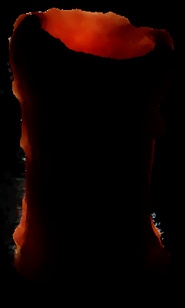} & \includegraphics[width=0.205\linewidth]{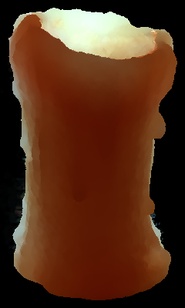} & \includegraphics[width=0.205\linewidth]{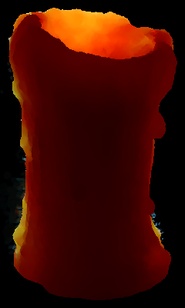} \\
\end{tabular}
\caption{\textbf{Editing the medium of the candle without retraining}, physics only.
Scaling the extinction $\boldsymbol{\sigma}_t$ changes how much light passes through the wax~(per channel: $\times(\frac12,1,2)$ in RGB).}
\label{fig:editcandle}

\end{figure}
}

\newcommand{\TabEditPattern}{
\begin{table}[t]\centering\small\setlength{\tabcolsep}{3pt}
\caption{\textbf{Recovered medium and edits} on the synthetic objects: recovered over true extinction~(RGB), and object PSNR~(dB) against the edited Blender truth, mean over four extinction edits, of our unedited physics, the best global change~(unedited truth, one scale per channel and view) and ours~(unedited truth plus our change, scaled alike); cosine between our change and the true one.}
\label{tab:editpattern}\label{tab:sigmagt}
\resizebox{\linewidth}{!}{
\begin{tabular}{l c l ccc c}
\toprule
 & $\boldsymbol{\sigma}_t$ recovered / true & & Unedited & Uniform & Ours & Cosine \\
\midrule
Soap   & \multirow{2}{*}{0.91 / 0.97 / 1.08} & front-lit & 27.24 & \textbf{35.13} & 33.89 & 0.88 \\
       &                                     & back-lit  & 24.31 & 29.51 & \textbf{30.01} & 0.31 \\
Candle & \multirow{2}{*}{0.70 / 0.51 / 0.26} & front-lit & 20.56 & 26.69 & \textbf{29.18} & 0.33 \\
       &                                     & back-lit  & 24.32 & 21.88 & \textbf{27.22} & 0.77 \\
Statue & \multirow{2}{*}{0.46 / 0.40 / 0.40} & front-lit & 22.67 & 31.52 & \textbf{32.18} & 0.17 \\
       &                                     & back-lit  & 26.66 & 30.77 & \textbf{31.07} & 0.15 \\
\bottomrule
\end{tabular}}
\end{table}
}

\definecolor{cvprblue}{rgb}{0.21,0.49,0.74}
\usepackage[pagebackref,breaklinks,colorlinks,allcolors=cvprblue]{hyperref}

\newcommand{\projectpage}{\url{https://hectorandrade.xyz/jade/}}

\title{\lucent: Translucency Needs a Medium\\[3pt]{\large Editable Subsurface Scattering on Radiant Foam}}

\author{
Guillermo Figueroa-Araneda$^{1,*}$ \qquad
H\'ector Andrade-Loarca$^{2,3,*}$ \qquad
Shrisudhan Govindarajan$^{4}$\\
Manny Ko$^{5}$ \qquad
Andrea Tagliasacchi$^{4,6}$ \qquad
Daniel Cremers$^{2,3}$
}

\newcommand{\jadeaffiliations}{%
$^{1}$Centro Nacional de Inteligencia Artificial (CENIA) \quad
$^{2}$Technical University of Munich \qquad
$^{*}$Equal contribution\\
$^{3}$Munich Center for Machine Learning \quad
$^{4}$Simon Fraser University \quad
$^{5}$Independent Researcher \quad
$^{6}$University of Toronto\\
{\footnotesize\tt \{guillermo.figueroa, hector.andrade, cremers\}@tum.de \quad shrisudhan07@gmail.com \quad ataiya@sfu.ca}}
\makeatletter
\def\@maketitle{%
   \newpage
   \null
   \vskip .05in
   \begin{center}
      {\Large \bf \@title \par}
      \vspace*{8pt}{%
        \large
        \lineskip .5em
        \begin{tabular}[t]{c}\@author\end{tabular}%
        \par
      }%
      \vspace*{3pt}%
      {\small \jadeaffiliations \par}%
      \vspace*{4pt}%
   \end{center}
}
\makeatother

\begin{document}
\twocolumn[{
\renewcommand\twocolumn[1][]{#1}
\maketitle
\begin{center}
  \captionsetup{type=figure}
  \includegraphics[width=\textwidth]{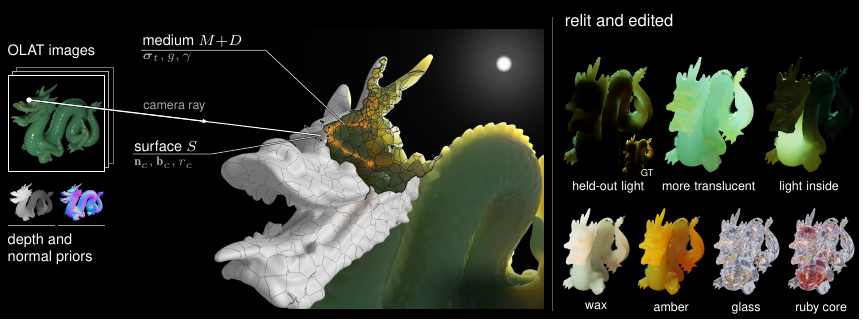}
  \caption{\lucent reconstructs a translucent object as a single radiant foam~(left; a quarter of the head is cut away, beside a back-lit render).
  A signed distance field splits its cells into a closed surface, which reflects light, and an interior medium, in which we simulate how light scatters and diffuses instead of learning it for each light.
  Because the medium is described by a few optical constants, the reconstruction can be relit and edited without retraining~(right, physics only): a held-out back light~(GT inset), a more translucent medium, a light placed inside, and new materials, with refraction added at render time for glass.
  We name our method after jade, a stone that glows because light scatters beneath its surface, the material the dragon is made of.}
  \label{fig:teaser}
\end{center}
}]

\begin{abstract}

Editing a captured object's translucency requires modeling how light travels inside it.
Relightable Gaussian splatting methods learn subsurface appearance without recovering a light- and view-independent interior.
We propose \lucent{}, which reconstructs a reflecting surface enclosing an editable scattering medium from multi-view, multi-light images.
We simulate light transport in the volumetric representation that renders it, a Radiant Foam.
Optical constants independent of light and view let us edit translucency without retraining.
After fixing the physical model, a learned correction captures unexplained appearance.
Under unseen lights, \lucent{} relights real captures as accurately as state-of-the-art Gaussian methods, and unlike them, its medium is editable: edits of the recovered extinction follow reference renders, while scaling the scattering parameters of a Gaussian baseline does not.

\end{abstract}

\section{Introduction}
\label{sec:intro}

Relighting a translucent object requires recovering how light reflects at its surface and scatters through its interior~\cite{deng2022translucent}.
Subsurface scattering spreads light beyond its entry point and gives materials such as marble and skin their characteristic appearance~\cite{jensen2001bssrdf}.
To edit that appearance after capture, a reconstruction must describe how the material \textit{transports} light.
The back-lit dragon and material edits in \Cref{fig:teaser} illustrate the control we seek.

Relightable variants of 3D Gaussian splatting~(3DGS), such as SSS-GS~\cite{dihlmann2024sssgs} and SSD-GS~\cite{zheng2026ssdgs}, reproduce translucent appearance under new lights.
They learn the subsurface radiance, or the scattering parameters used to compute it, from the light and view directions.
However, their Gaussians \textit{do not enclose a medium} through which light can propagate.
They thus fit the appearance of subsurface scattering without recovering optical constants that describe an interior independently of the lighting and viewpoint~(\cref{tab:capabilities}).

We propose \lucent{}, which reconstructs a translucent object from multi-view, multi-light images as a surface enclosing an editable scattering medium.
Our core idea is to \textit{simulate light transport} on the same volumetric representation that renders the object.
We build on Radiant Foam~\cite{govindarajan2025radfoam}, which partitions space into Voronoi cells with known neighbors and shared faces.
This partition gives us both paths to trace through the object and connections across which light can spread.
To distinguish the surface from the interior, we fit a signed distance field.
The faces separating inside and outside cells form the reflecting surface, while the inside cells form the scattering medium.

We \textit{recover optical constants} that describe light transport independently of lighting and viewpoint.
To render light that scatters once, we trace its path through the cells.
To approximate repeated scattering, we solve the light diffusion equation across neighboring cells, with one sparse linear system per light.
Each physical render then separates surface reflection, single scattering, and multiple scattering, as shown in \Cref{fig:decomposition}.
We can edit the recovered optical constants and solve again without retraining~(\cref{fig:editing}).

To retain this control while fitting details beyond the physical model, we learn an image residual after fixing the recovered surface and medium.
Sharing optical constants across the object makes the inverse problem better constrained, but leaves out material variations.
Our model also approximates repeated scattering and omits refraction during reconstruction.
We fit a per-object multilayer perceptron~(MLP) to account for these discrepancies, including printed patterns finer than the cells, variations of the material that one homogeneous medium cannot hold, and capture effects such as glare.
Switching off the residual exposes the physical render without changing the recovered medium.

Our experiments show that the recovered medium can be edited.
When we change its extinction, the rendered brightness moves with reference renders generated in Blender, and the predicted change captures their spatial pattern, while scaling the scattering parameters predicted by SSD-GS does not~(\cref{sec:exp:editing}).
This editability does not cost relighting accuracy: under held-out lights, \lucent achieves the highest mean peak signal-to-noise ratio among the compared methods on the real captures~(\cref{tab:objects}).
The residual uses fewer network weights than SSD-GS, leaves the medium unchanged and improves views from cameras unseen during its training~(\cref{sec:exp:share}).

\paragraph{Contributions.}
To summarize, our contributions are:
\begin{enumerate*}[label=(\roman*), itemjoin={{ }}]
\item we formulate surface reflection and volumetric scattering on one reconstructed foam, without an auxiliary transport grid;
\item we separate an editable medium, described by view- and light-independent optical constants, from a residual fitted with the physical model fixed; and
\item we show that edits of the recovered medium follow reference renders, at a relighting accuracy on par with state-of-the-art methods, and measure how much of each method's image its learned part carries.
\end{enumerate*}

\section{Related work}
\label{sec:related}

Radiant Foam~\cite{govindarajan2025radfoam} replaces Gaussian splats~\cite{kerbl20233dgs} with a Voronoi tessellation.
It renders by walking the Delaunay adjacency of the cells, which gives exact ray segments and differentiable ray tracing at rasterization speed.
Power Foam~\cite{govindarajan2026powerfoam} generalizes the cells to power diagrams, and SDFoam~\cite{rech2025sdfoam} couples them to a signed distance field.
These works use the adjacency to \emph{render}; we also use it to \emph{transport} light.

\TabCapabilities
\paragraph{Relightable translucent reconstruction.}
Neural scattering functions~\cite{yu2023osf} and relightable Gaussians~\cite{bi2024gs3,fan2025rng} learn light-dependent appearance directly.
SSS-GS~\cite{dihlmann2024sssgs} pairs a physically based surface with a learned subsurface residual conditioned on light and view.
SSD-GS~\cite{zheng2026ssdgs} evaluates a dipole per Gaussian, with a learned shadow term and with scattering and absorption predicted by a network from the light and view directions.
In these methods, the subsurface light, or the parameters that produce it, is the output of a network conditioned on the light and view.
GTSR~\cite{gtsr2026} targets translucent surface reconstruction, and volumetric ray-traced primitives~\cite{condor2025dsyg} render scattering media but do not recover them from images.\footnote{Concurrent with our work, VolS-GS~\cite{ahn2026volsgs} solves a finite-volume diffusion equation with an adjoint on a $64^3$ grid rasterized from a Gaussian scene, while a network conditioned on the light predicts scattering and absorption per Gaussian.}
In \lucent, the medium is a few constants on the same cells that render the object.
They do not depend on light or view, which makes the medium editable~(\cref{tab:capabilities}).

\paragraph{Physically based inverse subsurface scattering.}
Deng~\etal~\cite{deng2022translucent} fit a heterogeneous dipole BSSRDF on a mesh with differentiable rendering, the closest physically based precedent.
Volumetric path tracing~\cite{zhang2021pathspace} is exact but needs many samples per pixel.
Diffusion has been solved on a polygrid~\cite{wang2008polygrid} and with finite elements~\cite{arbree2011fem}, and neural radiative transfer~\cite{nsampi2026nrt} learns the operator itself.
The error of the dipole in thin geometry~\cite{dipolethin2026} motivates volumetric transport.
We solve diffusion on the cells of the reconstructed radiance field, with no mesh template or interior scaffold.
As for surfaces~\cite{gao2023relightable3dgs,kouros2026specgloss}, the parameterization decides what is identifiable: a global extinction is, a per-cell one is not.

\paragraph{Generative priors for geometry.}
Diffusion models fine-tuned for dense prediction~\cite{ke2024marigold,pavlovic2026marigoldv2,garcia2025e2eft} give depth and normals that regularize implicit surfaces~\cite{yu2022monosdf}, usually for sparse views.
We use them because photometric normals fail at highlights and glow.

\section{Method}
\label{sec:method}

We recover geometry, surface reflectance and subsurface scattering parameters from multi-view, multi-light images.
Our key idea is to use one volumetric foam twice: as the support of the object's surface, and as the medium in which we solve subsurface light transport.
We first reconstruct the geometry with a signed distance field supervised by monocular priors, which partitions the foam into surface and interior cells~(\cref{sec:method:geo}).
We then fit the appearance with a physically based model of surface reflection, single scattering and diffusion, all evaluated on the foam cells~(\crefrange{sec:method:relight}{sec:method:transport}).
Once the physics has converged, a small MLP corrects what the homogeneous medium cannot explain~(\cref{sec:method:residual}).
\Cref{fig:teaser}~(left) shows the surface and the medium on the foam.
Project page with video and code: \projectpage.

\subsection{Preliminaries: Radiant Foam}
\label{sec:method:prelim}
Radiant Foam~\cite{govindarajan2025radfoam} is a differentiable ray-traced representation: a Voronoi diagram of $N$ cells with sites $\mathbf{x}_i{\in}\mathbb{R}^3$, densities $\sigma_i{\ge}0$ and view-dependent colors $\mathbf{c}_i$.
Rays traverse the cells through the dual Delaunay adjacency.
The pixel color is the exact volume rendering of the traversed segments and is differentiable with respect to the sites:
\begin{equation}
  \begin{gathered}
  C(\mathbf{r}) = \sum_{k} T_k \big(1 - e^{-\sigma_{i_k}\delta_k}\big)\, \mathbf{c}_{i_k}(\mathbf{d}),\\
  T_k = \exp\Big(-\sum_{j<k} \sigma_{i_j}\delta_j\Big),
  \end{gathered}
  \label{eq:radfoam}
\end{equation}
where $i_k$ is the $k$-th cell the ray $\mathbf{r}$~(direction $\mathbf{d}$) traverses and $\delta_k$ the length of its segment in that cell.
We reuse these cells and their adjacency for light transport~(\cref{sec:method:transport}).

\subsection{Preliminaries: light transport}
The radiance leaving a translucent object is the sum of the light reflected at its interface, $L_{\text{surf}}$, and the light that enters the medium, scatters and exits, $L_{\text{sss}}$~\cite{jensen2001bssrdf,deng2022translucent}.
We shade the interface with a rough dielectric microfacet BRDF~\cite{walter2007ggx},
\begin{equation}
  f_r=\frac{\mathbf{b}}{\pi}+\frac{D_{\text{GGX}}\,G_{\text{Smith}}\,F_{\text{Schlick}}}{4\,(\mathbf{n}\cdot\mathbf{v})_+(\mathbf{n}\cdot\boldsymbol{\ell})_+},
  \label{eq:brdf}
\end{equation}
with normal $\mathbf{n}$, view and light directions $\mathbf{v}$ and $\boldsymbol{\ell}$, base color $\mathbf{b}{\in}[0,1]^3$, GGX width $r^2$ for roughness $r{\in}[0,1]$, and $F_0{=}0.04$.
Inside, the medium has scattering and absorption coefficients $\sigma_s$ and $\sigma_a$, extinction $\sigma_t{=}\sigma_s{+}\sigma_a$, albedo $\alpha_{\text{ss}}{=}\sigma_s/\sigma_t$ and phase anisotropy $g$, and attenuates light by $T{=}\exp(-\int\sigma_t\,dt)$.
$L_{\text{sss}}$ splits into single scattering $L^{(1)}$, light scattered once along the view ray, and multiple scattering $L^{\text{ms}}$~\cite{hanrahan1993,jensen2001bssrdf}, which we approximate by the diffusion of the fluence $\phi$~\cite{stam1995}:
\begin{equation}
  -\nabla\cdot(\kappa\nabla\phi) + \sigma_a\phi = Q, \quad
  \kappa = {1}/{(3\sigma_t')},
  \label{eq:diffusion}
\end{equation}
where $\sigma_t'{=}\sigma_a+(1-g)\,\sigma_s$ is the reduced extinction and the source $Q$ is the singly scattered light.
$L^{\text{ms}}$ is read out from $\phi$ at the boundary through a Fresnel boundary condition.
We solve both terms on the foam~(\cref{sec:method:transport}).

\begin{figure}[t]
  \centering
  \setlength{\unitlength}{0.49\linewidth}
  \begin{tabular}{@{}c@{\hspace{0.02\linewidth}}c@{}}
    \begin{picture}(1,0.7667)
      \put(0,0){\includegraphics[width=\unitlength]{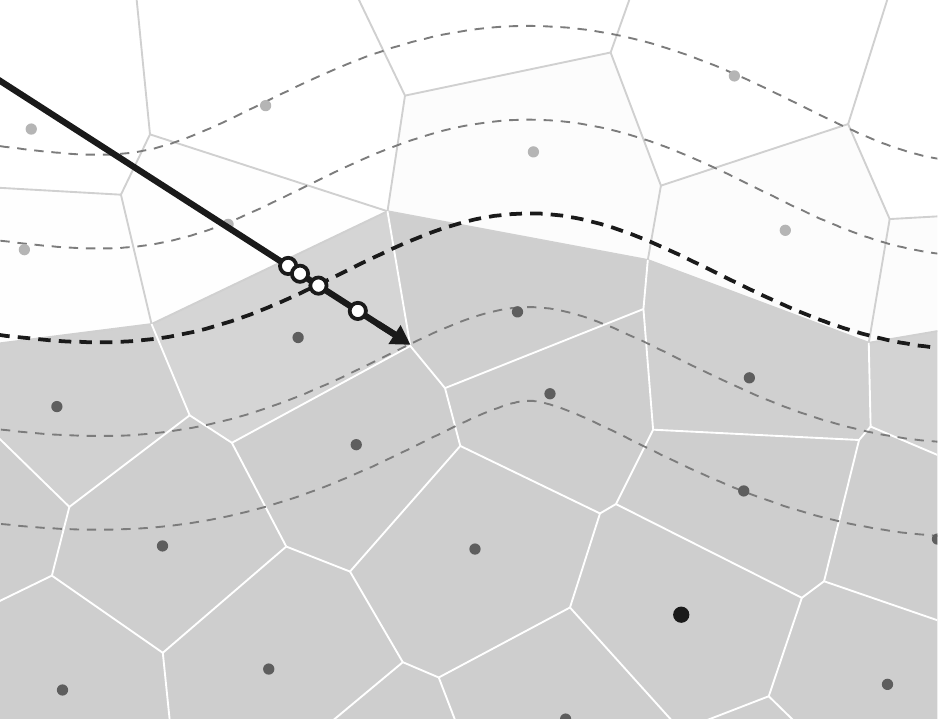}}
      \input{draft/fig3_sdf_foam/labels_a.tex}
    \end{picture} &
    \begin{picture}(1,0.7667)
      \put(0,0){\includegraphics[width=\unitlength]{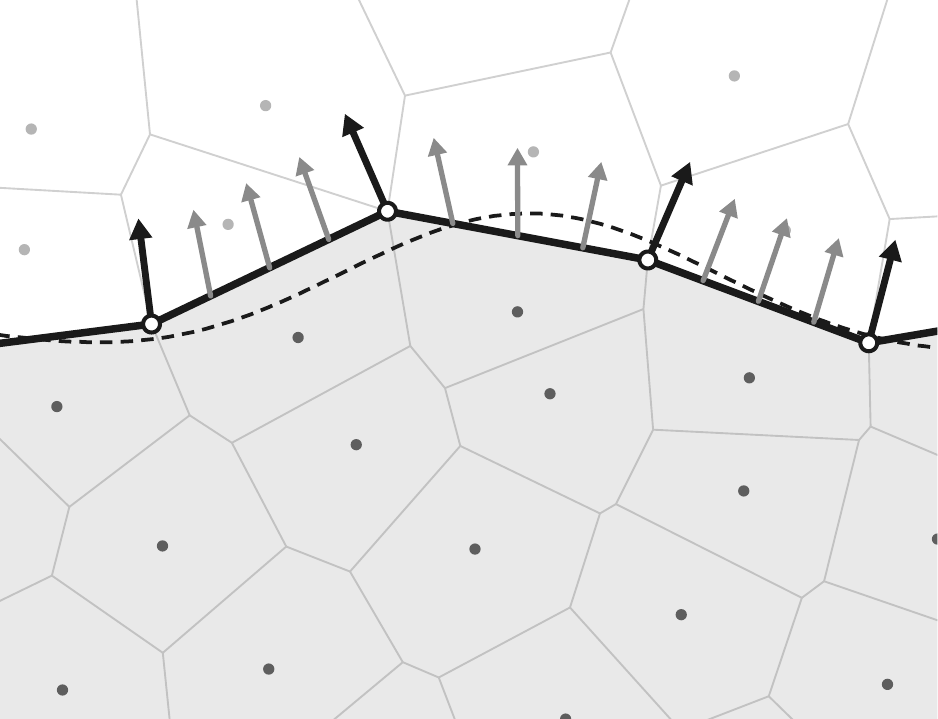}}
      \input{draft/fig3_sdf_foam/labels_b.tex}
    \end{picture} \\[1pt]
    {\small (left) Density} & {\small (right) Surface}
  \end{tabular}
  \caption{\textbf{The SDF in the foam.}
  (left) $f_\theta$~(dashed level sets), sampled at each site $\mathbf{x}_i$, sets one density $\sigma_i$ per cell~(shading); the depths $t_k$ of a camera ray gather where it becomes opaque.
  (right) The walls between inside and outside cells form the surface.
  Each corner takes a normal $\mathbf{n}_c$ from $\nabla f_\theta$~(black); $\mathbf{n}(\mathbf{x})$ is their interpolation~(gray).
  }
  \label{fig:foamroles}
\end{figure}

\subsection{Geometry}
\label{sec:method:geo}
Transport needs a closed surface with reliable normals and a connected interior.
A density fitted to the images gives neither: opacity spreads over several boundary cells, and cells behind the first opaque layer receive almost no gradient.
Following MonoSDF~\cite{yu2022monosdf}, we therefore drive the foam densities with a signed distance field supervised by monocular cues; in this stage the cell colors stay view-dependent as in \cref{eq:radfoam}.

\paragraph{SDF-driven density.}
An MLP $f_\theta:\mathbb{R}^3\to\mathbb{R}$, whose zero level set is the object surface, is evaluated at each site.
It uses the NeuS architecture~\cite{wang2021neus}, as in SDFoam~\cite{rech2025sdfoam}, and the cell density follows VolSDF~\cite{volsdf}: $\sigma_i {=} \beta^{-1}\Psi_\beta(-f_\theta(\mathbf{x}_i))$, where $\Psi_\beta$ is the Laplace CDF with learnable scale $\beta$.
Since $f_\theta$ is continuous, the surface normal is available anywhere as its normalized gradient.
For each ray, we take the $K$ depths $\{t_k\}$ at which the transmittance of \cref{eq:radfoam} crosses stratified levels, and render depth and normal as $\hat{Z}{=}\frac{1}{K}\sum_k t_k$ and $\hat{N}{=}\frac{1}{K}\sum_k\nabla f_\theta(\mathbf{r}(t_k))$, where $\mathbf{r}(t)$ is the ray.
Since the $K$ levels split the ray's weight into equal parts, $\hat{Z}$ is the expected termination depth of a ray that becomes opaque.

\paragraph{Monocular priors.}
We supervise $\hat{Z}$ and $\hat{N}$ with the depth and normal maps $\bar{Z}$ and $\bar{N}$ predicted by Marigold~\cite{ke2024marigold}.
The depth loss is $\mathcal{L}_d{=}\sum_{\mathbf{r}}\big(w\log\hat{Z}+q-\bar{Z}\big)^2$, with a per-image scale $w$ and shift $q$ fitted by least squares since the predicted depth has no absolute scale, plus a gradient-matching term as in MonoSDF~\cite{yu2022monosdf}.
The normal loss is $\mathcal{L}_n{=}\sum_{\mathbf{r}}\|\hat{N}-\bar{N}\|_1+\big(1-\hat{N}^\top\bar{N}\big)$ on normalized camera-space normals.
The geometry stage minimizes $\mathcal{L}_{\text{geo}}{=}\mathcal{L}_{\text{rgb}}+\lambda_d\mathcal{L}_d+\lambda_n\mathcal{L}_n+\lambda_{\text{eik}}\mathcal{L}_{\text{eik}}+\lambda_s\mathcal{L}_s+\lambda_O\mathcal{L}_O+\lambda_{\text{sil}}\mathcal{L}_{\text{sil}}$, where $\mathcal{L}_{\text{rgb}}$ is an $L_1$ photometric loss, $\mathcal{L}_{\text{eik}}$ the Eikonal loss~\cite{gropp2020igr}, $\mathcal{L}_s$ a normal smoothness term~\cite{oechsle2021unisurf}, $\mathcal{L}_O$ an $L_2$ loss between the accumulated opacity $O$ and the object mask, and $\mathcal{L}_{\text{sil}}$ a silhouette loss as in IDR~\cite{yariv2020idr}: a hinge on the minimum of $f_\theta$ along each ray, negative inside the mask and positive outside, so that the masks constrain the zero level set and not only the density.

\paragraph{Surface and medium extraction.}
The zero level set of $f_\theta$ defines the object's surface: a cell is \emph{inside} when $f_\theta(\mathbf{x}_i){<}0$ and \emph{outside} otherwise.
The Voronoi faces shared by an inside and an outside cell, which we call \emph{walls}, form a closed explicit surface, and the inside cells form the medium for subsurface transport~(\cref{fig:foamroles}).
Each wall is a polygon whose corners are Voronoi vertices.
Each corner $c$ stores the normal $\bar{\mathbf{n}}_c{=}\nabla f_\theta/\|\nabla f_\theta\|$ at the nearby zero crossing of $f_\theta$ along its edges.
Inside a wall, normals are interpolated from its corners with mean value coordinates~\cite{floater2003mvc}, a generalized barycentric interpolation~(\cref{sec:method:surface}).

\subsection{Relighting}
\label{sec:method:relight}
Under a point light, we render each pixel as the surface reflection at the wall the camera ray crosses plus the light that leaves the medium:
\begin{equation}
  C_{\text{phys}} = \mathbf{P}_m \odot \boldsymbol{\epsilon}_v \odot \Big[L_{\text{surf}}
  + (1-s)\,\big(L^{(1)} + L^{\text{ms}}\big)\Big].
  \label{eq:render}
\end{equation}
$L_{\text{surf}}$ is evaluated at the first wall the ray crosses~(\cref{sec:method:surface}), and $L^{(1)}$ and $L^{\text{ms}}$ are computed on the medium cells~(\cref{sec:method:transport}).
\Cref{fig:decomposition} shows the three terms on held-out views.
The learned share $s{\in}[0,1]$ splits the light that is not reflected specularly between a diffuse surface lobe and transmission into the medium.
$\mathbf{P}_m$ and $\boldsymbol{\epsilon}_v$ calibrate the RGB power of light $m$ and the exposure of camera pose $v$~(\cref{sec:method:training}).
The learned parameters are per-vertex normals, base colors and roughness of the walls; the extinction $\boldsymbol{\sigma}_t\in\mathbb{R}^3$, anisotropy $g$ and diffusion gain $\gamma$ of the medium; the share $s$; and the calibration.
We call $C_{\text{phys}}$ the physics-only render; the object residual of \Cref{sec:method:residual} is fitted on top of it once it has converged.

Each term is attenuated once along its own path to the light.
The surface term is attenuated by $V {=} \exp(-\boldsymbol{\sigma}_t\, l_{\text{in}})$, with $l_{\text{in}}$ the length of the shadow ray inside the medium, so shadows cast through thin parts are colored.

\subsection{Surface reflection}
\label{sec:method:surface}
We evaluate the BRDF $f_r$ at the point $\mathbf{x}$ where the camera ray enters the first inside cell.
Its normal and material are blended over the wall polygon from per-corner values with mean value coordinates~\cite{floater2003mvc}.
They therefore vary continuously across walls, and the shading ``hides'' the cell structure.

\paragraph{Corner normals.}
Each corner normal $\mathbf{n}_c$ is a tilt of the exported normal $\bar{\mathbf{n}}_c$ in its tangent plane, given by two learned offsets bounded by a $\tanh$ and initialized at zero.
Three regularizers keep each corner normal close to the normal of $f_\theta$ and to the normals of its neighbors.
Without them, the normals could bend away from the geometry to explain light that the medium should produce.
One weight is set from the photometric gradient and two are fixed, so none is tuned per object~(supplementary material).

\paragraph{Corner materials.}
Each corner stores a base color $\mathbf{b}_c$ and a roughness $r_c$; the roughness is a convex combination of four roughness levels shared by the object.
The specular lobe is normalized to unit directional albedo, and each light is modeled as a sphere light of small fixed radius.
With the visibility $V$ and the diffuse share $s$ of \cref{eq:render}, the surface term is~$L_{\text{surf}}{=}\frac{I_\ell}{d^2}\,V\,\big(s\,\mathbf{b}/\pi+f_{\text{spec}}\big)(\mathbf{n}\cdot\boldsymbol{\ell})_+$.

\subsection{Subsurface transport}
\label{sec:method:transport}
Both subsurface terms are evaluated on the inside cells.
Cells are \textit{homogeneous}, so the transmittance along any path inside the medium is an exact product over its segments, as in \cref{eq:radfoam}, with no Monte Carlo sampling~\cite{condor2025dsyg}.
Unlike a dipole~\cite{jensen2001bssrdf, zheng2026ssdgs}, which assumes a flat, semi-infinite medium, the transport follows the actual volume of the object.

\paragraph{Medium.}
The medium is the set of inside cells, and their stage-1 densities $\sigma_i$ remain frozen.
Light transport does not use these densities but its own extinction $\boldsymbol{\sigma}_t{\in}\mathbb{R}^3$, one value per color channel for the whole object.
It is bounded to a fixed range divided by $D_{\text{obj}}$, the extent of the medium, so the range scales with the object instead of being tuned per object.
The albedo is not a free parameter.
As in production renderers~\cite{chiang2016practical}, the base color acts as the multiple-scattering albedo, and each inside cell takes the single-scattering albedo $\mathbf{a}_i {=} \mathcal{M}(\mathbf{b}_{w(i)})$ of its graph-nearest wall $w(i)$, where $\mathcal{M}$ is the albedo inversion of~\cite{chiang2016practical}.
Editing the base color therefore recolors the surface and the medium together.
The anisotropy $g$ is one value per object, and~$\boldsymbol{\sigma}_{s,i}{=}\mathbf{a}_i\boldsymbol{\sigma}_t$, $\,\,\boldsymbol{\sigma}_{a,i}{=}(1-\mathbf{a}_i)\,\boldsymbol{\sigma}_t$.

\paragraph{Single scattering.}
Along the camera ray, each medium cell scatters the light once toward the camera:
\begin{equation}
  L^{(1)} = \sum_i T_i \odot \big(1-e^{-\boldsymbol{\sigma}_t\delta_i}\big) \odot
  \mathbf{a}_i\, f_p(\boldsymbol{\ell},\mathbf{v})\, V_i\, \tau_i\, F_t\,\frac{I_\ell}{d_i^2},
  \label{eq:single}
\end{equation}
where $f_p$ is the Henyey--Greenstein phase function with anisotropy $g$, $d_i$ the distance from cell $i$ to the light, $T_i$ the transmittance of $\boldsymbol{\sigma}_t$ along the camera ray, $V_i {=} \exp(-\boldsymbol{\sigma}_t\, l_{\text{in},i})$ the transmittance toward the light inside the medium, $F_t {=} 1-F_{\text{Schlick}}(\mathbf{n}\cdot\boldsymbol{\ell})$ the Fresnel transmittance at the wall, and $\tau_i$ the exact attenuation inside the cell of light that entered through the wall with normal $\mathbf{n}$.

\FigDecomposition

\paragraph{Multiple scattering.}
We approximate multiple scattering by diffusion \cref{eq:diffusion} and solve it \textit{directly} on the foam.
Each inside cell stores one fluence value $\phi_i$ per color channel.
Neighboring cells exchange light through their shared face, with face areas and volumes approximated from the cell sizes.
Discretizing \cref{eq:diffusion} with finite volumes on these cells gives one sparse linear system per light,
\begin{equation}
  \mathbf{K}\boldsymbol{\phi} = \mathbf{q},
  \label{eq:fv}
\end{equation}
where $\mathbf{K}$ spreads light between neighbors, removes the light absorbed in each cell, and lets light escape near the surface through a Robin boundary condition~\cite{jensen2001bssrdf}.
The source~$\mathbf{q}$ is the light that reaches each cell through the medium.
Since $\mathbf{K}$ is symmetric positive-definite, we solve the system with preconditioned conjugate gradients, and the backward pass is one more solve with the same matrix.
The source does not depend on the camera, so one solve per light serves every view.
Each cell then emits light $L^{\text{ms}}$ proportional to its albedo and fluence, scaled by one learned gain $\gamma$ per object.
A pixel reads this emission at the wall it crosses, interpolated from the cells around it.

\subsection{Object residual}
\label{sec:method:residual}
Some appearance lies outside the physical model.
Printed decals and patterns are finer than the medium cells, refraction through glass is not modeled, diffusion on the cells only approximates the transport, and real captures add camera glare.
Once the physical model has converged, we therefore freeze it and fit an MLP $R_\psi$ that corrects its render.
With $S$, $M$ and $D$ the surface, single-scattering and diffusion terms of \cref{eq:render}~(\cref{fig:decomposition}), each with its calibration applied,
\begin{equation}
  C {=} S \odot e^{\mathbf{g}_S} + (M+D) \odot e^{\mathbf{g}_V} + \mathbf{a},
  \,\, (\mathbf{o}_S, \mathbf{o}_V, \mathbf{a}) {=} R_\psi(\mathbf{z}),
  \nonumber
\end{equation}
with per-channel gains $\mathbf{g} {=} 3\tanh(\mathbf{o}/3)$, so that $R_\psi$ starts from $C{=}C_{\text{phys}}$.
Its input $\mathbf{z}$ encodes, at the surface point a pixel sees, the position, the light and view directions relative to the normal, the shadow features of the light, and the logarithms of the physical terms and of the opacity.
$R_\psi$ has three hidden layers of $256$ units~($150.8$k weights), fewer than the $193$k weights of the SSD-GS networks; the size is selected on validation lights~(supplementary material).
It is fitted for $150$k iterations to the training images of the same lights: $64$ per camera on the real captures, and $56$ of $112$ on the synthetic objects.
The foam, the surface and the medium stay fixed.
We evaluate it under held-out lights, at the same cameras on the real captures and at unseen ones in the supplementary material.
Switching it off returns exactly the physics-only model.

\subsection{Optimization}
\label{sec:method:training}
The appearance stage minimizes $\mathcal{L}_{\text{img}}(C_{\text{phys}})+\mathcal{L}_{\text{corner}}+\lambda_b\mathcal{L}_b$.
Here $\mathcal{L}_{\text{img}}(C){=}\|C-\bar{C}\|_1+0.2\,(1-\mathrm{SSIM}(C,\bar{C}))$ is computed on $32\times32$ patches of the target $\bar{C}$, $\mathcal{L}_{\text{corner}}$ are the corner-normal regularizers, and $\mathcal{L}_b$ is an edge-aware smoothness of the rendered base color~(supplementary material).
Real capture rigs differ in LED power and camera exposure, so we also learn an RGB power $\mathbf{P}_m$ per LED and an RGB exposure $\boldsymbol{\epsilon}_v$ per pose, with zero mean in log space, from the training images only.

\providecommand{\nreal}{20}
\section{Experiments}
\label{sec:experiments}

\TabObjects
\FigQual
\FigEditing
\FigLearned
\subsection{Setup}
\label{sec:exp:setup}

We use ten of the \nreal{} real captures of the SSS-GS OLAT dataset~\cite{dihlmann2024sssgs}\footnote{\url{https://huggingface.co/datasets/CGTuebingen/SSS-GS}}, with 158 camera poses and 114--168 LEDs per object and images $800$ pixels wide.
Each pose is trained with half of its LEDs and tested with the other half.
These are the ten objects on which we retrained both baselines with and without their networks; a baseline leads on three of them.
We also use four synthetic Cycles scenes~(soap, candle, dragon, statue), which hold out both cameras and lights.
We compare with SSS-GS~\cite{dihlmann2024sssgs} and SSD-GS~\cite{zheng2026ssdgs}, retrained on the same images; SSS-GS gets a one-line fix of its sRGB conversion that leaves every rendered value unchanged~(supplementary material).
Neither baseline calibrates LED power or exposure; in our earlier runs, giving \lucent and SSD-GS the best brightness per test view lowered our lead on the turbid real objects by about $0.5$\,dB~(supplementary material).
We report PSNR, SSIM and LPIPS on sRGB images on black, over $256$ views per real object~($32$ cameras, each under $8$ held-out lights) and $32$ per synthetic object~(one light per camera).
The geometry stage runs for 20k iterations on up to 262k sites~(2M for the statue) and the appearance stage for 10k iterations with Adam, each on patches of one image under one light; the residual is fitted afterwards.
Otherwise, one configuration is used for every object; on the joyful yell, five cameras that the dataset mis-registers are left out of our training~(all methods are scored on the same views).
Training takes $4.7$--$6.1$\,h per real object on one shared A100, longer than either baseline~(supplementary material), but a medium edit needs no retraining, only a new solve~(\solvetime{} per light).

\subsection{Relighting}

\label{sec:exp:baselines}

\Cref{tab:objects} and \Cref{fig:qual} compare \lucent to SSS-GS and SSD-GS on ten real captures, including glass, liquid in glass and a soft rubber ball, and on the four synthetic objects~(per object in the supplementary material).
\textbf{On the real captures, \lucent is the most accurate in PSNR.}
With its object residual, it has the best mean PSNR~($35.68$ against $33.95$\,dB for SSD-GS and $33.79$\,dB for SSS-GS) and leads on seven of the ten objects.
It is second in SSIM~($0.685$ against $0.692$ for SSD-GS and $0.633$ for SSS-GS) and second in LPIPS~($0.028$ against $0.032$ for SSD-GS), while SSS-GS, whose network keeps finer detail, has the best LPIPS~($0.019$).
Where it trails, on the blue bottle, the physalis and the clear crystal, it is within $0.8$\,dB of the best method.
\textbf{On the synthetic objects, SSD-GS leads.}
These hold out cameras as well as lights; SSD-GS is ahead in all three metrics~($39.09$ against $38.51$\,dB) and on three of the four objects; \lucent leads on the soap and is ahead of SSS-GS in PSNR and SSIM~($36.95$\,dB).
They have no decals or glare, so there the residual's gain~($3.8$--$11.2$\,dB) measures the error of the physical model itself, part of it at the silhouette~(supplementary material).

\subsection{What is physics and what is learned}
\label{sec:exp:share}

\textbf{The physics carries the transport.}
In \Cref{fig:decomposition}, the surface of the car carries the highlights and the volume the body glow; over the $256$ test views its light is $40\%$ surface, $57\%$ diffusion and $4\%$ single scattering.
The residual adds $3.5$--$7.9$\,dB to the physics-only render on the real objects, mostly in back-lit views, where much of the missing light is camera glare from LEDs near the optical axis~(supplementary material).
More scattering orders do not replace diffusion: with $K{=}8$ octaves and no diffusion, the candle is $3.8$\,dB below \lucent on back-lit views~(\cref{tab:ablation} in the supplementary material, with the geometry ablations).

\textbf{The residual adds accuracy without changing the medium.}
We switch off the learned part of each method, on the trained checkpoint and by retraining without it~(\cref{fig:learned_share}; supplementary material).
For \lucent, both give the same model, since the residual is fitted after the medium is fixed.
Without its network, SSS-GS loses the object's color~(up to $12.0$\,dB).
Retrained without its networks, SSD-GS keeps $0.5$--$3.4$M per-Gaussian values, an accurate reflectance model at held-out lights, and on the real objects it is more accurate than our physics alone~($32.8$ against $30.1$\,dB, \cref{tab:objects}), but it has no subsurface transport left to edit.
Our physics is ahead of SSS-GS without its network~($28.6$\,dB), and our residual, with fewer weights than the $193$k of SSD-GS's networks, recovers the accuracy.
On cameras it never saw, it still adds $5.8$\,dB on average and gains on each of the eight objects tested~(supplementary material).

\subsection{Editing the medium}
\label{sec:exp:editing}

\textbf{Only \lucent has a medium to edit.}
Editing it means changing a few numbers and solving again; the baselines have no medium independent of light and view to edit.
In \Cref{fig:editing}, lowering $\boldsymbol{\sigma}_t$ makes the soap and the dragon glow under back light.
Over a sweep of the extinction, the back-lit brightness of \lucent changes in the same direction as the Blender reference on all three synthetic objects, with an overshoot on the statue.
Scaling the scattering coefficients predicted by SSD-GS changes it the opposite way or not at all.
On the soap at $\boldsymbol{\sigma}_t\times\frac14$, the back-lit views of the reference brighten $1.7\times$ on average and those of \lucent $1.4\times$, while SSD-GS darkens.
\Cref{fig:editcandle} edits the medium of the candle with front and back light.
Because the transport is linear in the light, the same model relights under environment maps without retraining~(\cref{fig:envmap_main}); these, glass and interior edits are in the supplementary material.

\FigEditCandle
\TabEditPattern
\begin{figure}[t]
\centering\setlength{\abovecaptionskip}{3pt}
\setlength{\tabcolsep}{0pt}\renewcommand{\arraystretch}{0.5}\scriptsize
\begin{tabular}{@{}c@{}}
\includegraphics[width=0.27264\linewidth]{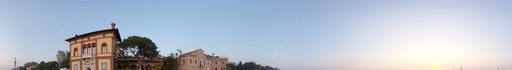}\hspace{0.01604\linewidth}\includegraphics[width=0.27264\linewidth]{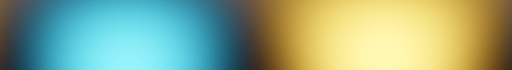}\hspace{0.01604\linewidth}\includegraphics[width=0.27264\linewidth]{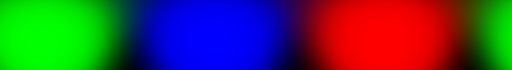} \\[1pt]
\includegraphics[width=0.85\linewidth]{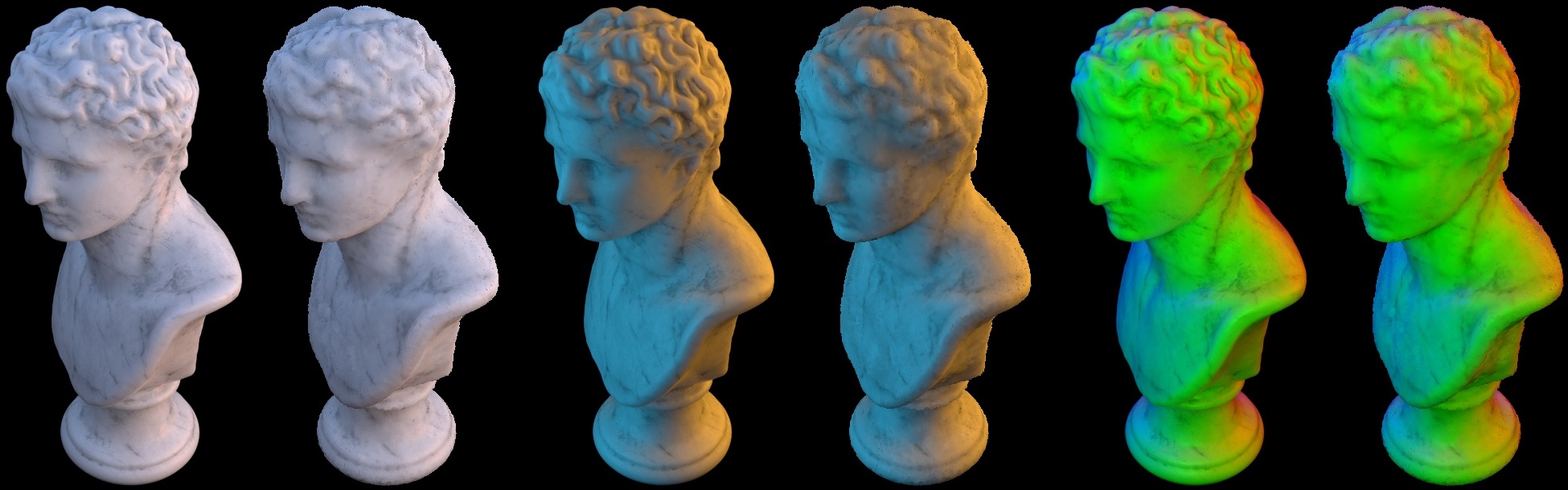}
\end{tabular}
\caption{\textbf{Relighting under environment maps without retraining}: reference~(left of each pair) and \lucent, physics only~(right).}
\label{fig:envmap_main}
\end{figure}
\textbf{The edits change the right pixels, and by the right amount where the extinction is recovered.}
We render each medium edit of the three synthetic objects in Blender.
On the soap, whose extinction is recovered within $10\%$~(\cref{tab:sigmagt}), seven of the eight edits bring the render closer to the edited truth in PSNR than leaving it unedited.
On the candle and the statue, whose extinction is too low in optically thick channels, the edits are not quantitative, but their pattern is right: with one fitted scale per channel, our predicted change beats the best uniform change in $19$ of $24$ edits~(\cref{tab:editpattern}).

\section{Conclusion}
\label{sec:conclusion}

\lucent reconstructs a translucent object as a surface enclosing a medium and simulates the light inside it on the cells of the radiant foam that renders it.
Unlike the Gaussian methods, its medium can be edited and relit with a new solve, and its edits move with reference renders.
This does not cost relighting accuracy: on ten real objects, it has the best mean PSNR and is second in SSIM and LPIPS.
Its learned part, smaller than the SSD-GS networks, can be switched off.

\paragraph{Limitations.}
Rays are not refracted during reconstruction, so glass is modeled only approximately.
In media that are optically thick at the cell size, the recovered constants and edits are right in direction but not in magnitude.
The residual absorbs heterogeneity, glare and transport error; SSD-GS remains more accurate on the synthetic objects; and we train more slowly than the baselines.

\clearpage
\section*{LLM Usage Statement}

The authors wrote this paper.
We used Claude~(Anthropic) to edit and review the text, to write analysis, scoring and plotting scripts for the experiments and figures, to compose the layout of the teaser figure, and to check the bibliography against publisher records.
The method, experiments and results are the authors' own; the authors reviewed and verified all content and take full responsibility for it.

\section*{Acknowledgments}

We thank Florian Hofherr for his great insights and his help reviewing the paper.
This work was supported by the ERC Advanced Grant SIMULACRON and the Munich Center for Machine Learning~(MCML).
In addition, the work of G.~Figueroa-Araneda was also supported by Universidad T\'ecnica Federico Santa Mar\'ia and Centro Nacional de Inteligencia Artificial in Chile.
This work was also supported in part by the Natural Sciences and Engineering Research Council of Canada~(NSERC) Discovery Grant, NSERC Collaborative Research and Development Grant, Digital Research Alliance of Canada, and the SFU Visual Computing Research Chair program.

{
    \small
    \bibliographystyle{ieeenat_fullname}
    \bibliography{main}

\begin{thebibliography}{37}
\providecommand{\natexlab}[1]{#1}
\providecommand{\url}[1]{\texttt{#1}}
\expandafter\ifx\csname urlstyle\endcsname\relax
  \providecommand{\doi}[1]{doi: #1}\else
  \providecommand{\doi}{doi: \begingroup \urlstyle{rm}\Url}\fi

\bibitem[Ahn and Choo(2026)]{ahn2026volsgs}
Junyeong Ahn and Jaegul Choo.
\newblock {VolS-GS}: Relightable {G}aussian splatting with volumetric subsurface scattering.
\newblock \emph{arXiv preprint arXiv:2610.04007}, 2026.

\bibitem[Alpay and Basaran(2026)]{dipolethin2026}
Faruk Alpay and Baris Basaran.
\newblock Dipole diffusion error in thin geometry: Optical thickness laws for grid-free subsurface scattering.
\newblock \emph{arXiv preprint arXiv:2606.29387}, 2026.

\bibitem[Arbree et~al.(2011)Arbree, Walter, and Bala]{arbree2011fem}
Adam Arbree, Bruce Walter, and Kavita Bala.
\newblock Heterogeneous subsurface scattering using the finite element method.
\newblock \emph{IEEE Transactions on Visualization and Computer Graphics}, 17\penalty0 (7):\penalty0 956--969, 2011.

\bibitem[Bi et~al.(2024)Bi, Zeng, Zeng, Pei, Feng, Zhou, and Wu]{bi2024gs3}
Zoubin Bi, Yixin Zeng, Chong Zeng, Fan Pei, Xiang Feng, Kun Zhou, and Hongzhi Wu.
\newblock {GS}$^3$: Efficient relighting with triple {G}aussian splatting.
\newblock In \emph{SIGGRAPH Asia 2024 Conference Papers}, 2024.

\bibitem[Chiang et~al.(2016)Chiang, Kutz, and Burley]{chiang2016practical}
Matt Jen-Yuan Chiang, Peter Kutz, and Brent Burley.
\newblock Practical and controllable subsurface scattering for production path tracing.
\newblock In \emph{ACM SIGGRAPH Talks}, 2016.

\bibitem[Condor et~al.(2025)Condor, Speierer, Bode, Bo{\v{z}}i{\v{c}}, Green, Didyk, and Jarabo]{condor2025dsyg}
Jorge Condor, S{\'e}bastien Speierer, Lukas Bode, Alja{\v{z}} Bo{\v{z}}i{\v{c}}, Simon Green, Piotr Didyk, and Adri{\'a}n Jarabo.
\newblock Don't splat your {G}aussians: Volumetric ray-traced primitives for modeling and rendering scattering and emissive media.
\newblock \emph{ACM Transactions on Graphics}, 44\penalty0 (1):\penalty0 10:1--10:17, 2025.

\bibitem[Deng et~al.(2022)Deng, Luan, Walter, Bala, and Marschner]{deng2022translucent}
Xi Deng, Fujun Luan, Bruce Walter, Kavita Bala, and Steve Marschner.
\newblock Reconstructing translucent objects using differentiable rendering.
\newblock In \emph{ACM SIGGRAPH 2022 Conference Proceedings}, 2022.

\bibitem[Dihlmann et~al.(2024)Dihlmann, Majumdar, Engelhardt, Braun, and Lensch]{dihlmann2024sssgs}
Jan-Niklas Dihlmann, Arjun Majumdar, Andreas Engelhardt, Raphael Braun, and Hendrik P.~A. Lensch.
\newblock Subsurface scattering for {G}aussian splatting.
\newblock In \emph{Advances in Neural Information Processing Systems (NeurIPS)}, pages 121765--121789, 2024.

\bibitem[Fan et~al.(2025)Fan, Luan, Yang, Ha{\v{s}}an, and Wang]{fan2025rng}
Jiahui Fan, Fujun Luan, Jian Yang, Milo{\v{s}} Ha{\v{s}}an, and Beibei Wang.
\newblock {RNG}: Relightable neural {G}aussians.
\newblock In \emph{Proceedings of the IEEE/CVF Conference on Computer Vision and Pattern Recognition (CVPR)}, pages 26525--26534, 2025.

\bibitem[Floater(2003)]{floater2003mvc}
Michael~S. Floater.
\newblock Mean value coordinates.
\newblock \emph{Computer Aided Geometric Design}, 20\penalty0 (1):\penalty0 19--27, 2003.

\bibitem[Gao et~al.(2024)Gao, Gu, Lin, Li, Zhu, Cao, Zhang, and Yao]{gao2023relightable3dgs}
Jian Gao, Chun Gu, Youtian Lin, Zhihao Li, Hao Zhu, Xun Cao, Li Zhang, and Yao Yao.
\newblock Relightable {3D} {G}aussians: Realistic point cloud relighting with {BRDF} decomposition and ray tracing.
\newblock In \emph{European Conference on Computer Vision (ECCV)}, 2024.

\bibitem[Govindarajan et~al.(2025)Govindarajan, Rebain, Yi, and Tagliasacchi]{govindarajan2025radfoam}
Shrisudhan Govindarajan, Daniel Rebain, Kwang~Moo Yi, and Andrea Tagliasacchi.
\newblock Radiant foam: Real-time differentiable ray tracing.
\newblock In \emph{IEEE/CVF International Conference on Computer Vision (ICCV)}, 2025.

\bibitem[Govindarajan et~al.(2026)Govindarajan, Rebain, Verbin, Yi, Prabhu, and Tagliasacchi]{govindarajan2026powerfoam}
Shrisudhan Govindarajan, Daniel Rebain, Dor Verbin, Kwang~Moo Yi, Anish Prabhu, and Andrea Tagliasacchi.
\newblock Power foam: Unifying real-time differentiable ray tracing and rasterization.
\newblock \emph{arXiv preprint arXiv:2604.24994}, 2026.

\bibitem[Gropp et~al.(2020)Gropp, Yariv, Haim, Atzmon, and Lipman]{gropp2020igr}
Amos Gropp, Lior Yariv, Niv Haim, Matan Atzmon, and Yaron Lipman.
\newblock Implicit geometric regularization for learning shapes.
\newblock In \emph{International Conference on Machine Learning (ICML)}, 2020.

\bibitem[Hanrahan and Krueger(1993)]{hanrahan1993}
Pat Hanrahan and Wolfgang Krueger.
\newblock Reflection from layered surfaces due to subsurface scattering.
\newblock In \emph{Proceedings of the 20th Annual Conference on Computer Graphics and Interactive Techniques (SIGGRAPH)}, pages 165--174, 1993.

\bibitem[Jensen et~al.(2001)Jensen, Marschner, Levoy, and Hanrahan]{jensen2001bssrdf}
Henrik~Wann Jensen, Stephen~R. Marschner, Marc Levoy, and Pat Hanrahan.
\newblock A practical model for subsurface light transport.
\newblock In \emph{Proceedings of the 28th Annual Conference on Computer Graphics and Interactive Techniques (SIGGRAPH)}, pages 511--518, 2001.

\bibitem[Ke et~al.(2024)Ke, Obukhov, Huang, Metzger, Daudt, and Schindler]{ke2024marigold}
Bingxin Ke, Anton Obukhov, Shengyu Huang, Nando Metzger, Rodrigo~Caye Daudt, and Konrad Schindler.
\newblock Repurposing diffusion-based image generators for monocular depth estimation.
\newblock In \emph{IEEE/CVF Conference on Computer Vision and Pattern Recognition (CVPR)}, 2024.

\bibitem[Kerbl et~al.(2023)Kerbl, Kopanas, Leimk{\"u}hler, and Drettakis]{kerbl20233dgs}
Bernhard Kerbl, Georgios Kopanas, Thomas Leimk{\"u}hler, and George Drettakis.
\newblock 3{D} {G}aussian splatting for real-time radiance field rendering.
\newblock \emph{ACM Transactions on Graphics}, 42\penalty0 (4), 2023.

\bibitem[Kouros et~al.(2026)Kouros, Wu, and Tuytelaars]{kouros2026specgloss}
Georgios Kouros, Minye Wu, and Tinne Tuytelaars.
\newblock Spec-gloss surfels and normal-diffuse priors for relightable glossy objects.
\newblock In \emph{IEEE/CVF Winter Conference on Applications of Computer Vision (WACV)}, 2026.

\bibitem[Martin~Garcia et~al.(2025)Martin~Garcia, Knaebel, Schmidt, de~Geus, Hermans, and Leibe]{garcia2025e2eft}
Gonzalo Martin~Garcia, Karim Knaebel, Christian Schmidt, Daan de Geus, Alexander Hermans, and Bastian Leibe.
\newblock Fine-tuning image-conditional diffusion models is easier than you think.
\newblock In \emph{IEEE/CVF Winter Conference on Applications of Computer Vision (WACV)}, 2025.

\bibitem[Nsampi et~al.(2026)Nsampi, Djeacoumar, Seidel, Ritschel, and Leimk{\"u}hler]{nsampi2026nrt}
Ntumba~Elie Nsampi, Adarsh Djeacoumar, Hans-Peter Seidel, Tobias Ritschel, and Thomas Leimk{\"u}hler.
\newblock Volumetric inverse rendering via neural radiative transfer.
\newblock \emph{Computer Graphics Forum}, 45\penalty0 (4):\penalty0 e70541, 2026.

\bibitem[Oechsle et~al.(2021)Oechsle, Peng, and Geiger]{oechsle2021unisurf}
Michael Oechsle, Songyou Peng, and Andreas Geiger.
\newblock {UNISURF}: Unifying neural implicit surfaces and radiance fields for multi-view reconstruction.
\newblock In \emph{IEEE/CVF International Conference on Computer Vision (ICCV)}, 2021.

\bibitem[Pavlovic et~al.(2026)Pavlovic, Wandel, Obukhov, Bartolomei, Davydov, Tosi, Poggi, S{\"u}sstrunk, and Dai]{pavlovic2026marigoldv2}
Igor Pavlovic, Thiemo Wandel, Anton Obukhov, Luca Bartolomei, Andrey Davydov, Fabio Tosi, Matteo Poggi, Sabine S{\"u}sstrunk, and Dengxin Dai.
\newblock Marigold {V2}: Revisiting diffusion transformers for monocular depth estimation.
\newblock \emph{ACM Transactions on Graphics (Proc. SIGGRAPH Asia)}, 45\penalty0 (6), 2026.

\bibitem[Rech et~al.(2025)Rech, Conci, and Garau]{rech2025sdfoam}
Antonella Rech, Nicola Conci, and Nicola Garau.
\newblock {SDF}oam: Signed-distance foam for explicit surface reconstruction.
\newblock \emph{arXiv preprint arXiv:2512.16706}, 2025.

\bibitem[Stam(1995)]{stam1995}
Jos Stam.
\newblock Multiple scattering as a diffusion process.
\newblock In \emph{Rendering Techniques '95 (Proceedings of the Eurographics Workshop on Rendering)}, pages 41--50. Springer Vienna, 1995.

\bibitem[Sun et~al.(2025)Sun, Gao, Xie, Yang, and Wang]{sun2025svgir}
Hanxiao Sun, Yupeng Gao, Jin Xie, Jian Yang, and Beibei Wang.
\newblock {SVG-IR}: Spatially-varying {G}aussian splatting for inverse rendering.
\newblock In \emph{Proceedings of the IEEE/CVF Conference on Computer Vision and Pattern Recognition (CVPR)}, pages 16143--16152, 2025.

\bibitem[Walter et~al.(2007)Walter, Marschner, Li, and Torrance]{walter2007ggx}
Bruce Walter, Stephen~R. Marschner, Hongsong Li, and Kenneth~E. Torrance.
\newblock Microfacet models for refraction through rough surfaces.
\newblock In \emph{Rendering Techniques 2007 (Proceedings of the Eurographics Symposium on Rendering)}, pages 195--206, 2007.

\bibitem[Wang et~al.(2008)Wang, Zhao, Tong, Lin, Lin, Dong, Guo, and Shum]{wang2008polygrid}
Jiaping Wang, Shuang Zhao, Xin Tong, Stephen Lin, Zhouchen Lin, Yue Dong, Baining Guo, and Heung-Yeung Shum.
\newblock Modeling and rendering of heterogeneous translucent materials using the diffusion equation.
\newblock \emph{ACM Transactions on Graphics}, 27\penalty0 (1), 2008.

\bibitem[Wang et~al.(2021)Wang, Liu, Liu, Theobalt, Komura, and Wang]{wang2021neus}
Peng Wang, Lingjie Liu, Yuan Liu, Christian Theobalt, Taku Komura, and Wenping Wang.
\newblock {NeuS}: Learning neural implicit surfaces by volume rendering for multi-view reconstruction.
\newblock In \emph{Advances in Neural Information Processing Systems (NeurIPS)}, 2021.

\bibitem[Yariv et~al.(2020)Yariv, Kasten, Moran, Galun, Atzmon, Basri, and Lipman]{yariv2020idr}
Lior Yariv, Yoni Kasten, Dror Moran, Meirav Galun, Matan Atzmon, Ronen Basri, and Yaron Lipman.
\newblock Multiview neural surface reconstruction by disentangling geometry and appearance.
\newblock In \emph{Advances in Neural Information Processing Systems (NeurIPS)}, 2020.

\bibitem[Yariv et~al.(2021)Yariv, Gu, Kasten, and Lipman]{volsdf}
Lior Yariv, Jiatao Gu, Yoni Kasten, and Yaron Lipman.
\newblock Volume rendering of neural implicit surfaces.
\newblock In \emph{Advances in Neural Information Processing Systems (NeurIPS)}, 2021.

\bibitem[Yu et~al.(2023)Yu, Guo, Fathi, Chang, Chan, Gao, Funkhouser, and Wu]{yu2023osf}
Hong-Xing Yu, Michelle Guo, Alireza Fathi, Yen-Yu Chang, Eric~Ryan Chan, Ruohan Gao, Thomas Funkhouser, and Jiajun Wu.
\newblock Learning object-centric neural scattering functions for free-viewpoint relighting and scene composition.
\newblock \emph{Transactions on Machine Learning Research}, 2023.

\bibitem[Yu et~al.(2022)Yu, Peng, Niemeyer, Sattler, and Geiger]{yu2022monosdf}
Zehao Yu, Songyou Peng, Michael Niemeyer, Torsten Sattler, and Andreas Geiger.
\newblock {MonoSDF}: Exploring monocular geometric cues for neural implicit surface reconstruction.
\newblock In \emph{Advances in Neural Information Processing Systems (NeurIPS)}, 2022.

\bibitem[Yuan and Zhao(2026)]{gtsr2026}
Youwen Yuan and Xi Zhao.
\newblock {GTSR}: Subsurface scattering awared {3D} {G}aussians for translucent surface reconstruction.
\newblock \emph{arXiv preprint arXiv:2603.22036}, 2026.

\bibitem[Zhang et~al.(2021)Zhang, Yu, and Zhao]{zhang2021pathspace}
Cheng Zhang, Zihan Yu, and Shuang Zhao.
\newblock Path-space differentiable rendering of participating media.
\newblock \emph{ACM Transactions on Graphics (Proc. SIGGRAPH)}, 40\penalty0 (4), 2021.

\bibitem[Zhang et~al.(2022)Zhang, Sun, He, Fu, Jia, and Zhou]{zhang2022invrender}
Yuanqing Zhang, Jiaming Sun, Xingyi He, Huan Fu, Rongfei Jia, and Xiaowei Zhou.
\newblock Modeling indirect illumination for inverse rendering.
\newblock In \emph{Proceedings of the IEEE/CVF Conference on Computer Vision and Pattern Recognition (CVPR)}, pages 18643--18652, 2022.

\bibitem[Zheng et~al.(2026)Zheng, Tang, Doronin, Teal, and Zhang]{zheng2026ssdgs}
Iris Zheng, Guojun Tang, Alexander Doronin, Paul Teal, and Fang-Lue Zhang.
\newblock {SSD-GS}: Scattering and shadow decomposition for relightable {3D} {G}aussian splatting.
\newblock In \emph{International Conference on Learning Representations (ICLR)}, 2026.

\end{thebibliography}
}

\clearpage
\clearpage
\maketitlesupplementary
\providecommand{\agent}[1]{#1}%
\renewcommand{\thesection}{S\arabic{section}}\setcounter{section}{0}
\renewcommand{\thefigure}{S\arabic{figure}}\setcounter{figure}{0}
\renewcommand{\thetable}{S\arabic{table}}\setcounter{table}{0}
\renewcommand{\theHsection}{S\arabic{section}}
\renewcommand{\theHfigure}{S\arabic{figure}}
\renewcommand{\theHtable}{S\arabic{table}}
\renewcommand{\topfraction}{0.95}\renewcommand{\bottomfraction}{0.6}\renewcommand{\textfraction}{0.05}
\renewcommand{\floatpagefraction}{0.5}\renewcommand{\dbltopfraction}{0.95}\renewcommand{\dblfloatpagefraction}{0.5}
\setcounter{topnumber}{3}\setcounter{dbltopnumber}{3}\setcounter{totalnumber}{5}

\paragraph{Guide to this supplement.}
It follows the structure of the main paper: each section extends the main-paper section named at its start, and where the main paper refers to the supplementary material, the result is found here.
\begin{itemize}\setlength{\itemsep}{1pt}
\item \textbf{\Cref{sec:supp:related}, related work}~(main \cref{sec:related}): related methods compared by capability~(\Cref{tab:capabilities}).
\item \textbf{\Cref{sec:supp:implgroup}, method details}~(main \cref{sec:method}): training schedule, regularizers and the finite-volume diffusion; the cost of every stage and of an edit~(\cref{tab:cost}); and the ablations of the geometry~(\cref{tab:geometry}) and of the light transport, with negative results~(\cref{tab:ablation,tab:negative}).
\item \textbf{\Cref{sec:supp:compgroup}, relighting}~(main \cref{sec:exp:setup,sec:exp:baselines}): baseline setup, including the SSS-GS numerical fix, and calibration; every object and metric behind main \Cref{tab:objects}~(\cref{tab:objects_full}); all qualitative views and the views at our median PSNR for every object; accuracy by lighting direction, camera glare, the silhouette of the synthetic objects and the glass failure cases; three objects that are not in the main paper~(\cref{sec:supp:newobjects}); and environment lighting, known and unknown.
\item \textbf{\Cref{sec:supp:sharegroup}, what is physics and what is learned}~(main \cref{sec:exp:share}): decompositions of more objects, the switch-off and retraining protocol for every method~(\cref{tab:share}), the choice of the residual size and the residual at cameras it never saw~(\cref{tab:residual_sizes,tab:unseen}).
\item \textbf{\Cref{sec:supp:appgroup}, editing the medium}~(main \cref{sec:exp:editing}): the check of medium edits against Blender~(\cref{tab:sigmagt_full,tab:editgt} and main \cref{tab:editpattern}); the editing protocol for SSD-GS and the full sweep behind main \cref{fig:editing}~(\cref{tab:editsweep}); edits of real captures and the direction of the edit on all ten~(\cref{sec:supp:realedit}); and surface, glass and interior edits~(\cref{sec:supp:editing}).
\item \textbf{\Cref{sec:supp:faq}, limitations and anticipated questions}~(main \cref{sec:conclusion}), with pointers to the answers.
\item \textbf{\Cref{sec:supp:code}, code}: the code we release, and retrains with it.
\end{itemize}
A narrated video of about three minutes is on the project page~(\projectpage), where we will also release the code.
The video shows the capture, the reconstruction and the light transport as animations, the relighting against the baselines, and the edits in motion; every result in it is also in this document or the main paper, except the dragon's wax, amber and ruby edits, which the video shows with the fix of \cref{sec:supp:greencells}.

\section{Related work}
\label{sec:supp:related}
\textit{Extends main \cref{sec:related}.}

\Cref{tab:capabilities} in the main paper compares relightable methods for translucent objects by what they represent.
We compare against SSS-GS~\cite{dihlmann2024sssgs} and SSD-GS~\cite{zheng2026ssdgs} because both are designed for subsurface scattering and can be retrained on our data: SSS-GS introduced the one-light-at-a-time captures we use, and SSD-GS is the most recent method for the same task.
GS$^3$~\cite{bi2024gs3} relights Gaussians with learned light-dependent appearance without a subsurface model; VolS-GS~\cite{ahn2026volsgs}, concurrent with this work, solves diffusion on an auxiliary grid; and Deng~\etal~\cite{deng2022translucent} fit a dipole on a mesh by differentiable rendering.
\lucent is the only one with a closed surface, an interior volume and transport on the representation itself, with a medium independent of light and view.

\label{sec:supp:capabilities}

\section{Method details}
\label{sec:supp:implgroup}
\textit{Extends main \cref{sec:method}: implementation details of each stage, its cost, and the ablations of the geometry and of the light transport.}

\subsection{Training, cost and surface regularizers}
\label{sec:supp:impl}

\begin{table}[!tbp]
\caption{\textbf{Cost} of training per real object and of an edit on the synthetic objects~($^\ast$), from the training logs of our final runs~(with $\mathcal{L}_{\text{sil}}$) and of the baselines; the GPU was shared with other jobs, so all times are approximate.
Our times are over the eight real objects that ran alone on one GPU; the marmalade and the hair clip shared a GPU with another job~($10.8$ and $9.7$\,h in total) and are left out of the range.
Monocular priors are computed once beforehand and not included.
\lucent trains slower than both baselines, but an edit of its medium costs one new solve instead of a new capture and training.}
\label{tab:cost}
\centering\small
\setlength{\tabcolsep}{4pt}
\resizebox{\linewidth}{!}{
\begin{tabular}{l l c}
\toprule
Method & Stage & Time, one A100 \\
\midrule
\lucent & geometry~(stage 1, 20k it.) & $1.8$--$2.1$\,h \\
        & appearance~(stage 2, 10k it., with export) & $2.4$--$3.3$\,h \\
        & object residual~(cache and 150k it.) & $0.4$--$0.8$\,h \\
        & \textbf{total} & $\mathbf{4.7}$--$\mathbf{6.1}$\,\textbf{h}~(median $4.9$\,h) \\
        & medium edit: new solve and render, per light$^\ast$ & $0.12$--$0.34$\,s \\
\midrule
SSS-GS~\cite{dihlmann2024sssgs} & training~(60k it.) & $1.4$--$5.7$\,h~(median $2.1$\,h) \\
SSD-GS~\cite{zheng2026ssdgs} & training~(60k it.) & $41$--$76$\,min~(median $50$\,min) \\
\bottomrule
\end{tabular}}
\par\smallskip\footnotesize$^\ast$Median time to set the edit, solve the diffusion for one light~(three color channels in one conjugate-gradient run) and render one $800{\times}800$ view, on the final checkpoints: $0.12$\,s on the candle, $0.19$\,s on the soap and $0.34$\,s on the statue~($2$M sites). Each light needs its own solve; the $0.19$--$0.23$\,s quoted in the main paper is the solve alone for one light of an earlier soap model~($1.08$M medium cells).
\end{table}

\paragraph{Corner-normal regularizers.}
Each corner normal $\mathbf{n}_c$ is a tilt of the exported normal $\bar{\mathbf{n}}_c$ in its tangent plane, parameterized by two learned offsets that are bounded by a $\tanh$ and initialized at zero.
Three regularizers keep the field on the geometry: an $L_2$ penalty on the offsets, a tether $\sum_c (1-\mathbf{n}_c^\top\bar{\mathbf{n}}_c)$, and a Huber penalty on the angle between each corner normal and the mean of its neighbors.
Only the weight of the last term is set from the photometric gradient: its gradient on the offsets matches the RMS of the photometric gradient over the first iterations.
The offset penalty and the tether have fixed weights~(appearance stage below), the same for every object, so no weight is tuned per object.

\paragraph{Geometry stage.}
Sites are seeded at $65{,}536$ points inside the visual hull of the masks, together with the seed point cloud shipped with each capture~($100$k points uniform in a cube for this dataset), and are densified by a factor of $1.15$ up to $262{,}144$ between iterations $1$k and $10$k, and frozen at $18$k.
The network $f_\theta$ takes a positional encoding with six frequencies and is pre-fitted for $2$k iterations to the signed distance of the visual hull, sampled on a $128^3$ grid; the scale $\beta$ starts at $0.1$ times the radius of the hull.
Cell colors are spherical harmonics of degree~$3$.
The geometry stage fits, for each camera pose, the mean of its images over the training lights.
Images are downsampled by $4$, $2$ and $1$ from iterations $0$, $150$ and $500$, and the losses are evaluated on $16\times16$ patches, $128$ of them per iteration for the prior terms.
The loss weights are $\lambda_d=0.1$, $\lambda_n=0.05$, $\lambda_{\text{eik}}=0.1$, $\lambda_s=0.005$ and $\lambda_O=1$.
The silhouette loss uses a margin of $\beta/4$ on either side of zero and $16{,}384$ rays per iteration; the minimum of $f_\theta$ is a soft minimum~(temperature $0.1\beta$) over $32$ samples around the smallest of $128$ uniform samples along the ray. Its weight $\lambda_{\text{sil}}$ is set once, over iterations $500$ to $600$, so that its gradient on $f_\theta$ matches the RMS of the gradient of all other terms, and the loss is applied from then on, so no weight is tuned per object.

The learning rate of $f_\theta$ decays from $5\times10^{-4}$ to $5\times10^{-5}$ after $500$ warm-up iterations.

\paragraph{Finite-volume diffusion.}
Let $j\sim i$ denote two interior cells that share a face.
Every quantity is per color channel, and the three channels are solved together as three right-hand sides of one conjugate-gradient run per light.
Each cell is approximated by a sphere of radius $r_i$, the mean half-distance to its Delaunay neighbors, with volume $|\Omega_i|=\frac43\pi r_i^3$ and face measure $A_{ij}=r_ir_j$; $d_{ij}=\max(\|\mathbf{x}_i-\mathbf{x}_j\|,10^{-3})$.
With $\sigma_{s,i}=a_i\sigma_t$, $\sigma_{a,i}=(1-a_i)\sigma_t$ and $\kappa_i=1/\big(3\max(\sigma_t(1-a_i g),10^{-2})\big)$, the system $\mathbf{K}\boldsymbol{\phi}=\mathbf{q}$ has
\begin{align}
K_{ij}&=-w_{ij},\qquad w_{ij}=\frac{2\kappa_i\kappa_j}{\kappa_i+\kappa_j}\,\frac{A_{ij}}{d_{ij}},\\
K_{ii}&=\textstyle\sum_{j\sim i}w_{ij}+\sigma_{a,i}|\Omega_i|+\beta_i ,
\end{align}
so a face conducts with the harmonic mean of the two diffusion coefficients.
Light escapes through a Robin~(partial-current) term $\beta_i=\pi r_i^2/(2\mathcal{A})$ with $\mathcal{A}=(1+F_{dr})/(1-F_{dr})\approx3.96$ for $\eta=1.5$, applied on a one-cell-thick band below the walls, at the extrapolated-boundary depth capped by the $90\%$ depth quantile of the interior cells; other cells have no flux to the outside.
The source is the light that reaches the cell through the medium and scatters there,
\begin{equation}
q_i=|\Omega_i|\,\sigma_{s,i}\,\frac{I_\ell}{\max(\|\mathbf{x}_i-\boldsymbol{\ell}\|^2,10^{-2})}\,e^{-\sigma_t L_i},
\end{equation}
with $L_i$ the length inside the object of the straight segment from $\mathbf{x}_i$ to the light.
$\mathbf{K}$ is symmetric positive definite; we solve it with Jacobi-preconditioned conjugate gradients~(relative residual $10^{-6}$), and the gradients come from one adjoint solve with the same matrix.
Cell $i$ emits $e_i=\frac{\gamma}{4\pi}a_i\phi_i$, with $\gamma\in[0.3,10]$~(it reaches the upper bound on four of the eight real objects of \cref{tab:unseen}).
The emission is averaged onto each wall corner over the interior cells that touch it, smoothed over the corners~(10 passes of a $\frac12$ neighbor average), and blended across the wall the camera ray crosses with mean value coordinates; the result, weighted by $1-s$, is $L^{\mathrm{ms}}$.
In the single-scattering term~(Eq.~5 of the main paper), with $\mu_\ell=\mathbf{n}\cdot\boldsymbol{\omega}_\ell$ and $\mu_v=\max(\mathbf{n}\cdot\mathbf{v},10^{-3})$ about the shading normal of the first wall the camera ray crosses, $k=\mu_v/\mu_\ell$ and $x=\sigma_t\delta_i$, $\tau_i=\frac{1-e^{-(1+k)x}}{(1+k)(1-e^{-x})}$~(zero when $\mu_\ell\le0$), the attenuation of the entering light integrated along the camera segment under a locally planar model; it tends to $\mu_\ell/(\mu_\ell+\mu_v)$ in optically thick cells and to $1$ in thin ones.

\paragraph{Appearance stage.}
Each iteration renders $250$k rays in $32\times32$ patches, so one diffusion solve serves the whole batch.
The base-color smoothness has weight $\lambda_b=0.037$, with $\mathcal{L}_b = \sum_{(p,p')}\|\mathbf{b}(p)-\mathbf{b}(p')\|_1\,e^{-\|\bar{C}(p)-\bar{C}(p')\|_1}$ over adjacent pixels, which keeps base-color edges where the image has edges.
Learning rates decay from $5\times10^{-3}$ to $5\times10^{-4}$ for the base color and the roughness and from $10^{-3}$ to $10^{-4}$ for the corner normals; the extinction is trained at $10^{-3}$.
The base colors stay at their initialization for the first $1$k iterations, and the anisotropy $g$ is released at $7$k.
Cells outside the surface but within two cell lengths of it also scatter, each with its own trained albedo; this is the only albedo of the model that the base color does not set~(\cref{sec:supp:greencells}).
The main paper describes each light as a sphere of small fixed radius; the released code, with which every result here was computed, treats it as a point light, with a $1/d^2$ falloff normalized at the scene center.
The extinction is bounded so that the optical thickness across the object, $\boldsymbol{\sigma}_t$ times $D_{\text{obj}}$, stays between $0.5$ and $1024$; the diffusion gain $\gamma$ is bounded to $[0.3, 10]$ and starts at $2$.
The boundary condition of the diffusion uses a refractive index of $1.5$, and the conjugate-gradient solve stops at a relative residual of $10^{-6}$ or after $4000$ iterations.
Of the corner-normal regularizers, the offset penalty has weight $10^{-4}$ and the tether $0.275$, and the Huber penalty has its transition at $20^\circ$.

\paragraph{Object residual.}
The input $\mathbf{z}$ of $R_\psi$ has $65$ values: the surface point and its positional encoding with four frequencies~($27$); the cosines $\mathbf{n}\cdot\boldsymbol{\ell}$, $\mathbf{n}\cdot\mathbf{v}$, $\mathbf{n}\cdot\mathbf{h}$, $\boldsymbol{\ell}\cdot\mathbf{v}$ and $\mathbf{h}\cdot\mathbf{v}$, with $\mathbf{h}$ the half vector~($5$); spherical harmonics of degree~$2$ of the light and view directions~($18$); the shadow features, which are the per-channel transmittance $\exp(-\boldsymbol{\sigma}_t\,l_{\text{in}})$ of the shadow ray and the logarithms of its length inside the medium and of the squared distance to the light~($5$); the logarithms of $S$, $M$ and $D$~($9$); and the opacity~($1$).
The hidden layers use LeakyReLU activations, and the last layer starts at zero.
We train with Adam on batches of $262{,}144$ pixels, with a learning rate of $10^{-3}$~($500$ warm-up steps, cosine decay to $5\%$) and a squared error on sRGB values in which every image is weighted by the inverse of its running error; a few of the training lights are set aside for validation, and we keep the weights with the best PSNR on them.

\subsection{Geometry ablations}
\label{sec:supp:geometry}
\label{sec:supp:ablgroup}
\textit{Extends main \cref{sec:method:geo}.}
\TabPriors

\begin{figure}[!tbp]
  \centering
  \newlength{\figMeshW}\setlength{\figMeshW}{0.245\linewidth}
  \setlength{\tabcolsep}{0.5pt}\renewcommand{\arraystretch}{0.6}\scriptsize
  \begin{tabular}{@{}cccc@{}}
  GT & Radiant Foam & SDFoam & \lucent \\
  \includegraphics[width=\figMeshW]{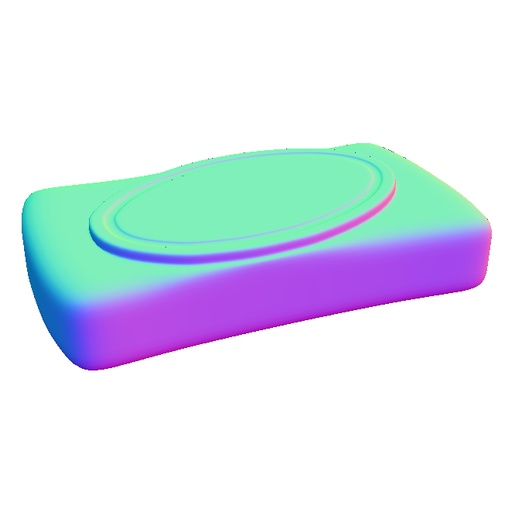} & \includegraphics[width=\figMeshW]{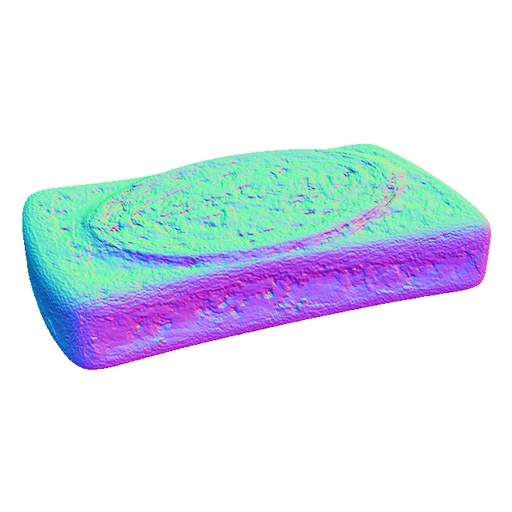} & \includegraphics[width=\figMeshW]{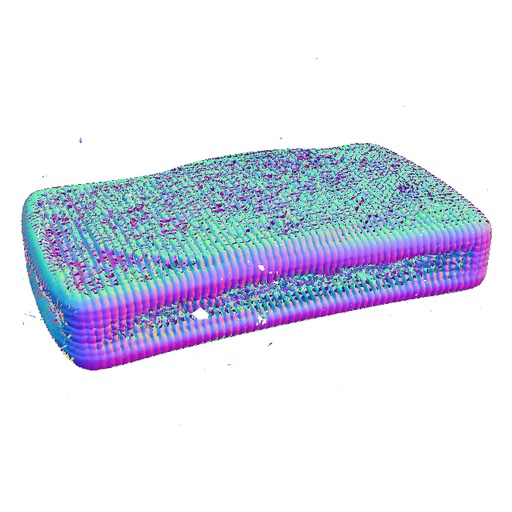} & \includegraphics[width=\figMeshW]{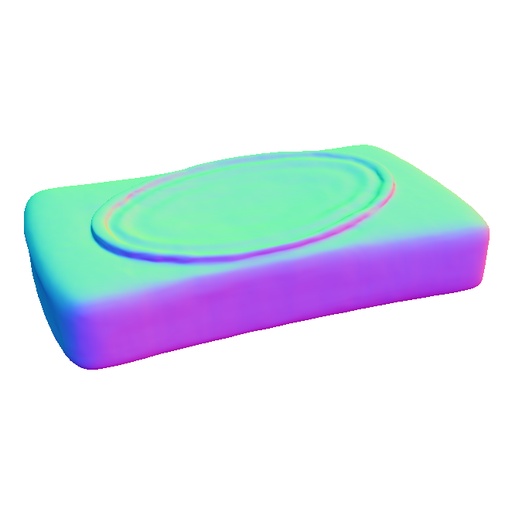} \\
  \includegraphics[width=\figMeshW]{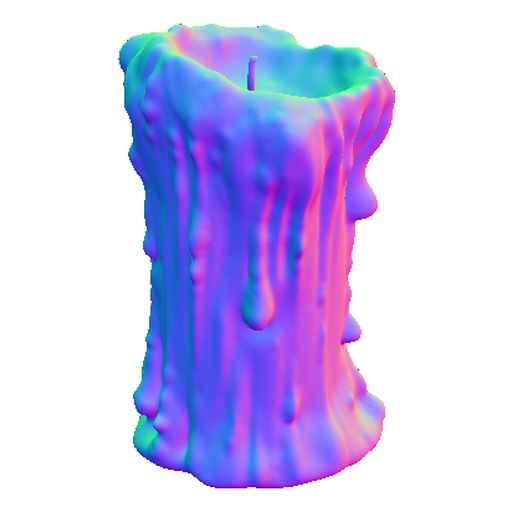} & \includegraphics[width=\figMeshW]{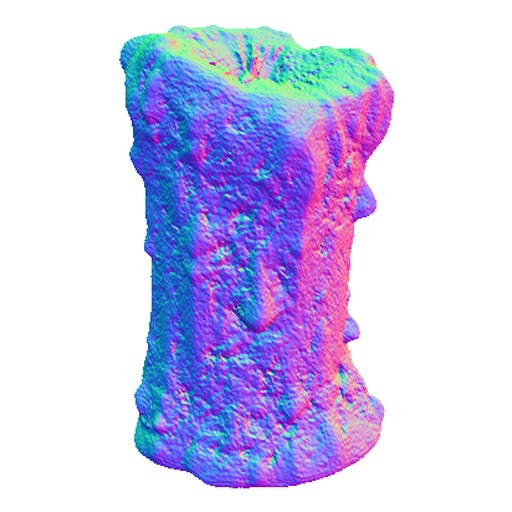} & \includegraphics[width=\figMeshW]{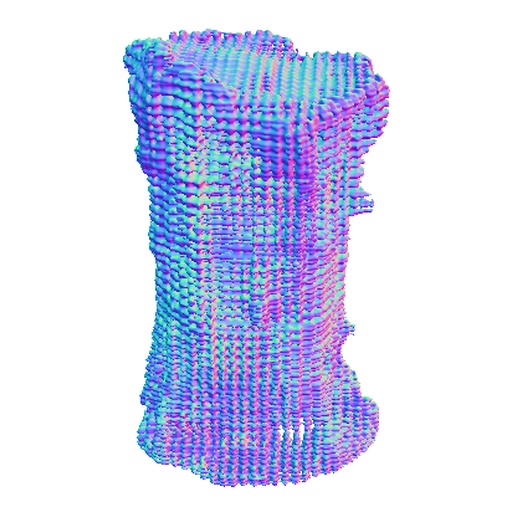} & \includegraphics[width=\figMeshW]{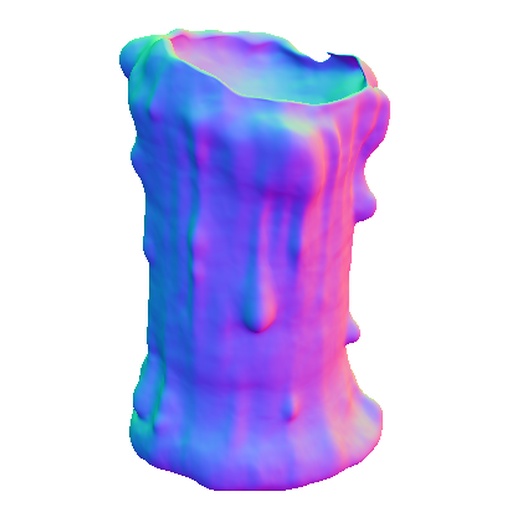} \\
  \includegraphics[width=\figMeshW]{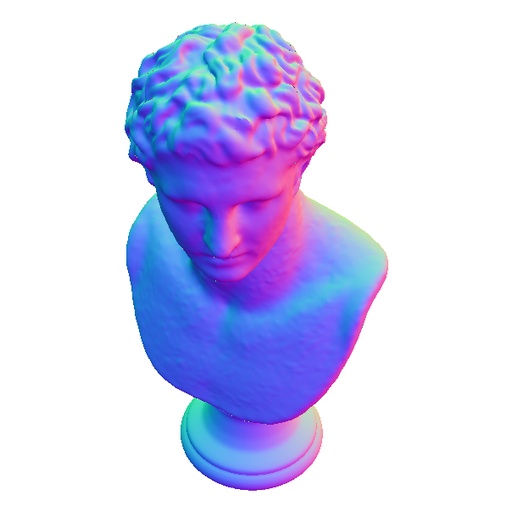} & \includegraphics[width=\figMeshW]{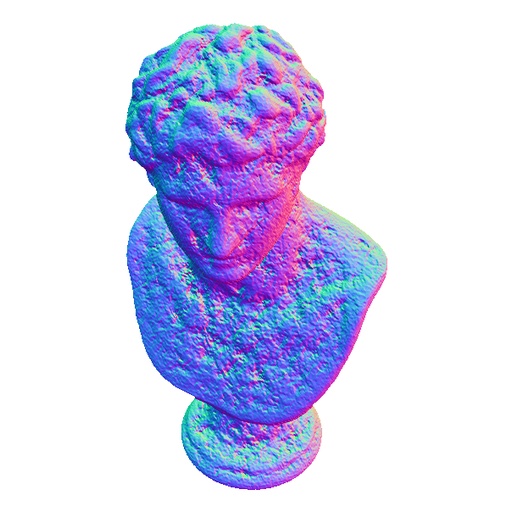} & \includegraphics[width=\figMeshW]{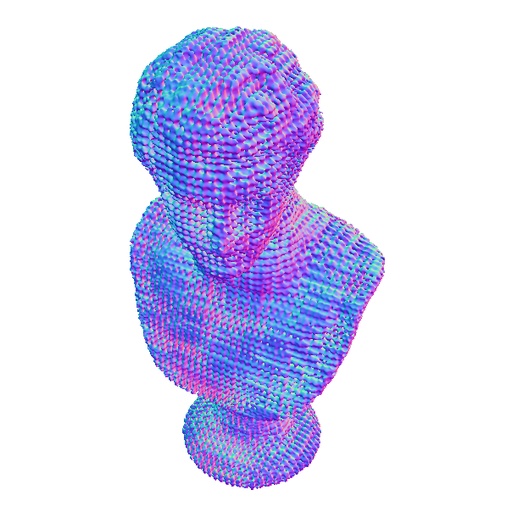} & \includegraphics[width=\figMeshW]{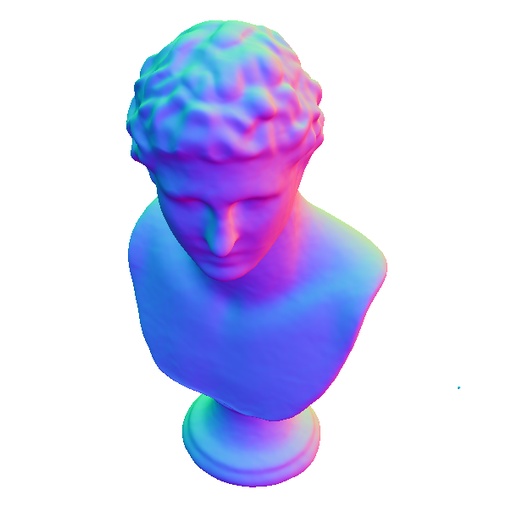} \\
  \end{tabular}
  \caption{\textbf{Shading normals} of the geometry stage against the foam it is built on; rows: soap, candle, statue.}
  \label{fig:geo_meshes}
\end{figure}

\begin{table}[!tbp]
\caption{\textbf{Quantitative results on rendered normals}, candle and statue, $11$ held-out views.
}
\label{tab:geometry}
\centering\small
\setlength{\tabcolsep}{2.5pt}
\begin{tabular}{l cccc}
\toprule
 & Normal MAE$\downarrow$ & PSNR$\uparrow$ & SSIM$\uparrow$ & LPIPS$\downarrow$ \\
\midrule
Radiant Foam~\cite{govindarajan2025radfoam} & 25.56 & 20.74 & 0.725 & 0.243 \\
SDFoam~\cite{rech2025sdfoam}                & 45.64 & 15.96 & 0.629 & 0.346 \\
\textbf{\lucent}                            & \textbf{11.39} & \textbf{25.05} & \textbf{0.866} & \textbf{0.094} \\
\bottomrule
\end{tabular}
\end{table}

\begin{figure}[!tbp]
  \centering
  \newlength{\figAblW}\setlength{\figAblW}{0.196\linewidth}
  \setlength{\tabcolsep}{0.5pt}\renewcommand{\arraystretch}{0.6}\scriptsize
  \begin{tabular}{@{}ccccc@{}}
  GT & SDF & \shortstack{SDF\\+ normal} & \shortstack{SDF\\+ depth} & \shortstack{\lucent\\(+ both)} \\
  \includegraphics[width=\figAblW]{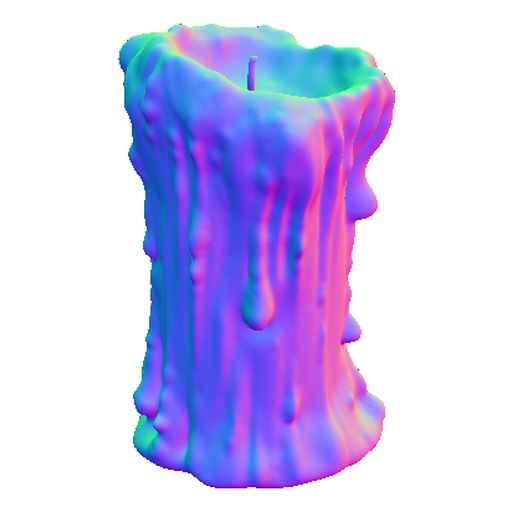} & \includegraphics[width=\figAblW]{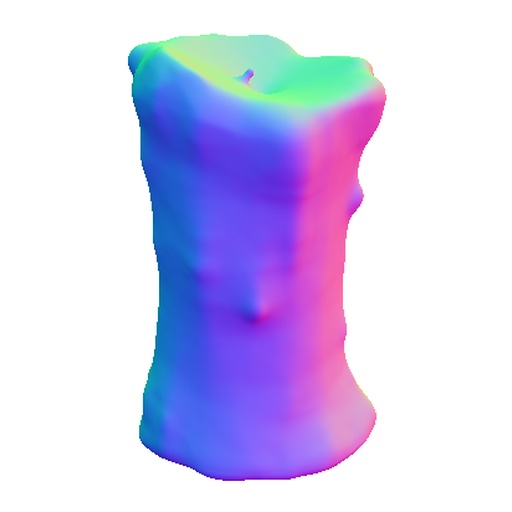} & \includegraphics[width=\figAblW]{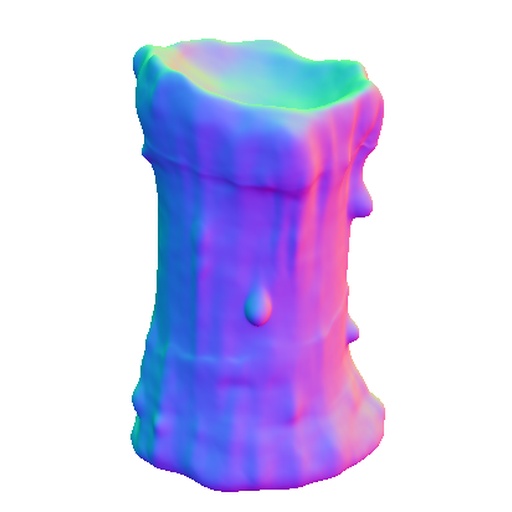} & \includegraphics[width=\figAblW]{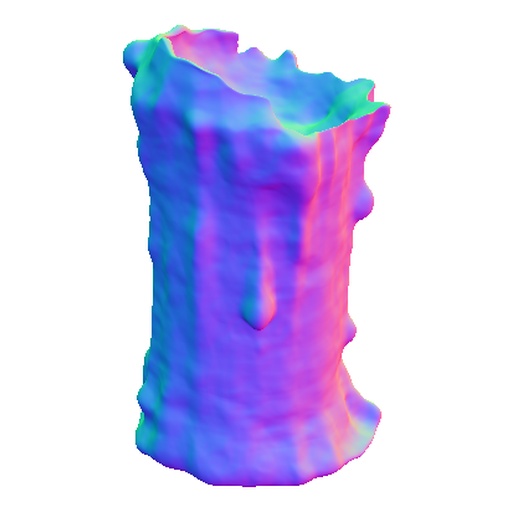} & \includegraphics[width=\figAblW]{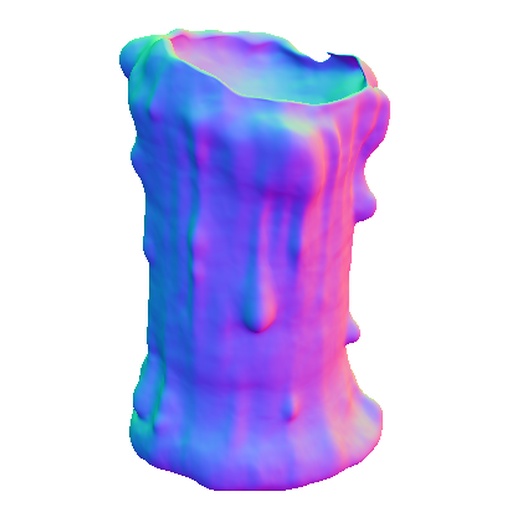} \\
  \end{tabular}
  \caption{\textbf{Geometry ablation: shading normals} on the candle.}
  \label{fig:abl_geo}
\end{figure}

\Cref{fig:geo_meshes,tab:geometry} compare the recovered geometry.

\Cref{fig:abl_geo,tab:priors} ablate the two priors of the SDF stage on the candle: the normal prior lowers the normal error from $21.40^\circ$ to $16.29^\circ$, the depth prior alone to $20.42^\circ$, and both together to $14.56^\circ$, with the best PSNR, SSIM and LPIPS; this ablation predates the silhouette loss, which lowers the error of the full geometry stage further, to $11.39^\circ$ over the candle and the statue~(\cref{tab:geometry}).

\subsection{Light-transport ablations and negative results}
\label{sec:supp:negative}
\textit{Extends main \cref{sec:method:transport}; the transport ablation cited in main \cref{sec:exp:share} is \cref{tab:ablation}.}
\begin{figure}[!tbp]
  \centering
  \includegraphics[width=\linewidth]{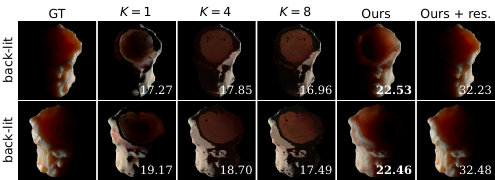}
  \caption{\textbf{More octaves vs.\ diffusion} on the synthetic candle, two held-out lights behind the object.
  $K$ columns: single scattering with $K$ octaves and no diffusion~(physics-only training, as in \cref{tab:ablation}).
  ``Ours'': \lucent's physics alone and with the object residual.
  Numbers: object PSNR~(dB); bold: best without the residual.
  More octaves do not carry light through the wax; diffusion does.}
  \label{fig:abl_octaves}
\end{figure}

\begin{table*}[!tbp]
\caption{\textbf{Ablations and negative results}, change against the corresponding reference run.}
\label{tab:negative}
\centering\small
\begin{tabular}{l ccc}
\toprule
Variant & $\Delta$ PSNR & $\Delta$ SSIM & $\Delta$ LPIPS \\
\midrule
Pin the diffusion gain to $(1-F_{dr})/4\pi$~(soap, back-lit) & $-1.6$ & $-0.157$ & $+0.012$ \\
Learned term before the background~(hair clip)               & $-1.17$ & $-0.023$ & $+0.002$ \\
Analytic Fresnel transmittance at the diffusion exit          & $0.00$ & $+0.001$ & $0.000$ \\
Learned entry/exit boundary networks around the solve         & $+0.24$ / $0.00$ & $+0.002$ / $0.000$ & $0.000$ / $0.000$ \\
Texture-resolution albedo field inside the physics            & $0.00$--$0.06$ & $-0.003$ to $+0.002$ & $-0.001$ to $0.000$ \\
Per-cell medium instead of one per object~(synthetic)        & $-2.18$ & -- & -- \\
\bottomrule
\end{tabular}
\end{table*}

\Cref{tab:ablation}, summarized in the main paper, compares diffusion with more scattering orders.
More octaves do not approach diffusion: $K{=}2$ is the closest variant~($1.1$\,dB below \lucent on back-lit views), while $K{=}4$ and $K{=}8$ are below single scattering over all views; the main paper quotes $K{=}8$.
With $K$ octaves, the single-scattering term becomes a sum over $k{=}0,\dots,K{-}1$ of the same term weighted by the albedo to the power $k{+}1$, with anisotropy $g/2^k$ and light visibility $V^{1/2^k}$, so that each octave stands for a higher scattering order with a broader lobe and a softer shadow.
\Cref{tab:negative} lists what we tried and did not keep.
Pinning the diffusion amplitude lets the extinction absorb the slack and miscolors back-lit views, so the gain stays a bounded learned scalar; a per-cell medium is not identifiable; and learned boundary networks or a finer albedo field gained at most $0.24$\,dB.
\FloatBarrier

\clearpage
\section{Relighting}
\label{sec:supp:compgroup}
\textit{Extends main \cref{sec:exp:setup,sec:exp:baselines}: the setup of the baselines, per-object results and failure cases, views at our median PSNR, objects that are not in the main paper, and relighting under environment light.}

\subsection{Baseline setup}
\paragraph{A numerical note on SSS-GS.}\label{sec:supp:sssgs} SSS-GS diverges because its sRGB conversion evaluates $x^{1/2.4}$ for every pixel, including pixels at or below zero that it then discards; the gradient there is infinite, and the masked selection turns it into NaN.
Clamping the argument before the power leaves every rendered value unchanged and removes the NaN.
All SSS-GS numbers in this paper use this fix: trained with it, SSS-GS runs 60k steps with no non-finite step and matches the earlier guarded run on the car~($31.42$ against $31.35$\,dB).

\paragraph{Calibration.}\label{sec:supp:calib}
The LED power $\mathbf{P}_m$ and the exposure $\boldsymbol{\epsilon}_v$ are learned in the appearance stage from the training images only, and looked up by LED and pose at test time; no scored view is a training image.
Every scored camera is a training camera, seen under other LEDs.
On eight of the ten real objects every scored LED is trained at other cameras; on the blue bottle and the physalis the scored LEDs never appear in training and keep $\mathbf{P}_m{=}1$.
At the cameras the residual never saw~(\cref{tab:unseen}), $\boldsymbol{\epsilon}_v{=}1$.
The synthetic objects use one global light color.
Their exposure is keyed by image file name rather than by pose, so one exposure is learned per training image~($11{,}200$ per object), and a test image whose file name matches a training image takes that image's exposure instead of $\boldsymbol{\epsilon}_v{=}1$.
SSS-GS has no such calibration, and the pose and light refinement of SSD-GS is off in our runs.
We train SSD-GS with its released configuration for the synthetic scenes of this dataset, which leaves both off; its release has no configuration for these real captures, and its script for other real data turns both on.
That script also trains longer~($100$k instead of $60$k iterations).
Retrained with it on the ten real captures~(\cref{tab:objects_full}), SSD-GS scores $33.92$\,dB on average against $33.95$\,dB with both off, between $-0.7$\,dB~(car) and $+0.7$\,dB~(blue bottle) per object, and our mean lead goes from $1.73$ to $1.76$\,dB.
On the blue bottle SSD-GS then passes \lucent~($34.22$ against $33.70$\,dB).
In the release's protocol, which these numbers follow, SSD-GS fits the camera pose and the LED position of each test view on the test image itself, which \lucent does not.
Without that fit, the same retrained model scores $33.99$\,dB with the corrections carried over from the training images and $33.87$\,dB with the test views uncorrected.

\paragraph{The scatter term of SSD-GS on the real captures.}\label{sec:supp:ssdgsscatter}
On most real captures, the scatter term of SSD-GS carries little or none of its image.
Switching it off on the same checkpoint leaves the car unchanged to every digit we report~(\cref{tab:share}).
Scaling its scattering coefficients~(\cref{sec:supp:ssdgsedit}) leaves the back-lit views of five of the ten objects exactly unchanged at every multiplier~(car, blue bottle, liqueur, joyful yell and physalis) and changes those of three more by at most $8\%$~(\cref{tab:direction}).
On these objects its diffuse and specular Gaussians carry its appearance, so the editing comparison shows that SSD-GS has no subsurface term left to edit there, not that its edit moves the wrong way.
We use its released configuration and did not tune it to make the scatter term active.

Giving both \lucent and SSD-GS the best brightness gain per test view lowers our lead on the turbid real objects~(car, marmalade, massage ball, joyful yell and physalis) from $2.4$ to $1.9$\,dB and changes nothing on the synthetic objects, so about $0.5$\,dB of that lead comes from per-view brightness.

\subsection{Per-object results and failure cases}
\label{sec:supp:errors}
\label{sec:exp:objects}
\TabObjectsFull

\Cref{tab:objects_full} gives every object and metric behind the means of \Cref{tab:objects}.
Retrained from scratch with two further random seeds~(both stages and the residual, with the same stored normal priors), \lucent scores $35.96\pm0.14$\,dB on the car and $42.75\pm0.13$\,dB on the soap with the residual, and $30.39\pm0.03$ and $31.52\pm0.18$\,dB physics only~(mean $\pm$ standard deviation over the three runs, including the one in \cref{tab:objects_full}).

\begin{figure*}[!tp]
  \centering
  \includegraphics[width=\textwidth,height=0.88\textheight,keepaspectratio]{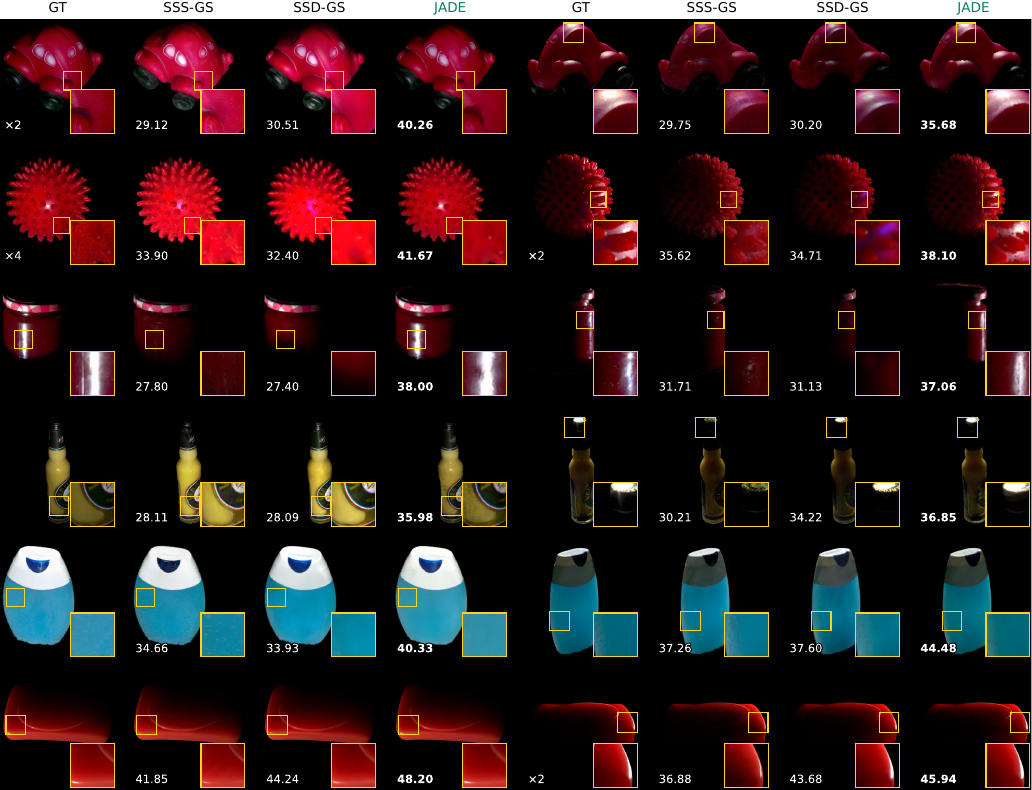}
  \caption{\textbf{Qualitative comparison, six objects}~(extends \cref{fig:qual}).
  Rows, top to bottom: car, massage ball, marmalade, liqueur and blue bottle~(real) and soap~(synthetic), each in one front-lit~(left) and one back-lit~(right) held-out view, selected among the views where \lucent leads both baselines.
  Insets magnify the boxed region.
  Numbers: PSNR~(dB) on black, best per view in bold; dark views brightened by the factor on the GT.}
  \label{fig:qual_full}
\end{figure*}

\Cref{fig:qual_full} extends the qualitative comparison of the main paper to six objects, and \Cref{fig:objects6} in \Cref{sec:supp:sharegroup} shows the physics-only decomposition of six of the ten real objects.

\begin{figure}[!tbp]
  \centering
  \includegraphics[width=\linewidth]{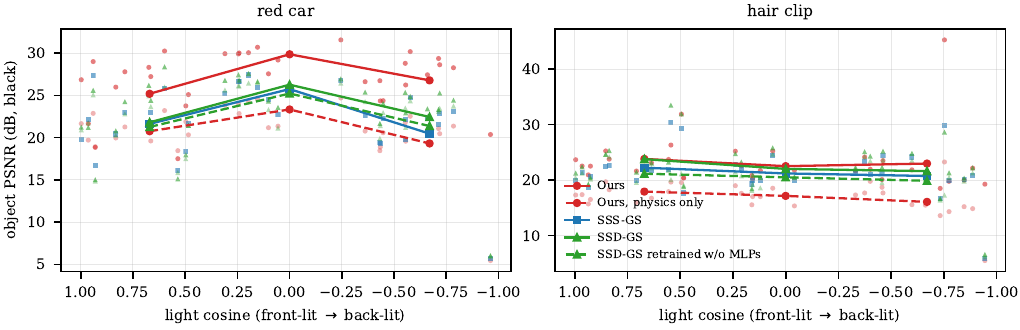}
  \caption{\textbf{Accuracy by lighting direction.}
  Object PSNR per view~(dots) and per bin~(lines: front-lit, cosine $>1/3$; side; back-lit, $<-1/3$; 12/8/12 views on the car, 12/7/13 on the clip); \lucent with the $150.8$k residual.
  The isolated views near cosine $-1$ are glare views, where every method scores lowest.}
  \label{fig:lightangle}
\end{figure}

\paragraph{By lighting direction.}
\Cref{fig:lightangle} bins the views by the angle between light and camera~(object PSNR), with the selected $150.8$k residual, on the $32$ views and the earlier $32$-light fits of \Cref{tab:share}.
On the car, \lucent leads SSD-GS, the stronger baseline, in every bin: by $3.4$\,dB front-lit~($25.19$ against $21.81$\,dB), $3.6$\,dB side-lit and $4.3$\,dB back-lit~($26.78$ against $22.45$\,dB).
On the clip it ties SSD-GS front-lit~($23.80$ against $23.76$\,dB) and leads by $0.5$\,dB side-lit and $1.4$\,dB back-lit.
The advantage grows where translucency is most visible.
The physics-only render, however, is furthest behind there~($3.1$\,dB below SSD-GS back-lit on the car, $5.5$\,dB on the clip), so this gain comes from the residual correcting the homogeneous model where it is weakest.
On the $256$ views of \Cref{tab:objects_full}, with the $64$-light fits and whole-image PSNR, \lucent also leads on the clip~($38.48$ against $38.00$\,dB for SSD-GS).

\paragraph{Camera glare.}
Much of the missing light is camera glare rather than transport.
It has the color of the LED rather than of the object~(blue as strong as red on the red ball), it is spread over the whole object and adds to it, and it grows as the LED approaches the optical axis~(correlation $-0.57$ to $-0.81$ over the views of each object).
Views with the LED within $15^\circ$ of the axis hold $36$--$68\%$ of the deficit~($91\%$ on the physalis); without them, the physics alone gives $0.40$--$0.86$ of the ground-truth radiance.
The glare belongs to the capture, not to the object, so it should not enter the optical constants: the residual fits it together with the transport the homogeneous medium misses, and switching the residual off removes it.
On the eight real objects other than the car and the clip~(those of \cref{tab:unseen}), the residual also fits it better than the baselines: in the $6$--$16$ views per object with the LED within $15^\circ$ of the axis, \lucent leads the best baseline by $1.5$--$10.4$\,dB on all eight.
Over all views of these eight objects in \cref{tab:objects_full}, \lucent leads on five, trails the best baseline on the blue bottle, the clear crystal and the physalis~(by $0.73$, $0.45$ and $0.50$\,dB), and leads on average by $1.7$\,dB~($35.30$ against $33.60$\,dB for both baselines).
Without the glare views it still wins the same five~($+1.4$ to $+3.2$\,dB) and trails on the same three~(by $0.99$, $0.56$ and $0.73$\,dB), and the mean lead shrinks from $1.7$ to $1.4$\,dB~($35.43$ against $34.00$\,dB for SSS-GS and $33.95$\,dB for SSD-GS).

\paragraph{Silhouette of the synthetic objects.}
Without the silhouette loss, most of the gap to SSD-GS on the synthetic objects lay on the silhouette, not inside the object where subsurface scattering acts: a band of $4$\,px on either side of the outline held $44\%$ of our error on the candle and $43\%$ on the statue, against $16\%$ and $13\%$ for SSD-GS, and with this band replaced by the ground truth for both methods the gap would fall from $3.5$ to $1.6$\,dB on the candle and from $2.6$ to $0.7$\,dB on the statue.
On the candle, about half of the band error came from coverage: our silhouette extended slightly beyond the true outline~(mean opacity error $0.15$, against $0.02$ for SSD-GS), while its rim was no softer than the ground truth.
This motivates $\mathcal{L}_{\text{sil}}$, which constrains the zero level set at the outline; with it, the gap is $1.6$\,dB on the candle and $0.5$\,dB on the statue.
Pixels that our render misses entirely~(\eg a lost wick) cover at most $0.1\%$ of the image and cost $0.1$\,dB on the candle and $0.6$\,dB on the statue in these runs.
Fitting the residual on all $112$ lights adds only $0.2$--$0.5$\,dB.

\begin{figure*}[!tp]
  \centering
  \newlength{\figFlW}\setlength{\figFlW}{0.185\textwidth}
  \setlength{\tabcolsep}{0.5pt}\renewcommand{\arraystretch}{0.6}\scriptsize
  \begin{tabular}{@{}c@{\hspace{2pt}}ccccc@{}}
   & GT & Ours~(physics) & Surface $S$ & Single scatt.\ $M$ & Diffusion $D$ \\
  \rotatebox{90}{\,clear crystal} & \includegraphics[width=\figFlW]{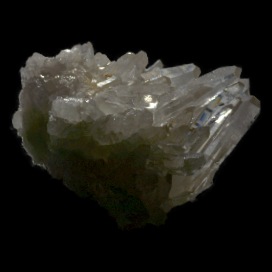} & \includegraphics[width=\figFlW]{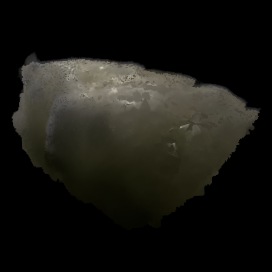} & \includegraphics[width=\figFlW]{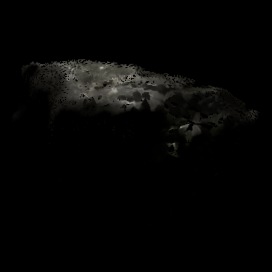} & \includegraphics[width=\figFlW]{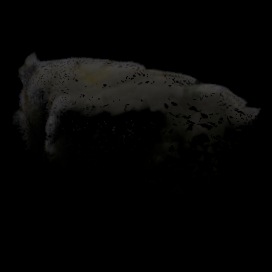} & \includegraphics[width=\figFlW]{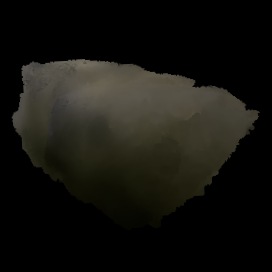} \\
  \rotatebox{90}{\,nail polish rem.} & \includegraphics[width=\figFlW]{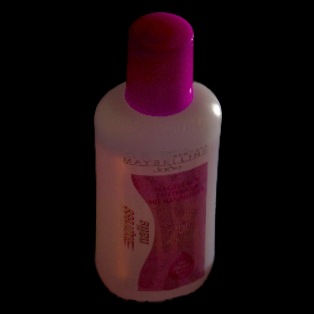} & \includegraphics[width=\figFlW]{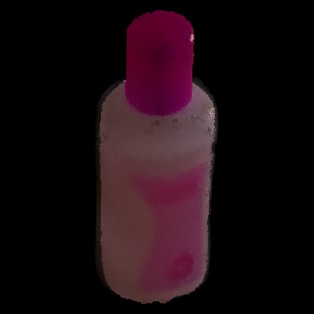} & \includegraphics[width=\figFlW]{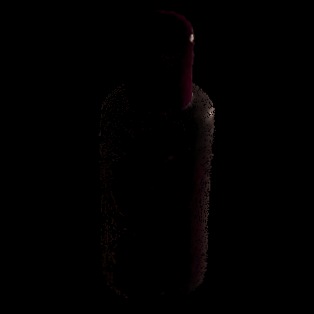} & \includegraphics[width=\figFlW]{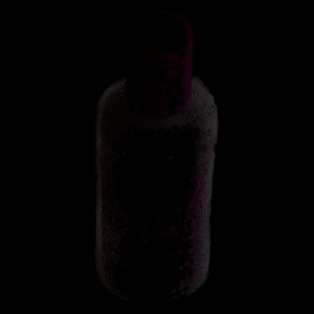} & \includegraphics[width=\figFlW]{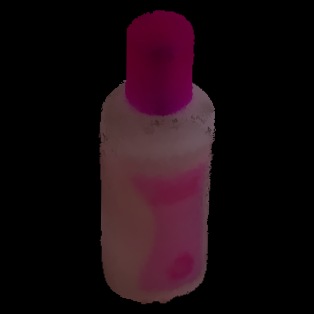} \\
  \end{tabular}
  \caption{\textbf{Failure cases.}
  The two glass objects, whose appearance is dominated by refraction, which a diffusion medium does not model~(held-out back-lit light); in PSNR, \lucent trails SSD-GS by $0.5$\,dB on the clear crystal and leads on the nail polish remover~(\cref{tab:objects_full}).}
  \label{fig:failures}
\end{figure*}

\Cref{fig:failures} shows the two glass objects, whose appearance is dominated by refraction, which a diffusion medium does not model.

\subsection{Views at our median PSNR}
\label{sec:supp:median}
\begin{figure*}[!tp]
\centering
\includegraphics[width=\textwidth,height=0.9\textheight,keepaspectratio]{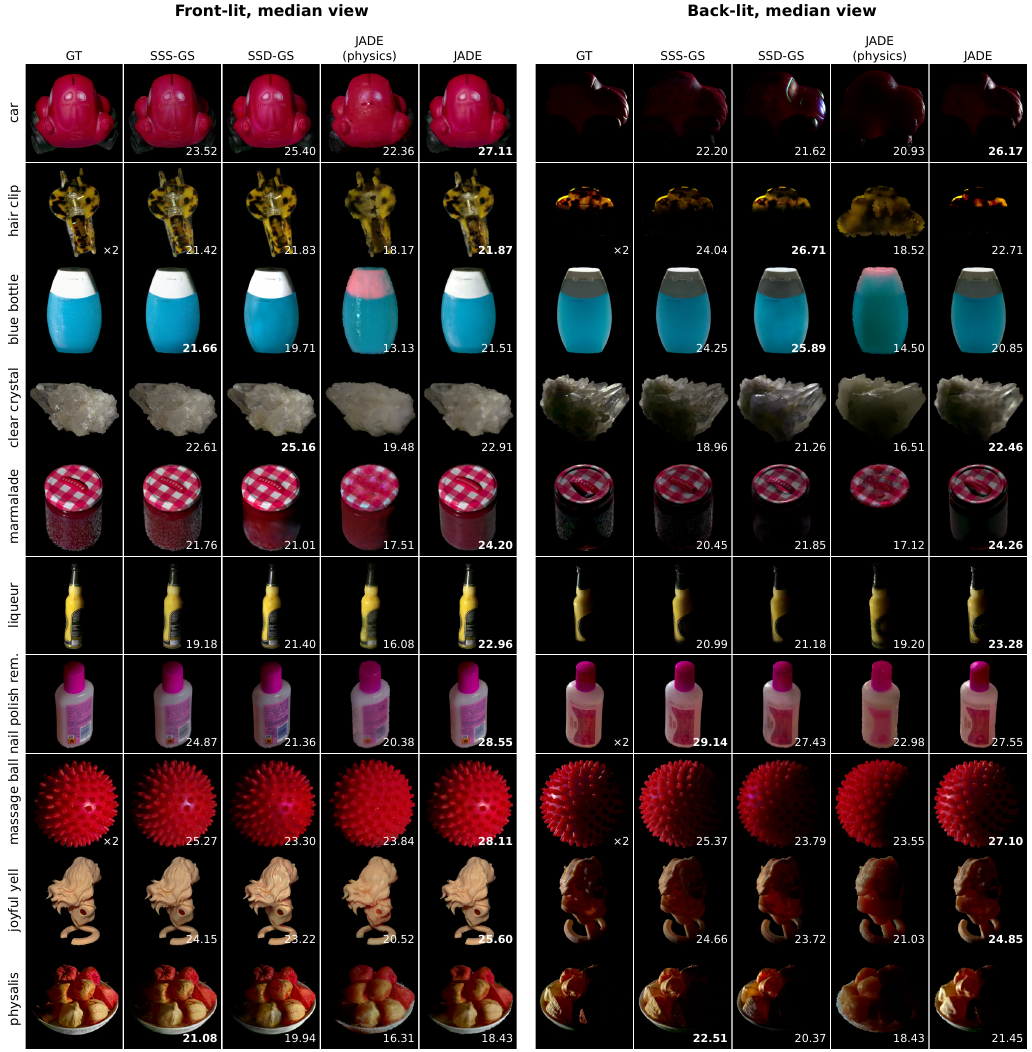}
\caption{\textbf{Real captures at our median PSNR}: for each object, the front-lit and the back-lit held-out view at the median whole-image PSNR of \lucent within that group. Numbers: object PSNR~(dB), best of the three full methods in bold; one display gain per view from the ground truth.}
\label{fig:median_real}
\end{figure*}

\begin{figure*}[!tp]
\centering
\includegraphics[width=\textwidth]{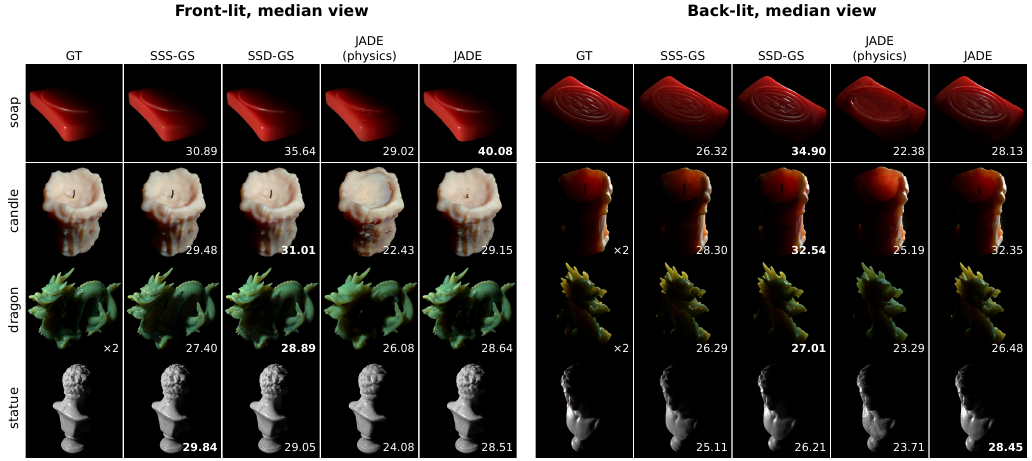}
\caption{\textbf{Synthetic objects at our median PSNR}, as in \cref{fig:median_real}.}
\label{fig:median_syn}
\end{figure*}

Main \cref{fig:qual} shows views where \lucent leads both baselines.
\Cref{fig:median_real,fig:median_syn} instead show, for every object, the held-out front-lit and back-lit view whose whole-image PSNR of \lucent is the median of its group, next to both baselines and to our physics-only render; the views are chosen by our score only.
Over all held-out views, \lucent has a higher whole-image PSNR than both baselines on $67\%$ of the views of the real objects~($32$--$97\%$ per object) and on $26\%$ of those of the synthetic objects~($9$--$62\%$), where SSD-GS leads on average~(main \cref{tab:objects}).
At the $20$ median views of the real objects, \lucent is first in whole-image PSNR in $11$, second in $6$ and third in $3$; at the $8$ of the synthetic objects, first in $2$, second in $3$ and third in $3$.
The numbers printed in \cref{fig:median_real,fig:median_syn} are object PSNR, by which the ranks at the same views differ~(real: first in $13$, second in $4$, third in $3$; synthetic: first in $2$, second in $4$, third in $2$).

\subsection{Objects not in the main paper}
\label{sec:supp:newobjects}
\begin{figure*}[!tp]
\centering
\includegraphics[width=0.92\textwidth]{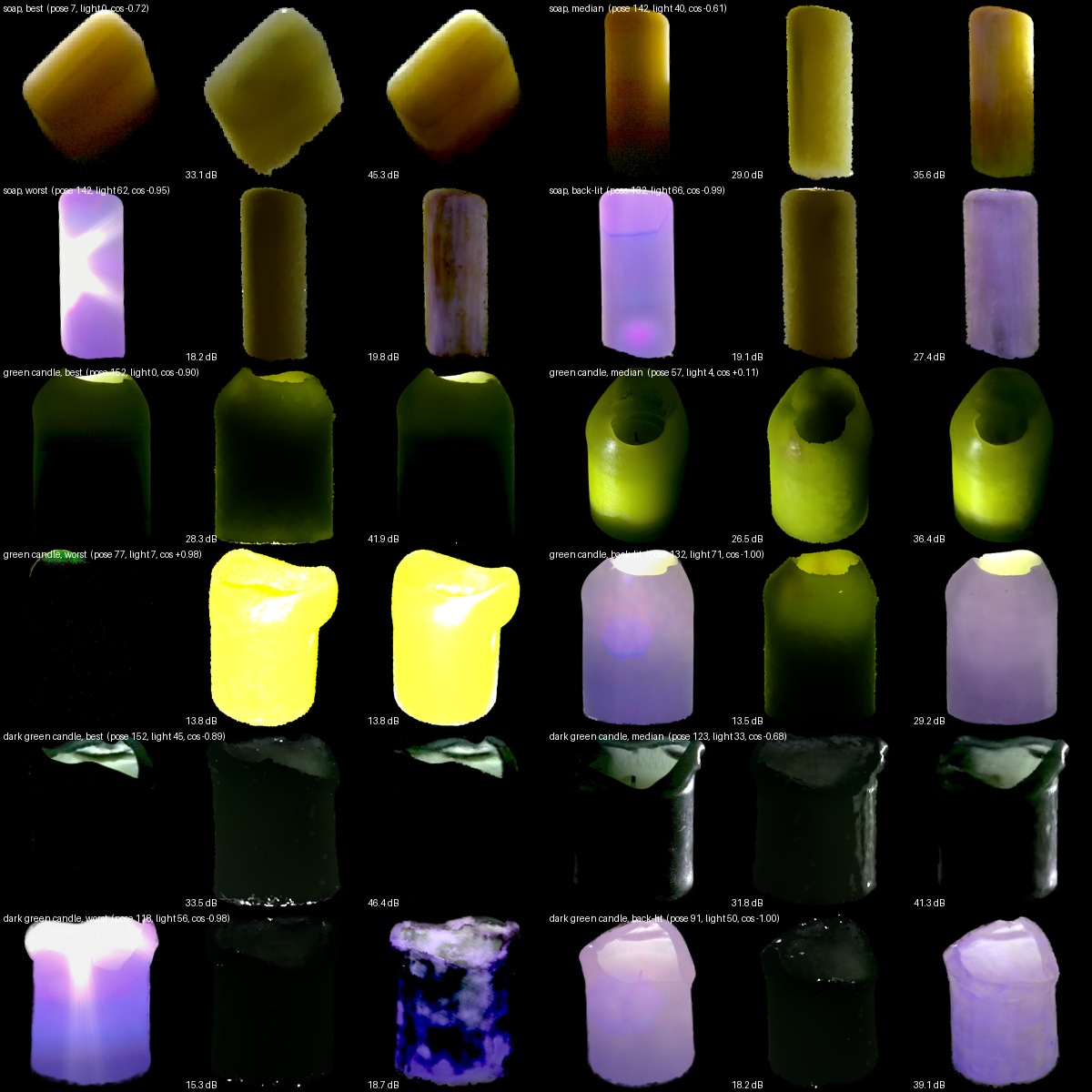}
\caption{\textbf{Objects not in the main paper}: for each object, four held-out views, chosen by the PSNR with the residual~(best, median, worst) and as the most back-lit view; each as photograph, \lucent physics only and \lucent with the residual. Numbers: PSNR~(dB) of the two renders; one display gain per view. The labels give the camera pose, the held-out light and the cosine between the directions to light and camera.
The purple wash in the photographs of the back-lit views is lens glare from the light in front of the camera, which the physics does not model. On the dark green candle, the recovered medium is almost opaque, so the physics-only render is much darker than the photograph and misses its white cut face. In the worst view of the green candle the photograph is almost black, which looks like a dim or failed light in that image.}
\label{fig:newobjects}
\end{figure*}

To test the configuration on objects it was not developed on, we trained three further real captures of the dataset of SSS-GS~\cite{dihlmann2024sssgs}: a soap~(a real capture, not the synthetic soap of the main paper), a green candle and a dark green candle.
Each was trained once, end to end, with the released code at its defaults~(physics, then the object residual); nothing was tuned, rerun or selected, and the only per-object settings concern data: one camera of the green candle is left out, and the residual of the dark green candle is fitted on its evaluation cameras~(below).
They are scored as the real objects of the main paper, on $256$ views~($32$ cameras, each under $8$ held-out lights) on black~(\cref{tab:newobjects,fig:newobjects}).
No baseline was trained on them, so there is no comparison.
For scale, the ten real objects of the main paper span $25.5$--$34.7$\,dB physics only and $29.0$--$40.2$\,dB with the residual~(\cref{tab:objects_full}).

\begin{table}[!tbp]\centering\small\setlength{\tabcolsep}{3pt}
\caption{\textbf{Objects not in the main paper}, trained once with the released code at its defaults: whole-image PSNR~(dB), SSIM and LPIPS on black over $256$ held-out-light views, and training time on one shared A100.
$^\dagger$Only $36$ cameras are usable, so its residual is fitted on the evaluation cameras~(under the training lights only); not the same test as the other rows.}
\label{tab:newobjects}
\resizebox{\linewidth}{!}{
\begin{tabular}{l ccc ccc c}
\toprule
 & \multicolumn{3}{c}{Physics only} & \multicolumn{3}{c}{+ residual} & \\
\cmidrule(lr){2-4}\cmidrule(lr){5-7}
Object & PSNR$\uparrow$ & SSIM$\uparrow$ & LPIPS$\downarrow$ & PSNR$\uparrow$ & SSIM$\uparrow$ & LPIPS$\downarrow$ & Time \\
\midrule
Soap~(real) & 31.03 & 0.656 & 0.024 & 35.80 & 0.785 & 0.010 & 4.5\,h \\
Green candle & 27.62 & 0.626 & 0.078 & 35.61 & 0.787 & 0.045 & 6.7\,h \\
Dark green candle & 32.00 & 0.408 & 0.041 & 39.45$^\dagger$ & 0.703$^\dagger$ & 0.023$^\dagger$ & 5.1\,h \\
\bottomrule
\end{tabular}}
\end{table}

All three fall within the range of the main paper, physics only and with the residual; on black, however, PSNR hides how dark the physics-only render of the dark green candle is.
What these runs show beyond the main paper:
\begin{itemize}\setlength{\itemsep}{1pt}
\item \textbf{The dark green candle is a failure of the physics.} Its recovered medium is almost opaque~(the diffusion gain ends at its lower bound, $0.3$), and the physics-only render is much darker than the photograph in the views of \cref{fig:newobjects}~(SSIM $0.41$); the residual adds $7.5$\,dB. Only $36$ of its cameras are usable, so its residual is fitted on the $32$ evaluation cameras, whose held-out lights it still never sees, and only four cameras are never seen; its with-residual numbers are therefore not the same test as on the other objects.
\item \textbf{Views straight into the light} are washed by lens glare in the photographs; the physics does not model it and scores $13$--$19$\,dB on these views, and the residual learns part of it. Views below $20$\,dB, physics only / with residual: soap $3/1$, green candle $10/6$, dark green candle $8/2$, of $256$ each.
\item \textbf{Bounds.} On the soap the diffusion gain ends at its upper bound~($10$), on the dark green candle at its lower bound.
\item \textbf{Data.} One camera of the green candle, whose pose is corrupted in the dataset, is left out of training, as five cameras of the joyful yell are in the main paper.
\end{itemize}

\subsection{Environment lighting}
\label{sec:supp:envgroup}
\label{sec:exp:envlight}
\textit{Extends the environment-map relighting of main \cref{sec:exp:editing}~(\cref{fig:envmap_main}).}

\begin{figure*}[!tp]
  \centering
  \newlength{\figEnvW}\setlength{\figEnvW}{0.105\textwidth}
  \newlength{\figEnvT}\setlength{\figEnvT}{0.316\textwidth}
  \setlength{\tabcolsep}{0.5pt}\renewcommand{\arraystretch}{0.6}\scriptsize
  \begin{tabular}{@{}c@{\hspace{4pt}}ccc@{\hspace{4pt}}ccc@{\hspace{4pt}}ccc@{}}
   & \multicolumn{3}{c}{\includegraphics[width=2\figEnvW]{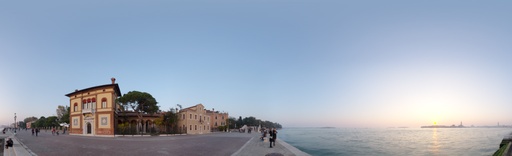}} & \multicolumn{3}{c}{\includegraphics[width=2\figEnvW]{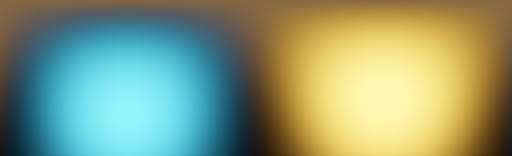}} & \multicolumn{3}{c}{\includegraphics[width=2\figEnvW]{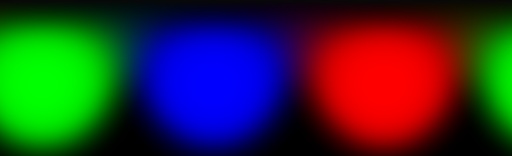}} \\
   & \multicolumn{3}{c}{sunset} & \multicolumn{3}{c}{warm cool} & \multicolumn{3}{c}{rgb tri} \\
  \cmidrule(lr){2-4}\cmidrule(lr){5-7}\cmidrule(lr){8-10}
   & GT & Physics & + residual & GT & Physics & + residual & GT & Physics & + residual \\[2pt]
  \rotatebox{90}{\hspace{2pt}Candle} & \includegraphics[width=\figEnvW]{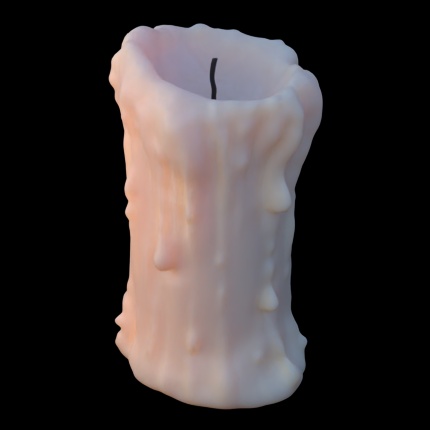} & \includegraphics[width=\figEnvW]{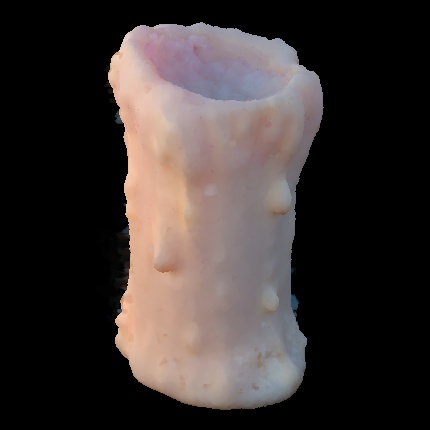} & \includegraphics[width=\figEnvW]{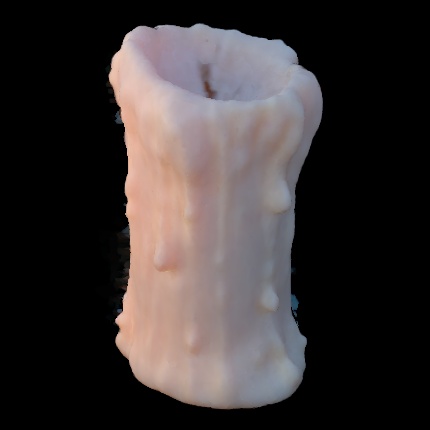} & \includegraphics[width=\figEnvW]{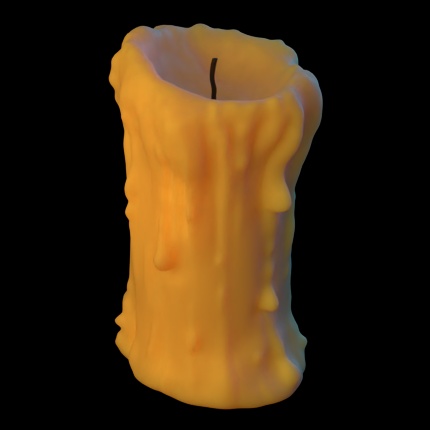} & \includegraphics[width=\figEnvW]{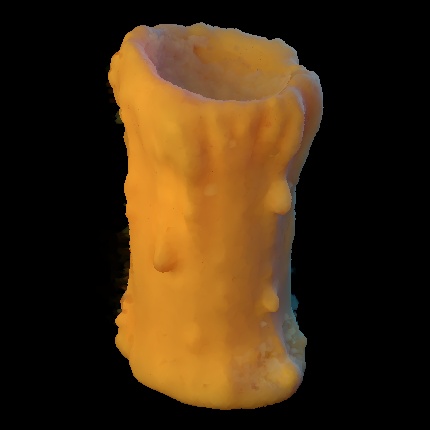} & \includegraphics[width=\figEnvW]{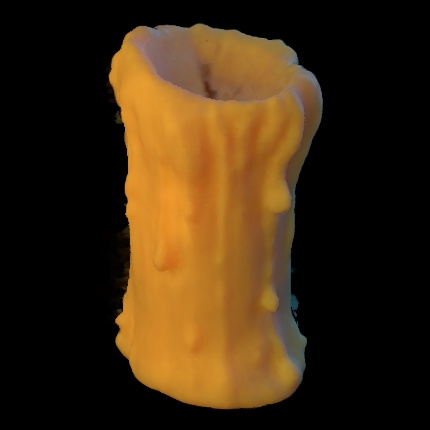} & \includegraphics[width=\figEnvW]{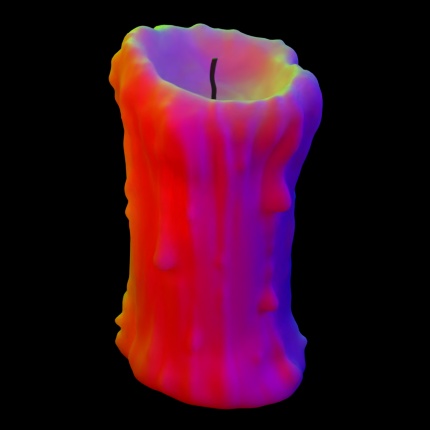} & \includegraphics[width=\figEnvW]{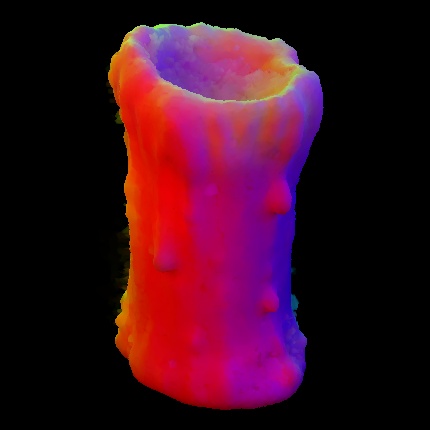} & \includegraphics[width=\figEnvW]{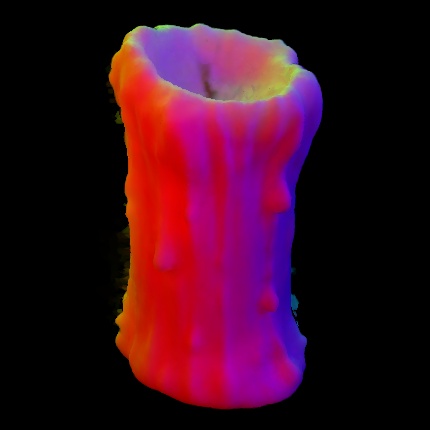} \\
   &  & 26.56 & 28.05 &  & 27.93 & 30.41 &  & 24.78 & 29.47 \\[2pt]
  \rotatebox{90}{\hspace{2pt}Soap} & \includegraphics[width=\figEnvW]{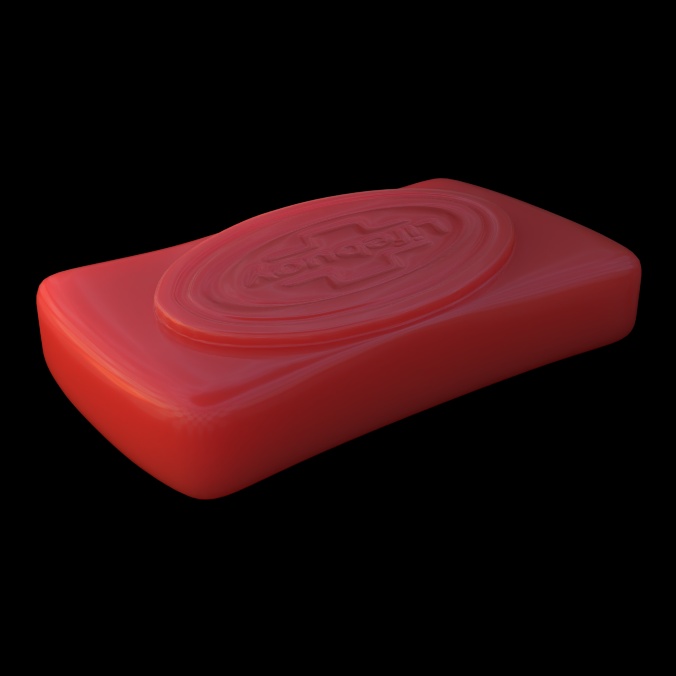} & \includegraphics[width=\figEnvW]{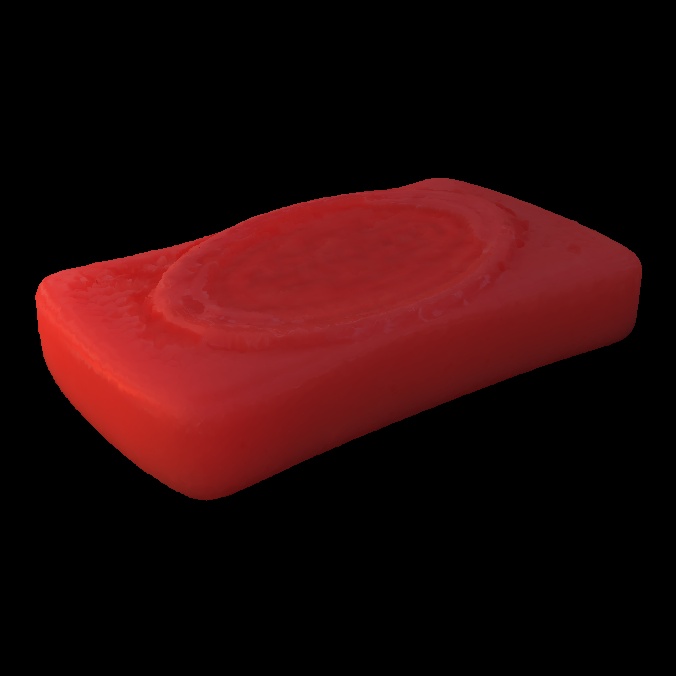} & \includegraphics[width=\figEnvW]{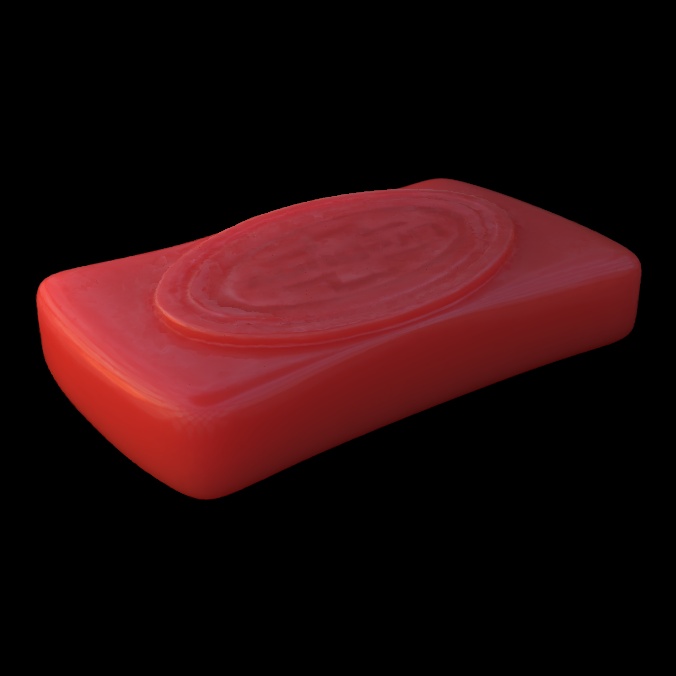} & \includegraphics[width=\figEnvW]{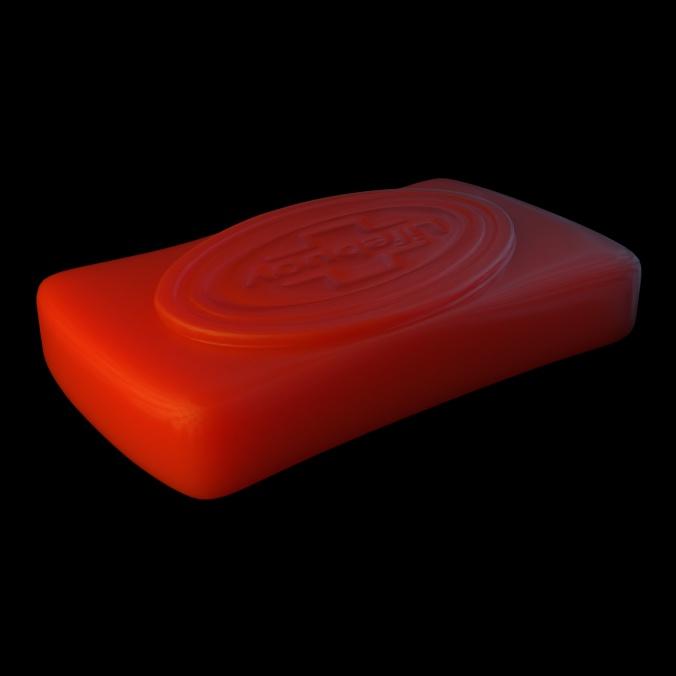} & \includegraphics[width=\figEnvW]{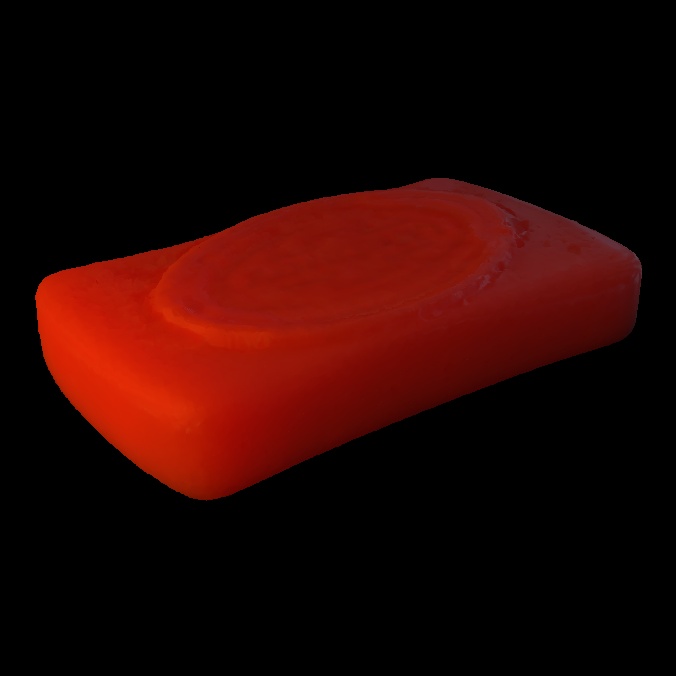} & \includegraphics[width=\figEnvW]{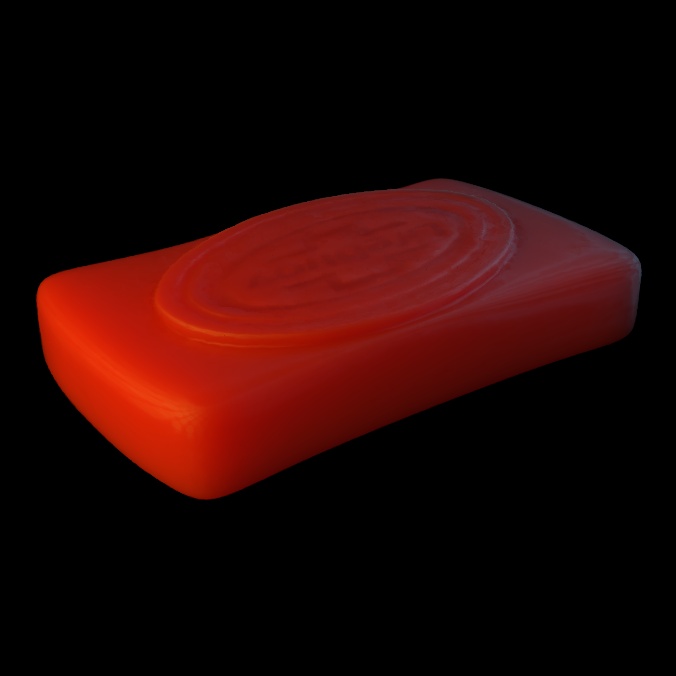} & \includegraphics[width=\figEnvW]{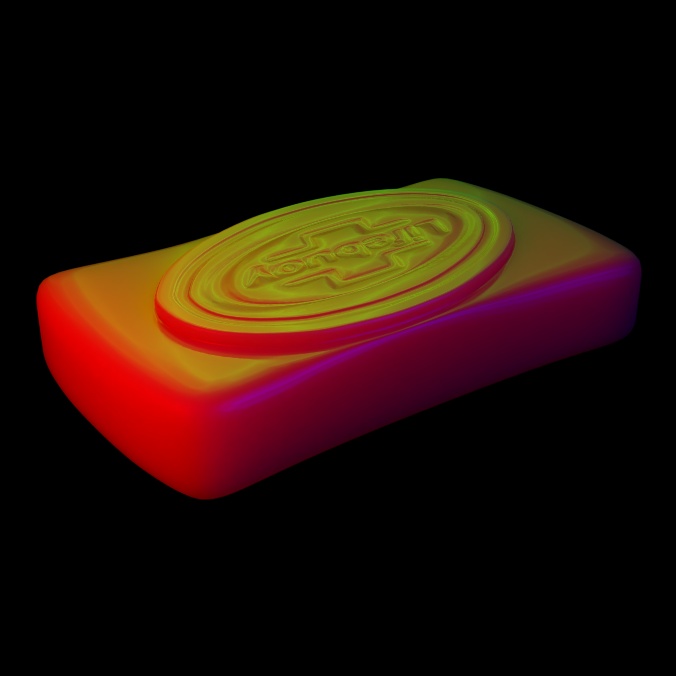} & \includegraphics[width=\figEnvW]{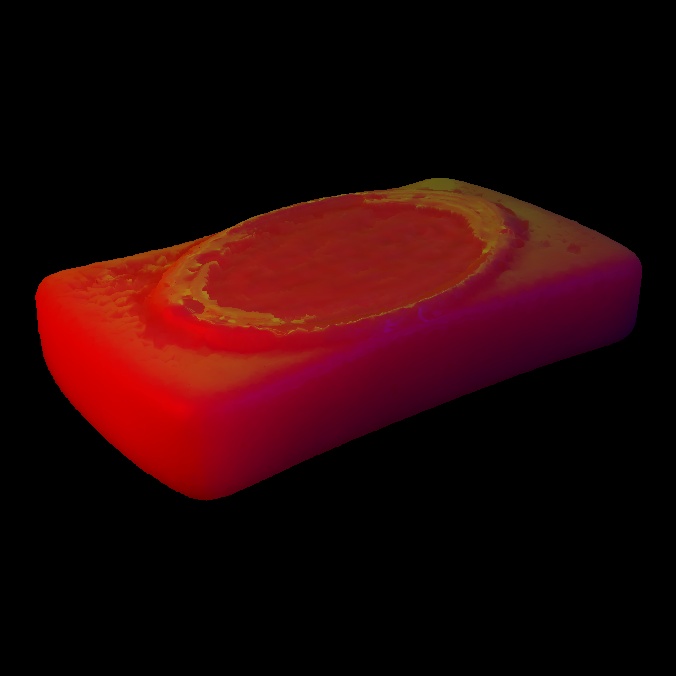} & \includegraphics[width=\figEnvW]{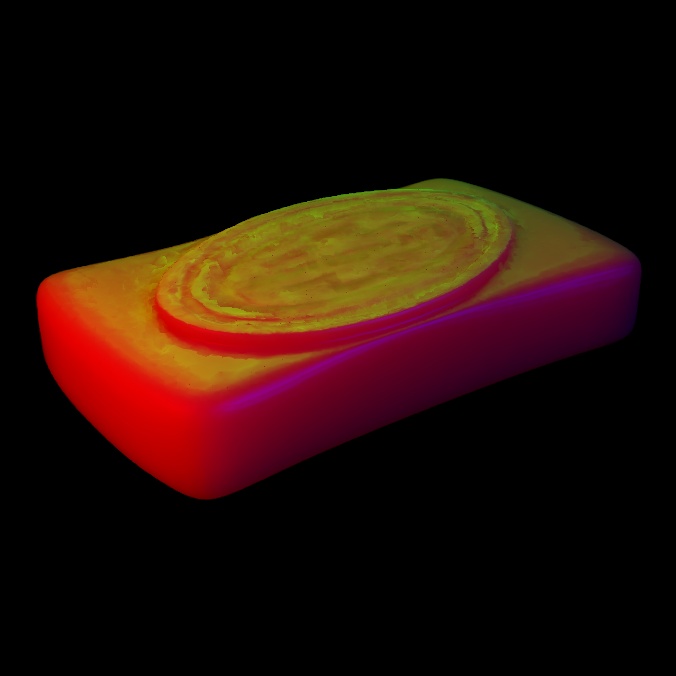} \\
   &  & 24.36 & 34.83 &  & 27.45 & 38.49 &  & 17.85 & 29.12 \\[2pt]
  \rotatebox{90}{\hspace{2pt}Dragon} & \includegraphics[width=\figEnvW]{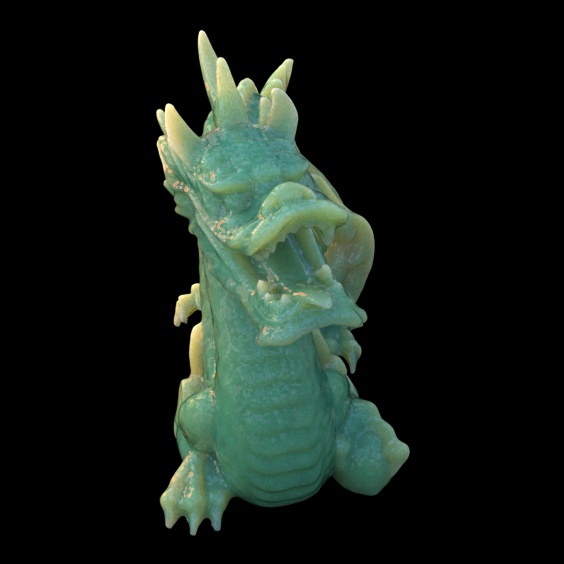} & \includegraphics[width=\figEnvW]{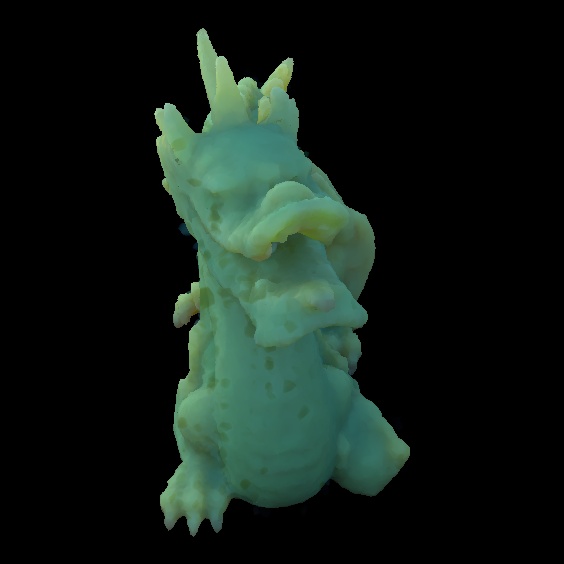} & \includegraphics[width=\figEnvW]{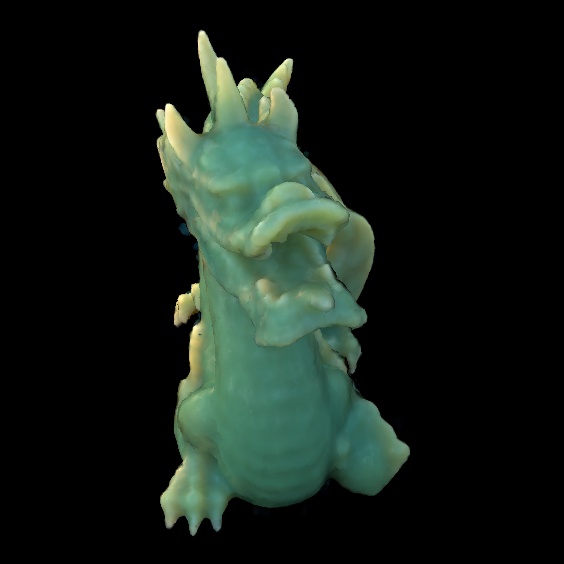} & \includegraphics[width=\figEnvW]{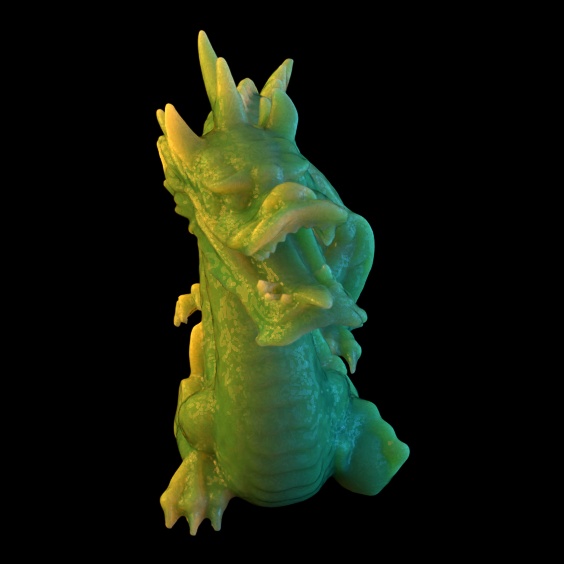} & \includegraphics[width=\figEnvW]{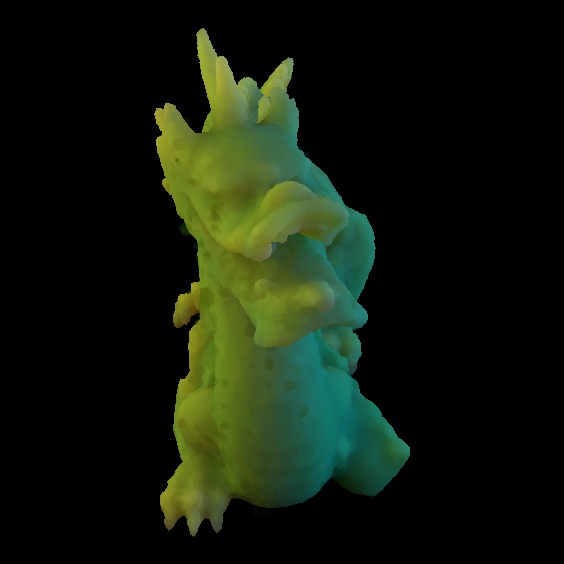} & \includegraphics[width=\figEnvW]{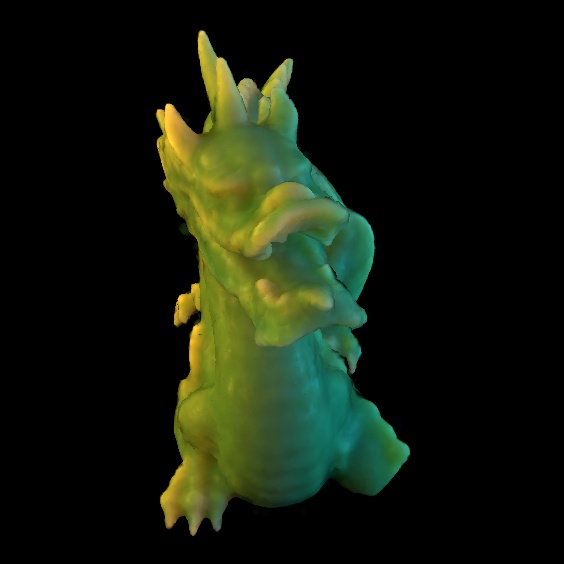} & \includegraphics[width=\figEnvW]{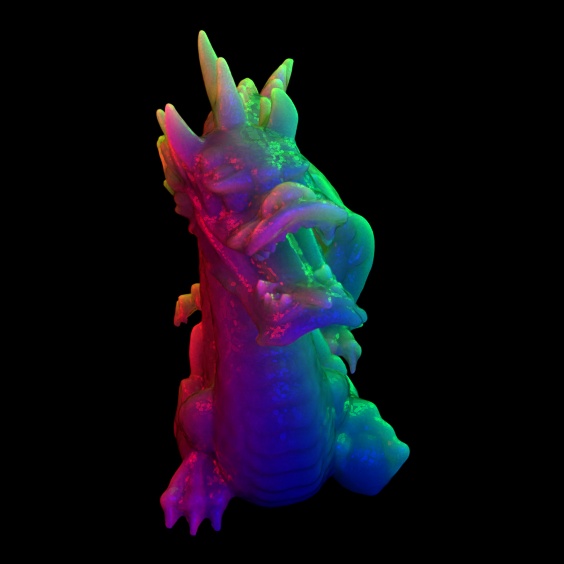} & \includegraphics[width=\figEnvW]{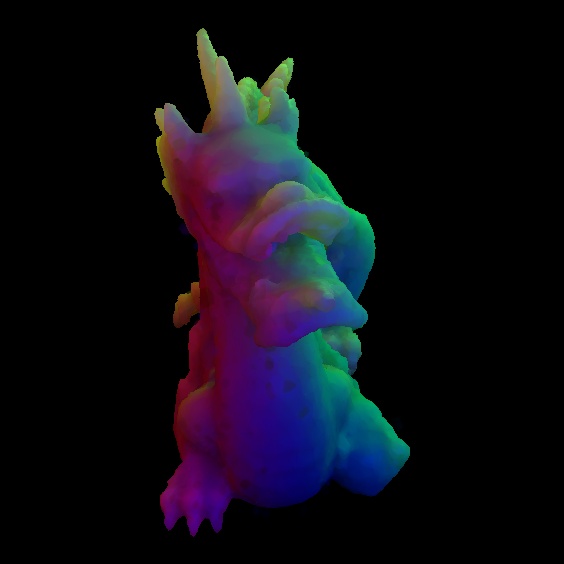} & \includegraphics[width=\figEnvW]{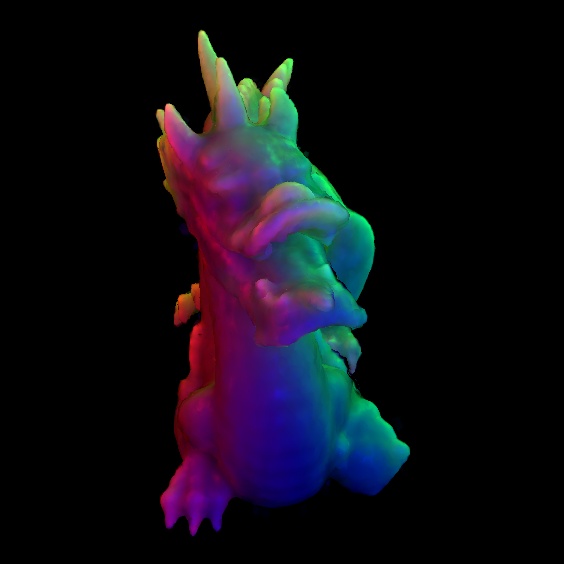} \\
   &  & 20.92 & 25.71 &  & 21.73 & 26.50 &  & 21.63 & 25.90 \\[2pt]
  \rotatebox{90}{\hspace{2pt}Statue} & \includegraphics[width=\figEnvW]{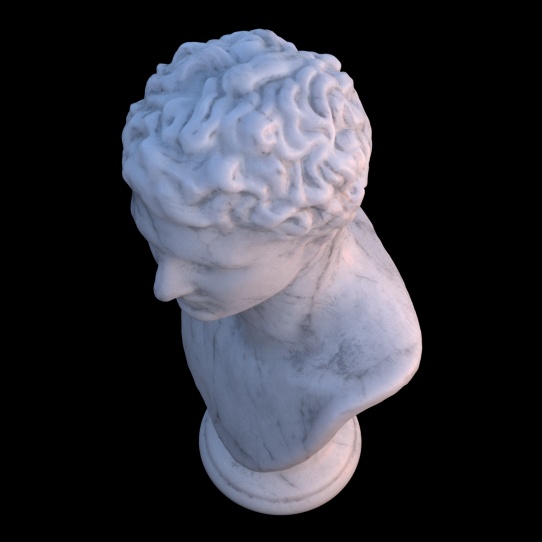} & \includegraphics[width=\figEnvW]{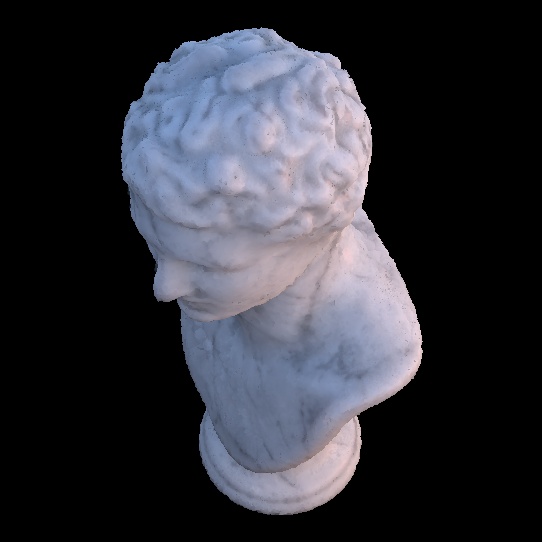} & \includegraphics[width=\figEnvW]{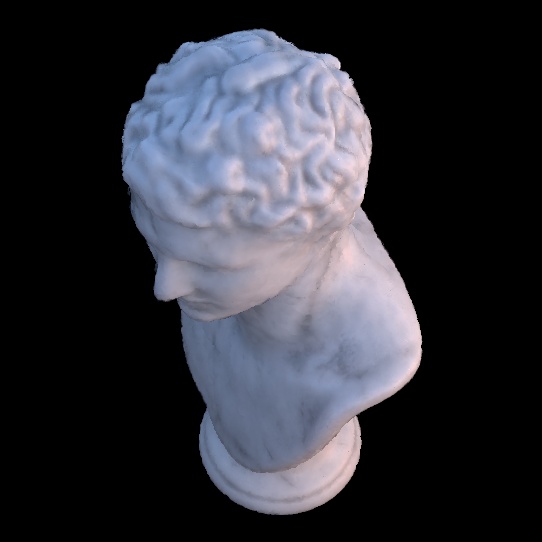} & \includegraphics[width=\figEnvW]{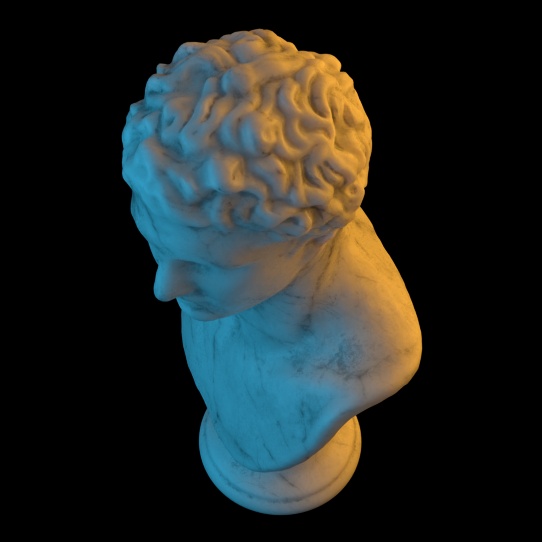} & \includegraphics[width=\figEnvW]{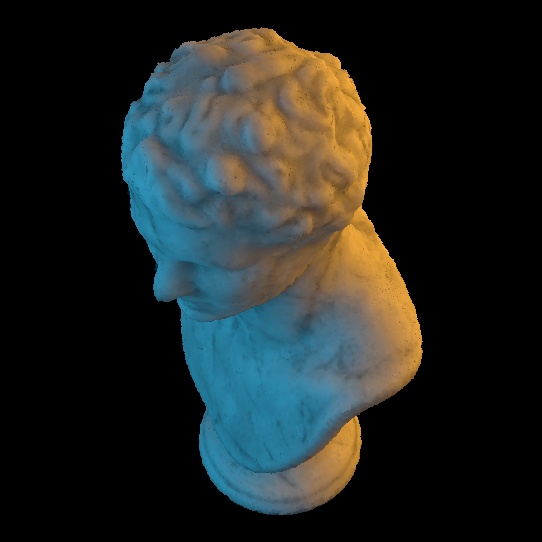} & \includegraphics[width=\figEnvW]{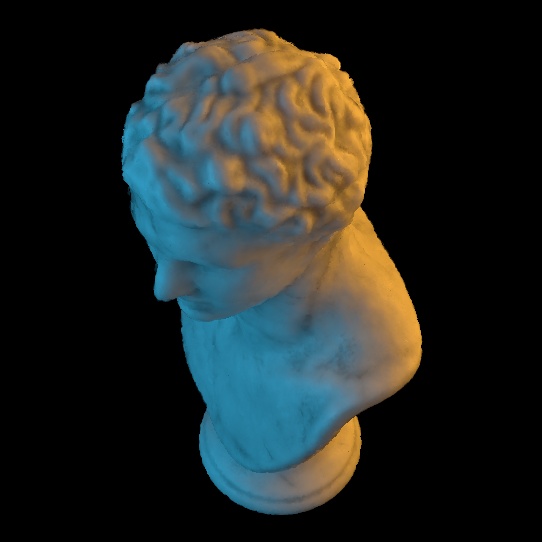} & \includegraphics[width=\figEnvW]{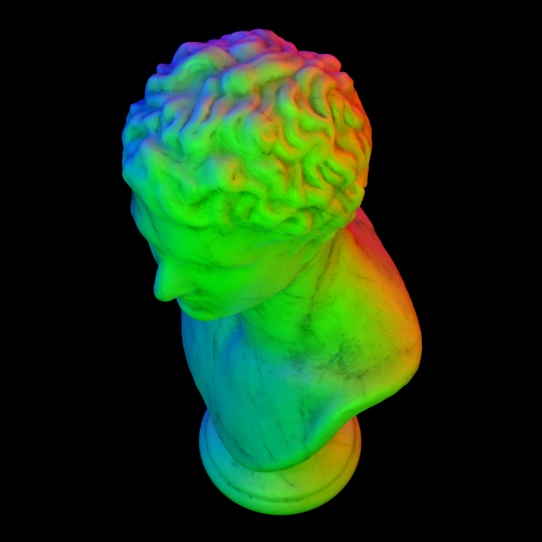} & \includegraphics[width=\figEnvW]{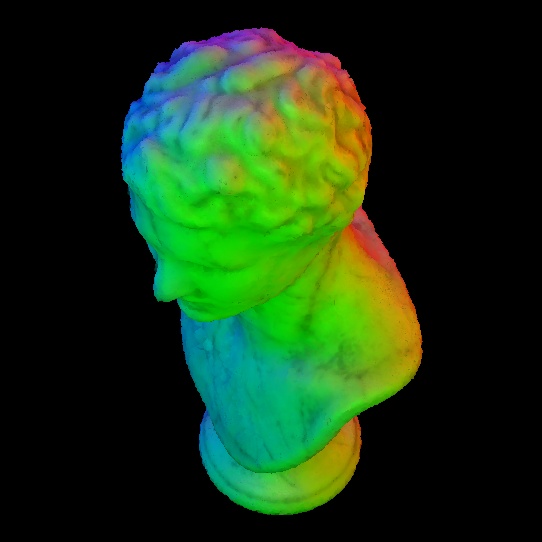} & \includegraphics[width=\figEnvW]{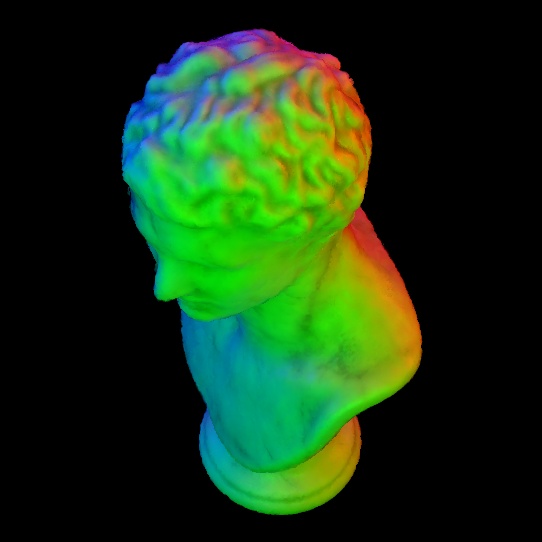} \\
   &  & 27.73 & 30.80 &  & 27.92 & 31.79 &  & 25.68 & 31.11 \\[2pt]
  \end{tabular}
  \caption{\textbf{Relighting under environment maps.}
  Held-out views; numbers are object PSNR~(dB).}
  \label{fig:envlight}
\end{figure*}

\begin{table*}[!tp]
\caption{\textbf{Relighting under environment maps}, synthetic objects.
Top: mean over objects; bottom: PSNR~(dB), SSIM and LPIPS per object.
}
\label{tab:envlight}
\centering\small
\setlength{\tabcolsep}{4pt}
\begin{tabular}{l ccc ccc ccc}
\toprule
 & \multicolumn{3}{c}{Point lights~(OLAT)} & \multicolumn{3}{c}{Captured HDRIs} & \multicolumn{3}{c}{Colored environments} \\
\cmidrule(lr){2-4}\cmidrule(lr){5-7}\cmidrule(lr){8-10}
Method & PSNR$\uparrow$ & SSIM$\uparrow$ & LPIPS$\downarrow$ & PSNR$\uparrow$ & SSIM$\uparrow$ & LPIPS$\downarrow$ & PSNR$\uparrow$ & SSIM$\uparrow$ & LPIPS$\downarrow$ \\
\midrule
\lucent~(physics only) & 31.88 & 0.791 & 0.0478 & 30.94 & 0.786 & 0.0522 & 31.29 & 0.792 & 0.0468 \\
\lucent~(+ residual) & \textbf{38.59} & \textbf{0.885} & \textbf{0.0239} & \textbf{37.39} & \textbf{0.867} & \textbf{0.0246} & \textbf{38.27} & \textbf{0.882} & \textbf{0.0189} \\
\bottomrule
\end{tabular}
\\[4pt]
\begin{tabular}{l l cc cc cc cc cc}
\toprule
 &  & \multicolumn{2}{c}{candle} & \multicolumn{2}{c}{soap} & \multicolumn{2}{c}{dragon} & \multicolumn{2}{c}{statue} & \multicolumn{2}{c}{mean} \\
\cmidrule(lr){3-4}\cmidrule(lr){5-6}\cmidrule(lr){7-8}\cmidrule(lr){9-10}\cmidrule(lr){11-12}
Lighting & & phys. & + res. & phys. & + res. & phys. & + res. & phys. & + res. & phys. & + res. \\
\midrule
Point lights~(OLAT) & PSNR$\uparrow$ & 31.82 & 38.21 & 32.78 & 44.36 & 32.05 & 35.49 & 30.86 & 36.30 & 31.88 & 38.59 \\
 & SSIM$\uparrow$ & 0.782 & 0.909 & 0.890 & 0.958 & 0.745 & 0.812 & 0.748 & 0.862 & 0.791 & 0.885 \\
 & LPIPS$\downarrow$ & 0.0438 & 0.0194 & 0.0460 & 0.0109 & 0.0493 & 0.0325 & 0.0522 & 0.0326 & 0.0478 & 0.0239 \\
\addlinespace[2pt]
Captured HDRIs & PSNR$\uparrow$ & 32.20 & 37.15 & 30.57 & 40.91 & 29.10 & 34.12 & 31.90 & 37.40 & 30.94 & 37.39 \\
 & SSIM$\uparrow$ & 0.882 & 0.942 & 0.840 & 0.926 & 0.625 & 0.723 & 0.798 & 0.877 & 0.786 & 0.867 \\
 & LPIPS$\downarrow$ & 0.0438 & 0.0179 & 0.0582 & 0.0145 & 0.0580 & 0.0371 & 0.0489 & 0.0291 & 0.0522 & 0.0246 \\
\addlinespace[2pt]
Colored environments & PSNR$\uparrow$ & 32.46 & 38.04 & 30.83 & 42.09 & 29.54 & 34.65 & 32.36 & 38.32 & 31.29 & 38.27 \\
 & SSIM$\uparrow$ & 0.862 & 0.943 & 0.852 & 0.939 & 0.651 & 0.749 & 0.805 & 0.897 & 0.792 & 0.882 \\
 & LPIPS$\downarrow$ & 0.0349 & 0.0128 & 0.0602 & 0.0130 & 0.0508 & 0.0299 & 0.0412 & 0.0200 & 0.0468 & 0.0189 \\
\bottomrule
\end{tabular}
\end{table*}

Light transport is linear, so the ground truth under an environment map is the weighted sum of the 112 single-light images of a view.
\Cref{fig:envlight,tab:envlight} light \lucent with the same weights.
Environment lighting costs about $1$\,dB compared with point lights~($0.9$\,dB physics only and $1.2$\,dB with the residual, for the captured HDRIs), and the object residual, fitted on point lights only, adds $6.5$--$7.0$\,dB on average under the environments, as much as under point lights~($6.7$\,dB).

\subsection{Unknown environment lighting}
\label{sec:exp:envmap_ir}

\begin{figure*}[!tp]
  \centering
  \newlength{\irH}\setlength{\irH}{0.46in}   %
  \newlength{\irM}\setlength{\irM}{0.66in}   %
  \newlength{\irC}\setlength{\irC}{0.82in}   %
  \newlength{\irB}\setlength{\irB}{1.95in}   %
  \newlength{\irE}\setlength{\irE}{1.90in}   %
  \newcommand{\irLab}[1]{\rotatebox[origin=c]{90}{#1}}
  \setlength{\tabcolsep}{0.5pt}\renewcommand{\arraystretch}{0.6}\scriptsize
  \begin{tabular}{@{}l@{}}
  \textbf{(a) Opaque: hotdog}, Synthetic4Relight~\cite{zhang2022invrender}, view 150 \\[3pt]
  \end{tabular}\\
  \begin{tabular}[t]{@{}c@{\hspace{3pt}}c@{\hspace{2pt}}cc@{}}
   & Light & Albedo & Roughness \\[3pt]
  \irLab{GT} & \raisebox{-.5\height}{\includegraphics[height=\irM]{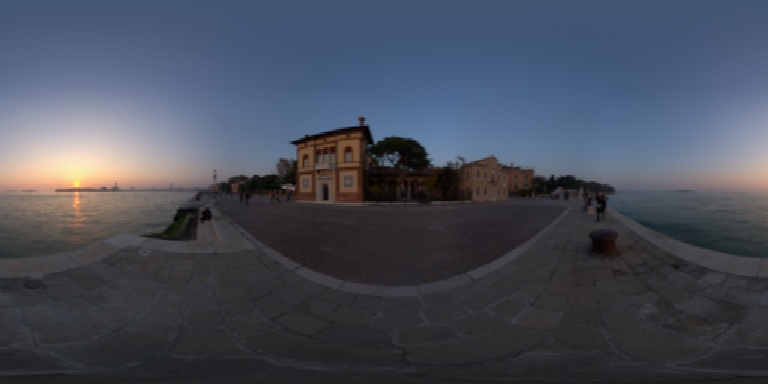}} & \raisebox{-.5\height}{\includegraphics[height=\irM]{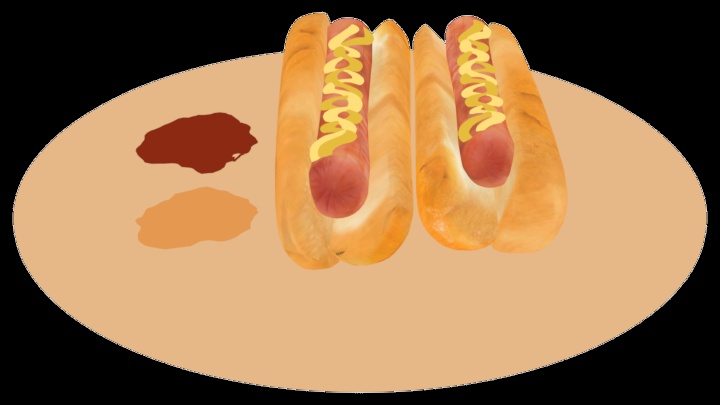}} & \raisebox{-.5\height}{\includegraphics[height=\irM]{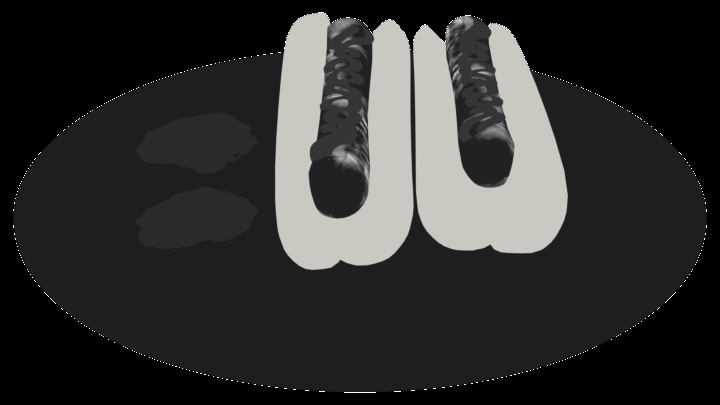}} \\[1pt]
  \irLab{Ours} & \raisebox{-.5\height}{\includegraphics[height=\irM]{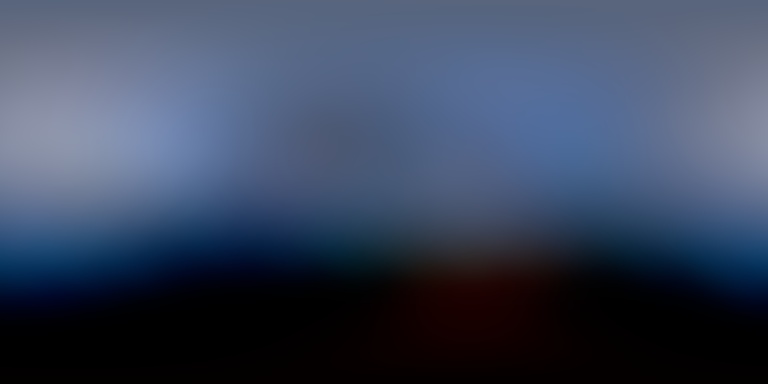}} & \raisebox{-.5\height}{\includegraphics[height=\irM]{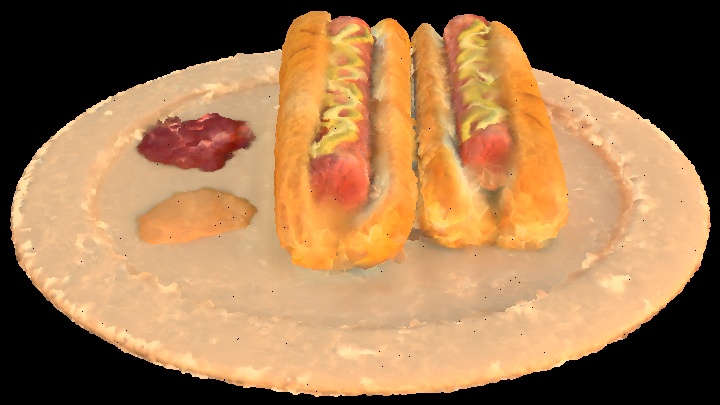}} & \raisebox{-.5\height}{\includegraphics[height=\irM]{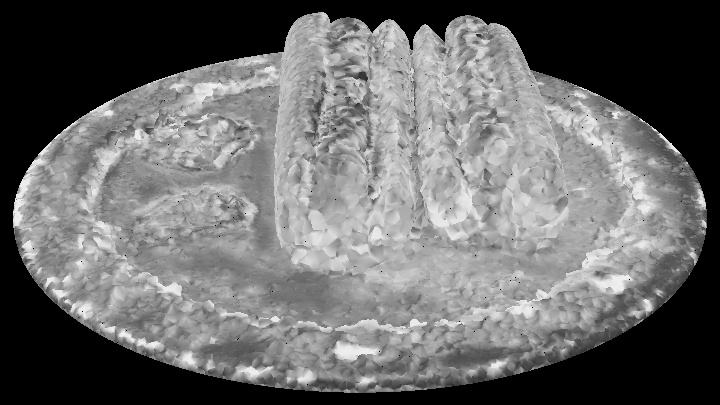}} \\
   \rule{0pt}{8pt}& & 21.89 & MSE 0.186 \\
  \end{tabular}%
  \hspace{9pt}%
  \begin{tabular}[t]{@{}c@{\hspace{3pt}}ccc@{}}
   & Novel view & Relit, env.~1 & Relit, env.~2 \\[3pt]
  \irLab{GT} & \raisebox{-.5\height}{\includegraphics[height=\irH]{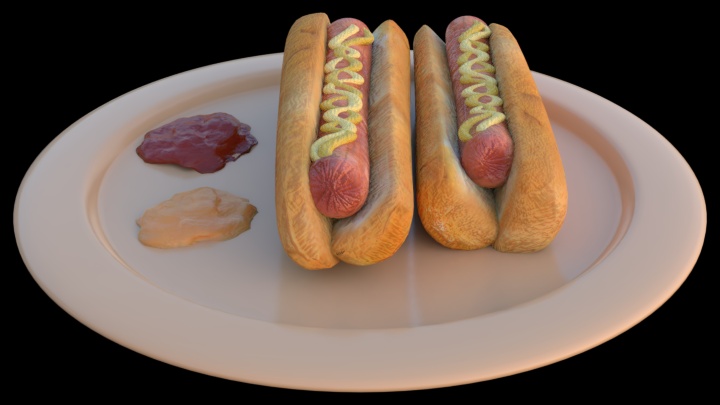}} & \raisebox{-.5\height}{\includegraphics[height=\irH]{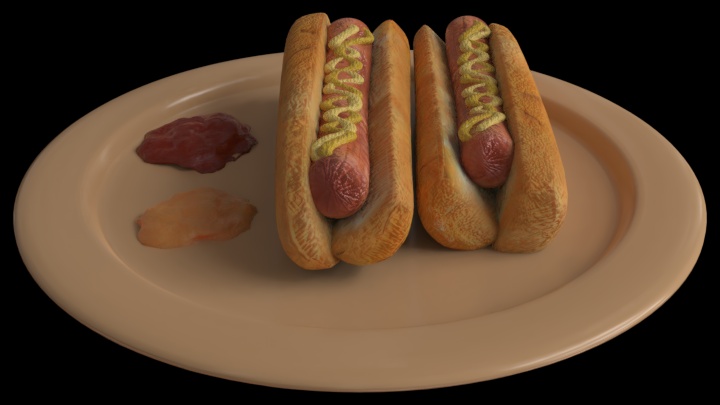}} & \raisebox{-.5\height}{\includegraphics[height=\irH]{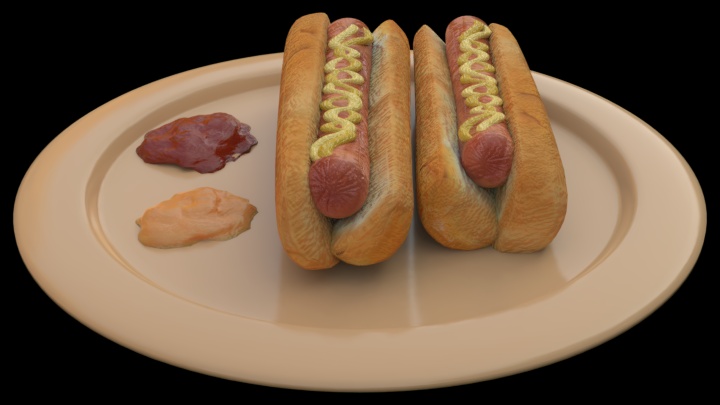}} \\[1pt]
  \irLab{Ours} & \raisebox{-.5\height}{\includegraphics[height=\irH]{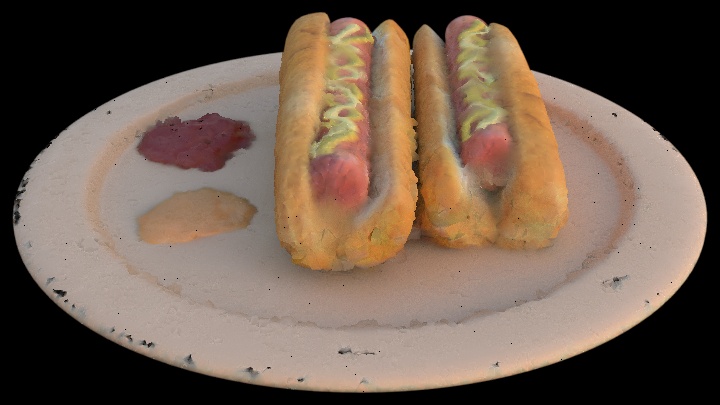}} & \raisebox{-.5\height}{\includegraphics[height=\irH]{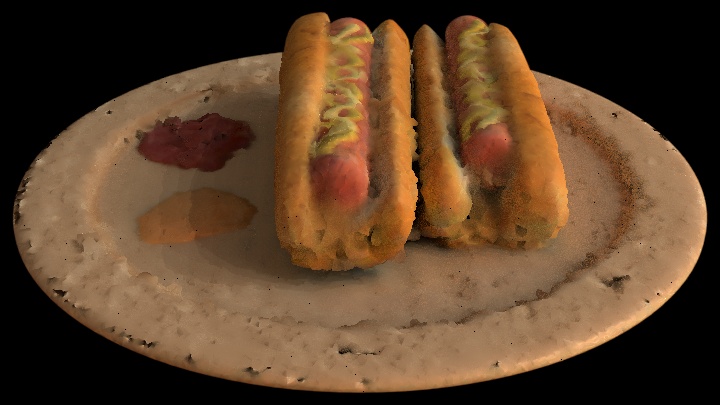}} & \raisebox{-.5\height}{\includegraphics[height=\irH]{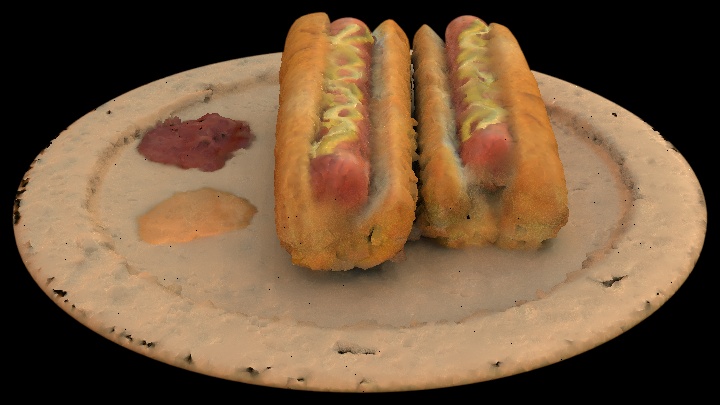}} \\
   \rule{0pt}{8pt}& 29.39 & 28.87 & 27.20 \\[3pt]
  \irLab{+ residual} & \raisebox{-.5\height}{\includegraphics[height=\irH]{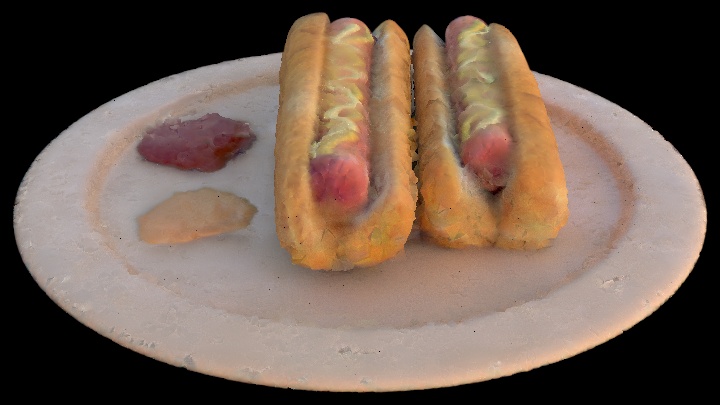}} & \raisebox{-.5\height}{\includegraphics[height=\irH]{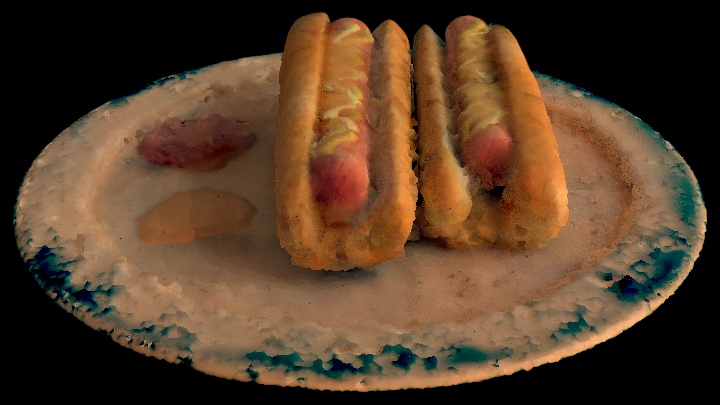}} & \raisebox{-.5\height}{\includegraphics[height=\irH]{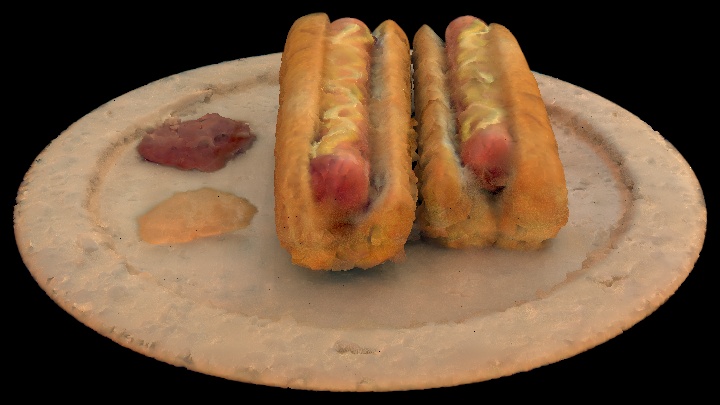}} \\
   \rule{0pt}{8pt}& 31.64 & 24.19 & 28.23 \\
  \end{tabular}\\[8pt]
  \begin{tabular}{@{}l@{}}
  \textbf{(b) Translucent: candle}, SSS-GS~\cite{dihlmann2024sssgs}, pose 72, light learned as order-2 SH \\[3pt]
  \end{tabular}\\
  \begin{tabular}[t]{@{}c@{\hspace{3pt}}c@{}}
   & Light~(city) \\[3pt]
  \irLab{GT} & \raisebox{-.5\height}{\includegraphics[width=\irE]{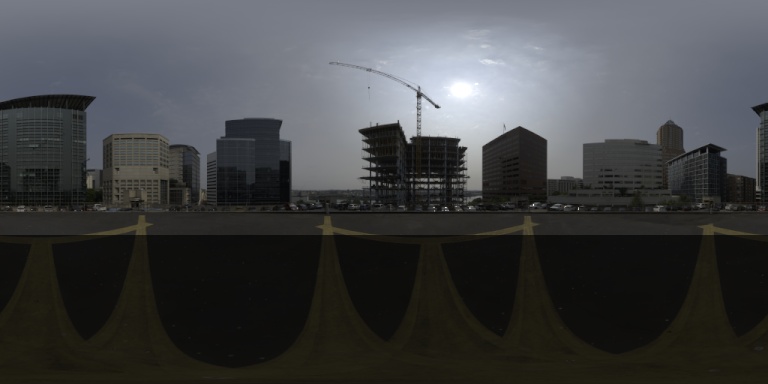}} \\[1pt]
  \irLab{Ours} & \raisebox{-.5\height}{\includegraphics[width=\irE]{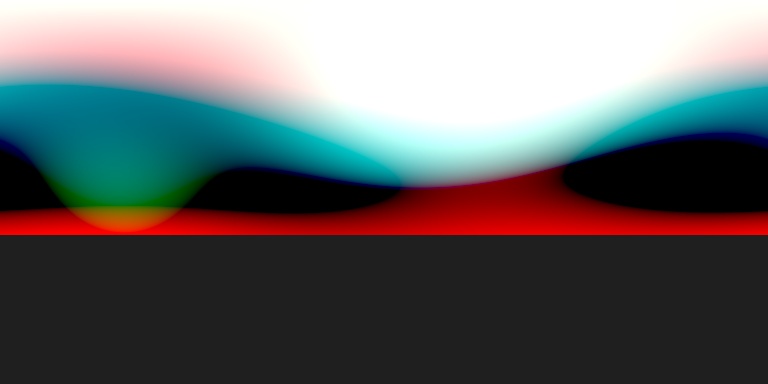}} \\
  \end{tabular}%
  \hspace{9pt}%
  \begin{tabular}[t]{@{}c@{\hspace{3pt}}cc@{}}
   & \shortstack{Base\\color} & Roughness \\[3pt]
  \irLab{Ours} & \raisebox{-.5\height}{\includegraphics[height=\irB]{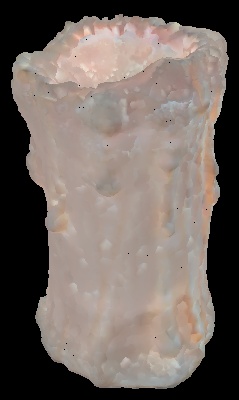}} & \raisebox{-.5\height}{\includegraphics[height=\irB]{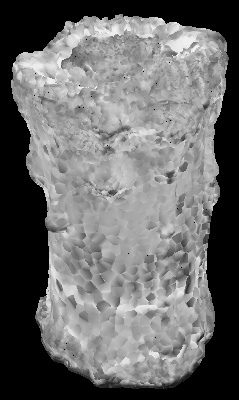}} \\
  \end{tabular}%
  \hspace{9pt}%
  \begin{tabular}[t]{@{}c@{\hspace{3pt}}ccc@{}}
   & Novel view & \shortstack{Relit,\\held-out env.} & \shortstack{Relit,\\point light} \\[3pt]
  \irLab{GT} & \raisebox{-.5\height}{\includegraphics[height=\irC]{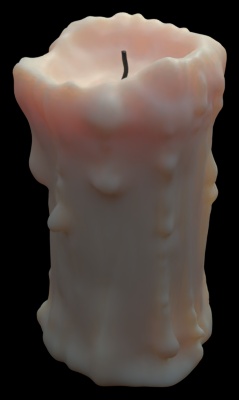}} & \raisebox{-.5\height}{\includegraphics[height=\irC]{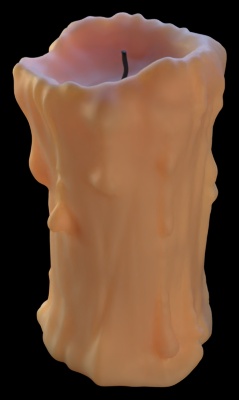}} & \raisebox{-.5\height}{\includegraphics[height=\irC]{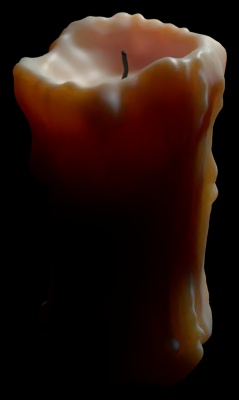}} \\[1pt]
  \irLab{Ours} & \raisebox{-.5\height}{\includegraphics[height=\irC]{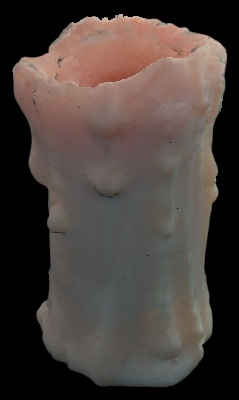}} & \raisebox{-.5\height}{\includegraphics[height=\irC]{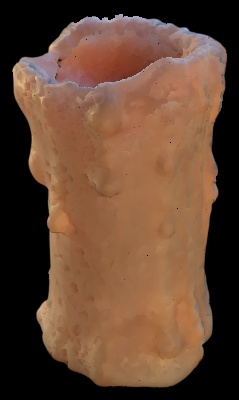}} & \raisebox{-.5\height}{\includegraphics[height=\irC]{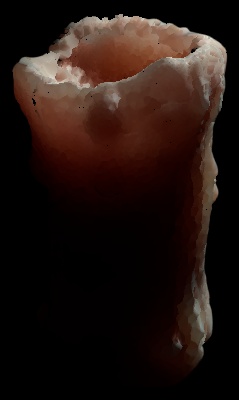}} \\
   \rule{0pt}{8pt}& 31.41 & 35.03 & 31.22 \\[3pt]
  \irLab{+ residual} & \raisebox{-.5\height}{\includegraphics[height=\irC]{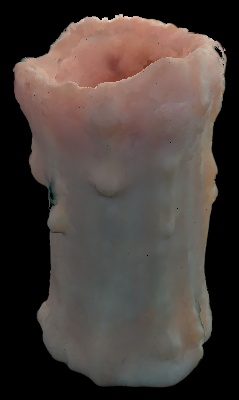}} & \raisebox{-.5\height}{\includegraphics[height=\irC]{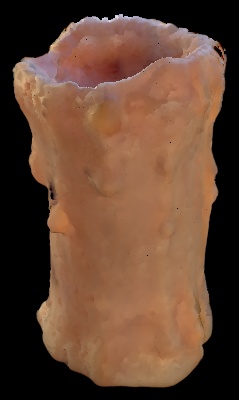}} & \raisebox{-.5\height}{\includegraphics[height=\irC]{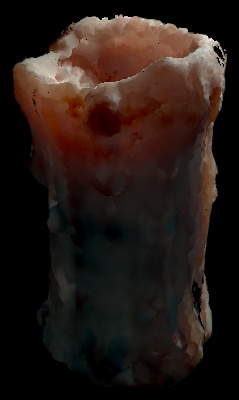}} \\
   \rule{0pt}{8pt}& 30.75 & 34.27 & 29.87 \\
  \end{tabular}
  \caption{\textbf{Inverse rendering under an unknown environment.}
  Rows: ground truth, ours, ours + residual.
  (a)~Hotdog: albedo, relighting and light after the albedo scale.
  (b)~Candle: relighting after a global RGB scale.
  PSNR~(dB), roughness MSE.}
  \label{fig:envmap_ir}
\end{figure*}

\begin{table*}[!tp]
\caption{\textbf{Inverse rendering under an unknown environment.}
(a)~Hotdog~\cite{zhang2022invrender}, 20 test views; albedo and relighting after the per-channel albedo scale.
(b)~Candle, 32 held-out poses.
SSIM and LPIPS over the whole image.
$^\dagger$After a global RGB scale fitted on the test images; unscaled in parentheses.}
\label{tab:envmap_ir}
\centering\small
\setlength{\tabcolsep}{4pt}
{\setlength{\tabcolsep}{2.3pt}
\begin{tabular}{l ccc ccc c cccc c}
\toprule
 & \multicolumn{3}{c}{Novel view} & \multicolumn{3}{c}{Albedo} & Roughness & \multicolumn{4}{c}{Relighting} & Light \\
\cmidrule(lr){2-4}\cmidrule(lr){5-7}\cmidrule(lr){8-8}\cmidrule(lr){9-12}\cmidrule(lr){13-13}
(a) Hotdog, light learned & PSNR$\uparrow$ & SSIM$\uparrow$ & LPIPS$\downarrow$ & PSNR$\uparrow$ & SSIM$\uparrow$ & LPIPS$\downarrow$ & MSE$\downarrow$ & PSNR$\uparrow$ & SSIM$\uparrow$ & LPIPS$\downarrow$ & unscaled$\uparrow$ & $r\uparrow$ \\
\midrule
Direct light & 28.74 & \textbf{0.934} & 0.117 & \textbf{22.34} & 0.923 & 0.089 & \textbf{0.127} & 27.39 & \textbf{0.918} & \textbf{0.133} & 24.59 & \textbf{0.83} \\
+ indirect light & 29.28 & 0.924 & 0.122 & 21.98 & \textbf{0.928} & \textbf{0.084} & 0.162 & \textbf{27.75} & 0.904 & 0.145 & \textbf{24.67} & 0.82 \\
\midrule
+ indirect light + residual & \textbf{32.50} & 0.932 & \textbf{0.102} & -- & -- & -- & -- & 26.84 & 0.901 & 0.146 & 24.01 & -- \\
\bottomrule
\end{tabular}}
\\[8pt]
\begin{tabular}{l l cccc}
\toprule
 & & 4 training envs. & 6 held-out envs. & Point light & Novel view \\
\cmidrule(lr){3-3}\cmidrule(lr){4-4}\cmidrule(lr){5-5}\cmidrule(lr){6-6}
(b) Candle, trained on & Light & PSNR$\uparrow$ & PSNR$\uparrow$ & PSNR$\uparrow$ & PSNR$\uparrow$ \\
\midrule
112 point lights & known & 32.15 & 33.53 & 31.31 & -- \\
4 environments & known & 35.07 & 35.17 & 31.04 & -- \\
4 environments & known, order-2 SH & 34.03 & 34.33 & 30.84 & -- \\
4 environments & learned SH, mean known & 34.38$^\dagger$~(27.04) & 34.62$^\dagger$~(28.09) & 31.04$^\dagger$~(28.49) & 28.96 \\
4 environments & learned SH & 34.77$^\dagger$~(24.78) & 35.22$^\dagger$~(25.98) & 31.33$^\dagger$~(27.22) & 29.05 \\
City~(1 of the 4) & learned SH & 34.65$^\dagger$~(24.67) & 34.48$^\dagger$~(25.87) & 30.75$^\dagger$~(27.05) & 27.77 \\
\midrule
4 environments & known, + residual & \textbf{37.24} & \textbf{36.37} & \textbf{31.47} & -- \\
4 environments & learned SH, + residual & 35.12$^\dagger$~(24.72) & 35.36$^\dagger$~(25.95) & 30.44$^\dagger$~(26.87) & \textbf{29.19} \\
\midrule
 & & SSIM$\uparrow$ / LPIPS$\downarrow$ & SSIM$\uparrow$ / LPIPS$\downarrow$ & SSIM$\uparrow$ / LPIPS$\downarrow$ & SSIM$\uparrow$ / LPIPS$\downarrow$ \\
\midrule
112 point lights & known & 0.979 / 0.043 & 0.981 / 0.039 & \textbf{0.972} / \textbf{0.044} & -- \\
4 environments & known & 0.985 / 0.035 & 0.984 / 0.035 & 0.968 / 0.047 & -- \\
4 environments & known, order-2 SH & 0.984 / 0.037 & 0.983 / 0.036 & 0.969 / 0.047 & -- \\
4 environments & learned SH, mean known & 0.985$^\dagger$ / 0.036$^\dagger$ & 0.983$^\dagger$ / 0.035$^\dagger$ & 0.969$^\dagger$ / 0.047$^\dagger$ & 0.982 / 0.043 \\
4 environments & learned SH & 0.985$^\dagger$ / 0.034$^\dagger$ & 0.984$^\dagger$ / 0.033$^\dagger$ & 0.970$^\dagger$ / 0.046$^\dagger$ & 0.983 / 0.037 \\
City~(1 of the 4) & learned SH & 0.984$^\dagger$ / 0.035$^\dagger$ & 0.982$^\dagger$ / 0.036$^\dagger$ & 0.967$^\dagger$ / 0.048$^\dagger$ & 0.983 / 0.034 \\
\midrule
4 environments & known, + residual & \textbf{0.989} / \textbf{0.027} & \textbf{0.985} / \textbf{0.028} & 0.958 / 0.059 & -- \\
4 environments & learned SH, + residual & 0.986$^\dagger$ / 0.029$^\dagger$ & 0.983$^\dagger$ / 0.029$^\dagger$ & 0.955$^\dagger$ / 0.063$^\dagger$ & \textbf{0.985} / \textbf{0.032} \\
\bottomrule
\end{tabular}
\end{table*}

\Cref{fig:envmap_ir,tab:envmap_ir} train \lucent on images lit by environment maps it is not given, and learn the light together with the materials.
On the opaque hotdog of Synthetic4Relight~\cite{zhang2022invrender}, the light is learned at 256 directions with the light priors of SVG-IR~\cite{sun2025svgir}.
On the candle, the training images are lit by four environments, each learned as order-2 spherical harmonics.
The learned light recovers the candle's materials up to one global color scale, close to training with the known light.
With the light learned, the object residual raises novel-view accuracy~($29.28$ to $32.50$\,dB on the hotdog) but lowers relighting under new light~($27.75$ to $26.84$\,dB on the hotdog, $31.33$ to $30.44$\,dB under the candle's point lights): fitted under the training environments, it does not carry over to new light, so relighting should use the physics alone.
No baseline was trained on this experiment.
Every run of it, the point-light reference included, computes the diffusion, single scattering and coverage on the volume bounded by the walls rather than on the foam cells~(\texttt{diffusion\_geometry: explicit} in the code).

\clearpage
\section{What is physics and what is learned}
\label{sec:supp:sharegroup}
\textit{Extends main \cref{sec:exp:share}; the transport ablation it cites is \cref{tab:ablation}, and the other transport ablations are in \cref{sec:supp:negative}.}

\subsection{Decomposition and learned share}

\begin{figure*}[!tp]
  \centering
  \includegraphics[width=\textwidth,height=0.88\textheight,keepaspectratio]{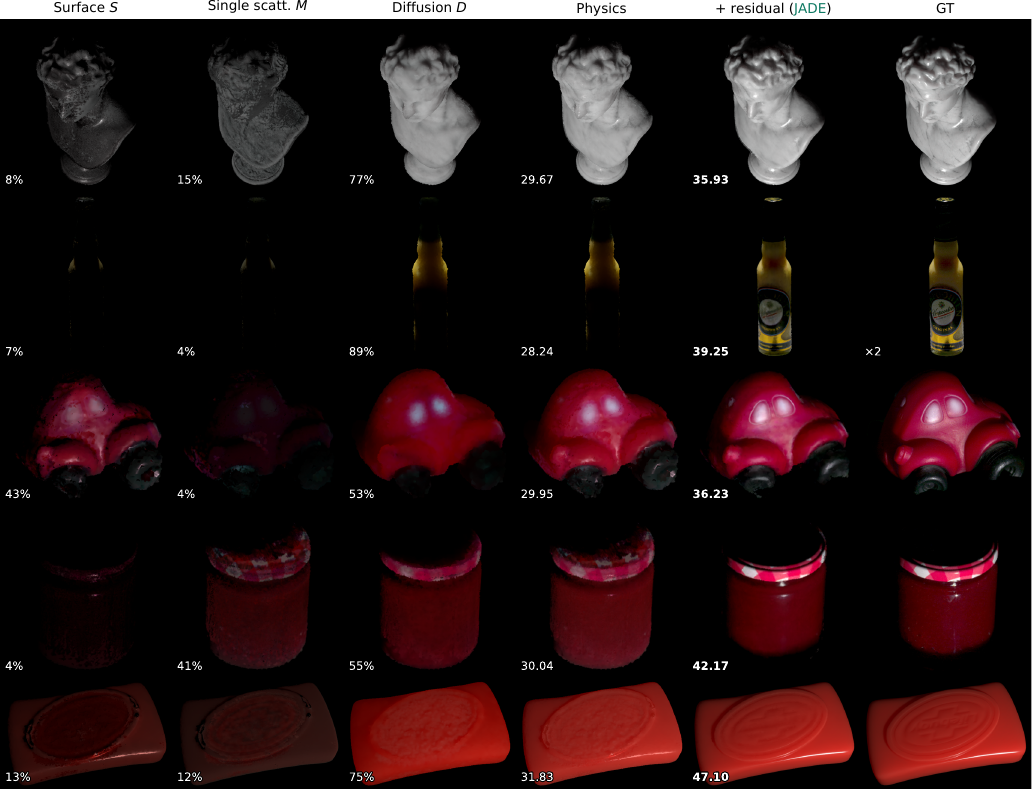}
  \caption{\textbf{Decomposition, five objects}~(extends \cref{fig:decomposition}): statue, liqueur, car, marmalade and soap.
  On the statue and the soap, the homogeneous synthetic objects here, the residual mainly corrects brightness and sharpens the shading; on the marmalade it adds the lid pattern, which one medium cannot hold.}
  \label{fig:decomp_full}
\end{figure*}

\begin{figure*}[!tp]
  \centering
  \includegraphics[width=\textwidth,height=0.88\textheight,keepaspectratio]{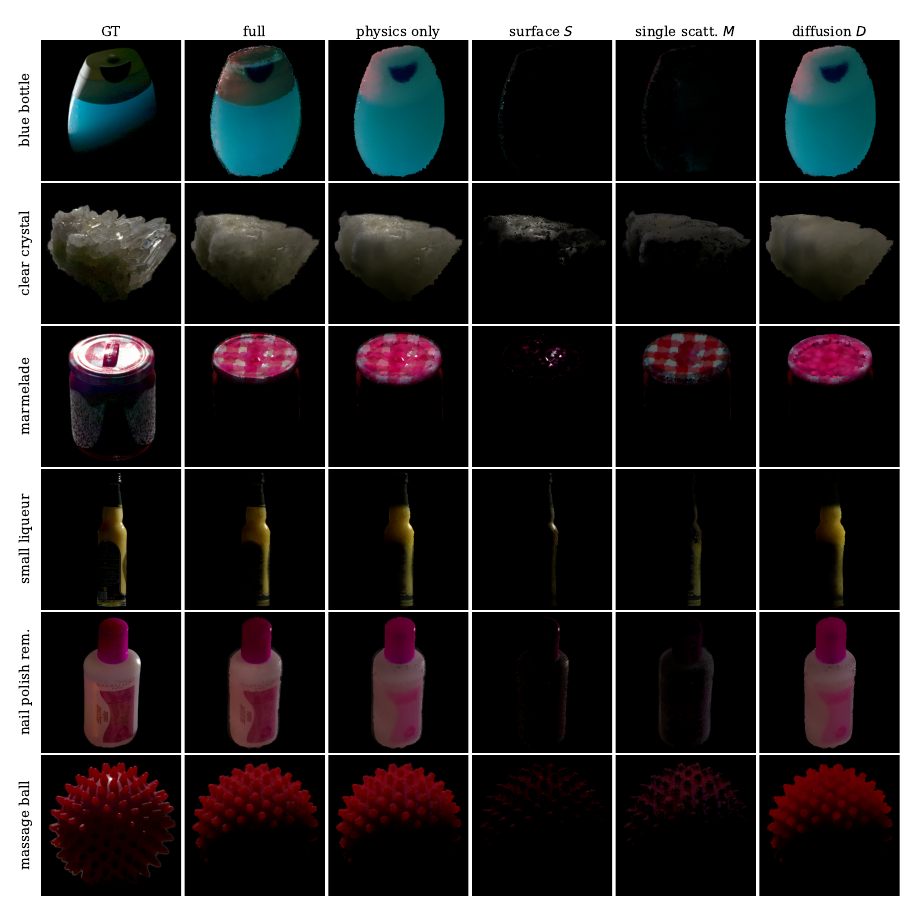}
  \caption{\textbf{Physics-only decomposition of six of the ten real objects}, one configuration, held-out back-lit lights~(light cosine $\approx-0.6$).
  Physics only is the full render without the residual, and $S$, $M$, $D$ are its surface, single-scattering and diffusion parts~(each tone-mapped).}

  \label{fig:objects6}
\end{figure*}

\Cref{fig:decomp_full} extends the decomposition of the main paper to five objects and \Cref{fig:objects6} to six of the ten real objects under back light; \Cref{fig:learned_share,tab:share} switch off each method's learned part.
SSS-GS keeps the object's color in its network; retrained without it, it loses up to $12.0$\,dB~(blue bottle).
On the real objects, SSD-GS retrained without its networks loses only $0.4$--$2.8$\,dB~(\cref{tab:objects_full}), and switching them off on the same checkpoint costs $2.3$\,dB on the car and $3.6$\,dB on the clip~(\cref{tab:share}), because its per-Gaussian reflectance absorbs what its networks carried; retrained without them, it fits $0.5$--$3.4$M values per object, about $45$ per Gaussian.
Our per-corner values are about as many~(up to $3.6$M), but they can only describe a BRDF, and our medium is a few constants.

\label{sec:supp:share}

\begin{table*}[!tbp]
\caption{\textbf{Switching off the learned part}, whole-image PSNR~(dB), SSIM and LPIPS, each method's own switches, on the $32$ standard views of the car and the clip~(one held-out light per camera) with our earlier $32$-light fits; \Cref{tab:objects_full} scores both objects on $256$ views.
``Network off'' for SSS-GS: subsurface weight 0 and incident light from its traced point light instead of the network.
``Networks off'' for SSD-GS: scatter and shadow/phase networks off, leaving its diffuse and anisotropic specular Gaussians.}
\label{tab:share}
\centering\small
\setlength{\tabcolsep}{5pt}
\begin{tabular}{ll ccc @{\hspace{10pt}} ccc}
\toprule
 & & \multicolumn{3}{c}{Red car} & \multicolumn{3}{c}{Hair clip} \\
\cmidrule(lr){3-5}\cmidrule(lr){6-8}
Method & Learned part & PSNR$\uparrow$ & SSIM$\uparrow$ & LPIPS$\downarrow$ & PSNR$\uparrow$ & SSIM$\uparrow$ & LPIPS$\downarrow$ \\
\midrule
\lucent & full~(physics + residual) & 35.57 & 0.744 & 0.041 & 38.80 & 0.605 & 0.014 \\
 & residual off~(= trained without it) & 29.48 & 0.585 & 0.073 & 32.16 & 0.217 & 0.039 \\
\midrule
SSS-GS & full & 31.42 & 0.632 & 0.029 & 37.21 & 0.552 & 0.009 \\
 & residual off, same checkpoint & 23.81 & 0.444 & 0.098 & -- & -- & -- \\
 & network off, same checkpoint & 24.13 & 0.387 & 0.093 & -- & -- & -- \\
 & retrained without network & 29.52 & 0.563 & 0.054 & 34.15 & 0.424 & 0.017 \\
\midrule
SSD-GS & full & 32.35 & 0.697 & 0.050 & 38.35 & 0.642 & 0.012 \\
 & scatter off, same checkpoint & 32.35 & 0.697 & 0.050 & 35.55 & 0.546 & 0.017 \\
 & networks off, same checkpoint & 30.08 & 0.676 & 0.052 & 34.80 & 0.475 & 0.020 \\
 & retrained without networks & 31.49 & 0.686 & 0.052 & 36.34 & 0.558 & 0.018 \\
\bottomrule
\end{tabular}
\end{table*}

\subsection{Size of the object residual and unseen cameras}
\label{sec:supp:residual}

\begin{table*}[!tp]
\caption{\textbf{Object-residual size.}
The residual is fitted on the frozen physics-only model at the training cameras, with $32$ lights per camera~(the main model uses $64$, \cref{tab:unseen}); whole-image PSNR~(dB), SSIM and LPIPS on black under held-out lights, on the 32 evaluation views~(``views'', cameras seen at other lights) and on cameras the residual never saw~(``unseen'': 32, 32, 11 cameras for car, clip and blue bottle).
``Val.
gain'': mean gain over the physics on validation lights of the training cameras, the only number used to choose the size.
The selected size~(MLP 256-256-256, 150.8k weights) is the largest MLP we tried and has the best validation gain among them; the gain still rises with size~($+7.40$ to $+8.05$\,dB from $50.6$k to $150.8$k weights), so a larger residual might gain more.
It has fewer weights than the networks of SSD-GS~($193$k).
Per-element parameters are not counted here: \lucent learns nine per surface corner for its base color, roughness and normal~($1.1$M on the car, $0.5$--$3.6$M over the real objects), SSD-GS about $45$ per Gaussian~($1.4$M on the car, $0.5$--$3.4$M over the real objects) and SSS-GS $36$~($10.8$M on the car).
Baselines are scored on the same $32$ views, as in \Cref{tab:share}.
The tri-plane row is a reference, not a candidate: validation lights lie on training cameras and cannot see viewpoint memorization, which the unseen cameras expose.
}
\label{tab:residual_sizes}
\centering\small
\setlength{\tabcolsep}{3pt}
\resizebox{\textwidth}{!}{
\begin{tabular}{l r c ccc ccc @{\hspace{8pt}} ccc ccc @{\hspace{8pt}} ccc ccc}
\toprule
 & & & \multicolumn{6}{c}{Red car} & \multicolumn{6}{c}{Hair clip} & \multicolumn{6}{c}{Blue bottle} \\
\cmidrule(lr){4-9}\cmidrule(lr){10-15}\cmidrule(lr){16-21}
 & & Val. & \multicolumn{3}{c}{views} & \multicolumn{3}{c}{unseen} & \multicolumn{3}{c}{views} & \multicolumn{3}{c}{unseen} & \multicolumn{3}{c}{views} & \multicolumn{3}{c}{unseen} \\
\cmidrule(lr){4-6}\cmidrule(lr){7-9}\cmidrule(lr){10-12}\cmidrule(lr){13-15}\cmidrule(lr){16-18}\cmidrule(lr){19-21}
Model & Weights & gain & PSNR & SSIM & LPIPS & PSNR & SSIM & LPIPS & PSNR & SSIM & LPIPS & PSNR & SSIM & LPIPS & PSNR & SSIM & LPIPS & PSNR & SSIM & LPIPS \\
\midrule
Physics only & -- & -- & 29.48 & 0.585 & 0.073 & 31.25 & 0.655 & 0.074 & 32.16 & 0.217 & 0.039 & 32.10 & 0.226 & 0.037 & 26.35 & 0.637 & 0.045 & 26.28 & 0.618 & 0.048 \\
\midrule
MLP 8 & 0.6k & +2.88 & 31.61 & 0.614 & 0.062 & 33.43 & 0.672 & 0.064 & 35.01 & 0.319 & 0.028 & 34.02 & 0.309 & 0.028 & 30.07 & 0.658 & 0.035 & 29.40 & 0.635 & 0.041 \\
MLP 16-16 & 1.5k & +3.81 & 32.11 & 0.625 & 0.059 & 34.08 & 0.680 & 0.061 & 35.65 & 0.371 & 0.026 & 34.57 & 0.343 & 0.026 & 31.28 & 0.695 & 0.030 & 30.67 & 0.689 & 0.036 \\
MLP 32-16-16 & 3.1k & +4.66 & 32.83 & 0.659 & 0.055 & 34.43 & 0.692 & 0.059 & 36.12 & 0.402 & 0.023 & 34.99 & 0.377 & 0.023 & 32.13 & 0.719 & 0.025 & 31.78 & 0.702 & 0.031 \\
MLP 64-32-32~(SSS-GS size) & 7.7k & +5.54 & 33.14 & 0.665 & 0.054 & 34.83 & 0.705 & 0.056 & 36.55 & 0.451 & 0.020 & 35.61 & 0.413 & 0.020 & 32.77 & 0.741 & 0.022 & 32.13 & 0.719 & 0.026 \\
MLP 128-96-64 & 27.6k & +6.82 & 34.29 & 0.703 & 0.048 & 35.37 & 0.724 & 0.053 & 37.47 & 0.519 & 0.018 & 36.21 & 0.462 & 0.017 & 33.67 & 0.757 & 0.016 & 33.28 & 0.741 & 0.022 \\
MLP 192-128-96 & 50.6k & +7.40 & 34.95 & 0.722 & 0.044 & \textbf{35.98} & \textbf{0.743} & \textbf{0.048} & 38.18 & 0.551 & 0.016 & 36.49 & 0.485 & 0.016 & 33.93 & \textbf{0.774} & 0.015 & 33.36 & 0.747 & 0.021 \\
\textbf{MLP 256-256-256} & 150.8k & +8.05 & 35.57 & 0.744 & 0.041 & 35.86 & 0.741 & 0.049 & 38.80 & 0.605 & 0.014 & \textbf{36.58} & \textbf{0.506} & \textbf{0.014} & 33.96 & 0.768 & 0.015 & \textbf{33.50} & \textbf{0.752} & \textbf{0.020} \\
MLP 256 + tri-planes~(reference) & 33.6M & +9.02 & \textbf{35.93} & \textbf{0.762} & \textbf{0.033} & 34.45 & 0.704 & 0.059 & \textbf{40.19} & \textbf{0.720} & \textbf{0.010} & 33.77 & 0.267 & 0.034 & \textbf{34.14} & \textbf{0.774} & \textbf{0.011} & 31.17 & 0.637 & 0.038 \\
\midrule
SSD-GS~\cite{zheng2026ssdgs}~(networks) & 193k & -- & 32.35 & 0.697 & 0.050 & -- & -- & -- & 38.35 & 0.642 & 0.012 & -- & -- & -- & 33.29 & 0.771 & 0.016 & -- & -- & -- \\
SSS-GS~\cite{dihlmann2024sssgs}~(network) & 7.1k & -- & 31.42 & 0.632 & 0.029 & -- & -- & -- & 37.21 & 0.552 & 0.009 & -- & -- & -- & 34.40 & 0.760 & 0.008 & -- & -- & -- \\
\bottomrule
\end{tabular}}
\end{table*}

\Cref{tab:residual_sizes} gives the residual sizes we considered and how we chose the $150.8$k-weight MLP of the main paper.
\TabAblation
\Cref{tab:ablation}, cited in main \cref{sec:exp:share}, compares diffusion with more scattering orders on the physics alone; \cref{sec:supp:negative} describes the octaves and the other transport ablations.

\begin{table*}[!tbp]
\caption{\textbf{The residual at cameras it has not seen}, for the real objects of \Cref{tab:objects_full} other than the car and the clip: whole-image PSNR~(dB), SSIM and LPIPS on black under held-out lights, physics only and with the residual, on $32$ evaluation views~(one held-out light per camera, at cameras seen under other lights, as in \cref{tab:residual_sizes}; \cref{tab:objects_full} uses $8$ lights per camera) and on the $n$ cameras the residual never saw~(the geometry and the medium were trained with them; exposure $1$), with the residual of the main paper~($64$ lights per camera).
Car and clip: \Cref{tab:residual_sizes}.}
\label{tab:unseen}
\centering\small
\setlength{\tabcolsep}{4pt}
\begin{tabular}{l ccc cc cc @{\hspace{10pt}} r ccc cc cc}
\toprule
 & \multicolumn{7}{c}{Evaluation views} & \multicolumn{8}{c}{Unseen cameras} \\
\cmidrule(lr){2-8}\cmidrule(lr){9-16}
 & \multicolumn{3}{c}{PSNR~(dB)$\uparrow$} & \multicolumn{2}{c}{SSIM$\uparrow$} & \multicolumn{2}{c}{LPIPS$\downarrow$} & & \multicolumn{3}{c}{PSNR~(dB)$\uparrow$} & \multicolumn{2}{c}{SSIM$\uparrow$} & \multicolumn{2}{c}{LPIPS$\downarrow$} \\
\cmidrule(lr){2-4}\cmidrule(lr){5-6}\cmidrule(lr){7-8}\cmidrule(lr){10-12}\cmidrule(lr){13-14}\cmidrule(lr){15-16}
Object & phys. & + res. & gain & phys. & + res. & phys. & + res. & $n$ & phys. & + res. & gain & phys. & + res. & phys. & + res. \\
\midrule
blue bottle & 26.29 & 33.51 & 7.22 & 0.626 & 0.771 & 0.043 & 0.015 & 11 & 26.04 & 34.22 & 8.18 & 0.601 & 0.757 & 0.046 & 0.019 \\
clear crystal & 30.31 & 34.54 & 4.23 & 0.410 & 0.566 & 0.043 & 0.030 & 24 & 31.99 & 35.62 & 3.62 & 0.455 & 0.565 & 0.039 & 0.028 \\
marmalade & 30.00 & 37.02 & 7.02 & 0.446 & 0.721 & 0.054 & 0.027 & 29 & 30.15 & 36.08 & 5.94 & 0.480 & 0.694 & 0.054 & 0.027 \\
small liqueur & 28.55 & 34.46 & 5.91 & 0.412 & 0.677 & 0.063 & 0.027 & 19 & 27.82 & 35.82 & 7.99 & 0.389 & 0.741 & 0.069 & 0.023 \\
nail polish rem. & 32.37 & 40.64 & 8.28 & 0.638 & 0.768 & 0.039 & 0.017 & 21 & 32.21 & 40.49 & 8.28 & 0.650 & 0.780 & 0.038 & 0.016 \\
massage ball~(red) & 34.38 & 39.51 & 5.12 & 0.569 & 0.736 & 0.037 & 0.015 & 31 & 35.67 & 40.03 & 4.36 & 0.626 & 0.752 & 0.035 & 0.015 \\
joyful yell & 27.63 & 33.36 & 5.73 & 0.545 & 0.719 & 0.083 & 0.042 & 31 & 27.55 & 33.31 & 5.77 & 0.539 & 0.723 & 0.083 & 0.043 \\
physalis & 23.90 & 28.34 & 4.44 & 0.308 & 0.525 & 0.086 & 0.062 & 32 & 24.23 & 26.63 & 2.40 & 0.349 & 0.471 & 0.082 & 0.065 \\
\midrule
mean & 29.18 & 35.17 & 5.99 & 0.494 & 0.685 & 0.056 & 0.029 &  & 29.46 & 35.27 & 5.82 & 0.511 & 0.686 & 0.056 & 0.030 \\
\bottomrule
\end{tabular}
\end{table*}

\Cref{tab:unseen} extends the unseen-camera columns of \Cref{tab:residual_sizes} to the other eight real objects, with the residual of the main paper.
On the cameras it never saw, its gain over the physics is positive on all eight objects, $5.8$\,dB on average against $6.0$\,dB on the evaluation views.

\clearpage
\section{Editing the medium}
\label{sec:supp:appgroup}
\textit{Extends main \cref{sec:exp:editing}: whether the edits are correct, the full extinction sweep, edits of real captures, and surface, glass and interior edits.}

\subsection{Are the edits correct?}
\label{sec:supp:editgt}
We render the ground truth of each edit in Cycles from the released scene~(subsurface radius $\times 1/m$ for an extinction scale $m$; the file's lights replaced by the capture's point light; $512$ samples, no denoising).
Re-rendering the unedited scene matches the released test images at $35.4$\,dB~(candle), $40.5$\,dB~(soap) and $39.3$\,dB~(statue), so the ground truth is reliable.
\Cref{tab:editgt} gives the object PSNR, SSIM and LPIPS of our edit against the true edit, and of our unedited render against the true edit~(``no edit'').
\Cref{tab:sigmagt_full} gives the true and recovered extinction, summarized in main \cref{tab:sigmagt}.
\TabSigma

On the soap, the edit is closer to the truth than no edit in PSNR in seven of the eight cases~(all but $\times2$ under front light), and under front light it predicts the true change~(cosine $0.71$--$0.95$ between our change image and the true one).
Under back light it does not~(cosine $0.14$--$0.56$): the true albedo of the soap is $0.165$ and $0.112$ in green and blue, so absorption dominates and diffusion does not hold.
On the candle, the green and blue extinctions are $2.0\times$ and $3.8\times$ too low, and on the statue all three are $2.2$--$2.5\times$ too low.
These channels are optically thick at the cell size, where the mean free path is shorter than a cell, and there the fit trades extinction for the learned gain~($7.0$ on the candle, $3.6$ on the soap).
Edits of the candle therefore change brightness in the right direction but not by the right amount, and a per-channel edit can shift the color the wrong way.
The true statue is dense enough that halving or doubling its extinction barely changes its appearance~($23$--$35$\,dB between the true edited and unedited renders); our edits change it more, so on the statue no edit scores higher than ignoring it under front light, and three of the four do under back light.

\begin{table}[!tbp]\centering\small\setlength{\tabcolsep}{3pt}
\caption{Edits against Cycles ground truth: object PSNR~(dB), SSIM and LPIPS of our edit / of no edit, both against the true edit; 19 front-lit and 5 back-lit of the 42 views; SSIM with a Gaussian window over the object mask eroded by $3$\,px, LPIPS on the whole image.
Chroma scales $\boldsymbol{\sigma}_t$ by $(\frac12,1,2)$ per channel.}
\label{tab:editgt}
\resizebox{\linewidth}{!}{
\begin{tabular}{ll cccc}
\toprule
 & & $\times\frac14$ & $\times\frac12$ & $\times 2$ & chroma \\
\midrule
\multicolumn{6}{@{}l}{\textit{Object PSNR~(dB)$\uparrow$}} \\
Candle & front-lit & 19.97 / 18.26 & 20.96 / 19.98 & 18.23 / 19.87 & 19.26 / 19.96 \\
       & back-lit  & 19.65 / 16.80 & 21.52 / 22.07 & 25.55 / 19.94 & 23.41 / 21.63 \\
Soap   & front-lit & 24.68 / 21.11 & 25.98 / 25.04 & 25.90 / 26.24 & 26.43 / 24.97 \\
       & back-lit  & 20.32 / 18.31 & 22.96 / 20.62 & 23.96 / 23.64 & 22.86 / 20.75 \\
Statue & front-lit & 21.19 / 22.16 & 22.56 / 23.15 & 20.54 / 21.85 & 21.64 / 22.41 \\
       & back-lit  & 24.79 / 21.53 & 26.71 / 25.77 & 25.88 / 25.89 & 26.30 / 25.88 \\
\midrule
\multicolumn{6}{@{}l}{\textit{SSIM$\uparrow$}} \\
Candle & front-lit & 0.680 / 0.597 & 0.691 / 0.655 & 0.641 / 0.659 & 0.640 / 0.639 \\
 & back-lit & 0.452 / 0.439 & 0.582 / 0.618 & 0.845 / 0.629 & 0.686 / 0.620 \\
Soap & front-lit & 0.825 / 0.830 & 0.849 / 0.859 & 0.843 / 0.861 & 0.830 / 0.834 \\
 & back-lit & 0.804 / 0.700 & 0.843 / 0.797 & 0.793 / 0.816 & 0.837 / 0.807 \\
Statue & front-lit & 0.721 / 0.651 & 0.734 / 0.708 & 0.613 / 0.677 & 0.654 / 0.666 \\
 & back-lit & 0.576 / 0.371 & 0.804 / 0.743 & 0.818 / 0.817 & 0.799 / 0.777 \\
\midrule
\multicolumn{6}{@{}l}{\textit{LPIPS$\downarrow$}} \\
Candle & front-lit & 0.056 / 0.064 & 0.050 / 0.053 & 0.055 / 0.054 & 0.053 / 0.053 \\
 & back-lit & 0.075 / 0.097 & 0.067 / 0.077 & 0.037 / 0.049 & 0.063 / 0.076 \\
Soap & front-lit & 0.092 / 0.089 & 0.080 / 0.076 & 0.072 / 0.078 & 0.088 / 0.082 \\
 & back-lit & 0.070 / 0.071 & 0.054 / 0.053 & 0.055 / 0.064 & 0.055 / 0.053 \\
Statue & front-lit & 0.068 / 0.069 & 0.058 / 0.059 & 0.072 / 0.063 & 0.073 / 0.075 \\
 & back-lit & 0.125 / 0.132 & 0.065 / 0.069 & 0.056 / 0.057 & 0.060 / 0.067 \\
\bottomrule
\end{tabular}}
\end{table}

Raw PSNR against the edited truth mixes two errors: that of the unedited physics~(the ``Unedited'' column of main \cref{tab:editpattern}) and that of the edit.
To isolate the edit, we give it and a uniform change the same freedom, one scale per color channel and view fitted to the true edited image.
The uniform change scales the true unedited image, the best that a global change of brightness or color can do; ours adds our predicted change, scaled, to the true unedited image.
Our change is closer to the truth in $19$ of the $24$ edits, by up to $7.8$\,dB on the back-lit candle, and loses on three front-lit soap edits, where the true change is close to a uniform brightening, and on the back-lit soap at $\times\frac14$ and the back-lit statue at $\times2$~(by $0.01$\,dB).
Where the recovered extinction is wrong in magnitude, the physics still predicts where the image changes, which no single global factor can.

\subsection{Editing the extinction: protocol and full sweep}
\label{sec:supp:ssdgsedit}
\paragraph{Protocol for SSD-GS.}
SSD-GS has no extinction to scale, so we scale the closest quantities it has.
For every Gaussian, light and view, its scatter network predicts the three parameters of a dipole profile: a distance $r$, a reduced scattering coefficient $\sigma_s'$ and an absorption coefficient $\sigma_a$; its scatter term is this profile times a per-Gaussian color and the inverse-square falloff of the light.
The network's outputs pass through a sigmoid and an affine map that bound $\sigma_s'$ and $\sigma_a$ to $[0.05,2.05]$ and $r$ to $[0.1,3.1]$.
We multiply $\sigma_s'$ and $\sigma_a$ by the same factor $m$ after this activation, so the scaled values may leave these bounds.
This scales the extinction of the dipole by $m$ and keeps its albedo, as our edit does~(the recovered $\boldsymbol{\sigma}_t$ times $m$, albedo unchanged, one new solve) and as the reference does~(subsurface radius divided by $m$ in Blender).
Everything else is unchanged: $r$, the per-Gaussian scatter color, the diffuse and specular terms and the shadow network; nothing is retrained or fitted.
We use the models of main \cref{tab:objects} and $m\in\{\frac18,\frac14,\frac12,2,4,8\}$.

\paragraph{Views and measure.}
Each synthetic object is rendered at $42$ views: three held-out cameras, each under $14$ lights.
A view is back-lit when the cosine between the directions from the scene center to the camera and to the light is below $-\frac13$~($5$ of the $42$ views; $6$ on the dragon) and front-lit when it is above $\frac13$~($19$; $21$ on the dragon).
The brightness change of a view is the mean linear radiance of the edited render over the object mask, divided by that of the unedited render of the same method; \cref{tab:editsweep} gives its mean over the back-lit and the front-lit views, and main \cref{fig:editing} plots the back-lit rows of the first three objects.
SSD-GS renders are clipped to $[0,1]$ in sRGB before linearization, as its training images are, so its changes at $\times4$ and $\times8$ are capped by the clip; the reference and \lucent are linear.
The dragon's reference is a re-created scene, because the dataset ships no Blender file for it: the Stanford dragon mesh registered to the dataset frame with one constant subsurface material fitted to the dataset images, without any of our recovered values~($38.8$\,dB against the dataset images, unedited); it is rendered at three multipliers only.

\begin{table}[!tbp]\centering\small\setlength{\tabcolsep}{3pt}
\caption{\textbf{Editing the extinction: brightness change} of the edited render against the unedited render of the same method, mean over the back-lit and the front-lit views. Reference: Blender, subsurface radius divided by the multiplier. \lucent: physics only, recovered $\boldsymbol{\sigma}_t$ times the multiplier. SSD-GS: its predicted $\sigma_s'$ and $\sigma_a$ times the multiplier. Nothing is retrained or fitted. $^\dagger$Re-created scene.}
\label{tab:editsweep}
\resizebox{\linewidth}{!}{
\begin{tabular}{ll cccccc}
\toprule
 & & $\times\frac{1}{8}$ & $\times\frac{1}{4}$ & $\times\frac{1}{2}$ & $\times2$ & $\times4$ & $\times8$ \\
\midrule
\multicolumn{8}{@{}l}{\textit{Back-lit views}} \\
Soap & Reference & 2.15 & 1.71 & 1.33 & 0.76 & 0.62 & 0.56 \\
 & \lucent & 1.35 & 1.38 & 1.24 & 0.75 & 0.51 & 0.33 \\
 & SSD-GS & 0.37 & 0.39 & 0.47 & 5.20 & 10.8 & 11.2 \\
\addlinespace[2pt]
Candle & Reference & 5.46 & 3.60 & 2.07 & 0.55 & 0.37 & 0.32 \\
 & \lucent & 3.77 & 3.00 & 1.94 & 0.40 & 0.15 & 0.10 \\
 & SSD-GS & 0.27 & 0.28 & 0.37 & 6.00 & 10.6 & 11.6 \\
\addlinespace[2pt]
Statue & Reference & 3.59 & 1.78 & 1.20 & 0.92 & 0.89 & 0.89 \\
 & \lucent & 4.22 & 2.32 & 1.45 & 0.70 & 0.47 & 0.30 \\
 & SSD-GS & 1.00 & 1.00 & 1.00 & 1.00 & 1.00 & 1.00 \\
\addlinespace[2pt]
Dragon & Reference$^\dagger$ & -- & 2.13 & 1.46 & 0.72 & -- & -- \\
 & \lucent & 2.85 & 2.44 & 1.70 & 0.53 & 0.29 & 0.21 \\
 & SSD-GS & 0.43 & 0.46 & 0.56 & 2.20 & 1.70 & 0.59 \\
\midrule
\multicolumn{8}{@{}l}{\textit{Front-lit views}} \\
Soap & Reference & 0.50 & 0.67 & 0.86 & 1.07 & 1.08 & 1.07 \\
 & \lucent & 0.58 & 0.72 & 0.86 & 1.10 & 1.00 & 0.68 \\
 & SSD-GS & 0.29 & 0.31 & 0.41 & 3.03 & 3.52 & 3.55 \\
\addlinespace[2pt]
Candle & Reference & 0.96 & 1.01 & 1.02 & 0.98 & 0.97 & 0.96 \\
 & \lucent & 1.13 & 1.20 & 1.15 & 0.82 & 0.68 & 0.60 \\
 & SSD-GS & 0.70 & 0.71 & 0.76 & 1.74 & 2.03 & 2.07 \\
\addlinespace[2pt]
Statue & Reference & 0.99 & 1.00 & 1.00 & 0.99 & 0.99 & 0.99 \\
 & \lucent & 0.76 & 0.88 & 0.98 & 0.88 & 0.68 & 0.49 \\
 & SSD-GS & 1.00 & 1.00 & 1.00 & 1.00 & 1.00 & 1.00 \\
\addlinespace[2pt]
Dragon & Reference$^\dagger$ & -- & 1.24 & 1.12 & 0.91 & -- & -- \\
 & \lucent & 1.09 & 1.22 & 1.19 & 0.73 & 0.52 & 0.42 \\
 & SSD-GS & 0.67 & 0.69 & 0.75 & 1.66 & 1.56 & 0.82 \\
\bottomrule
\end{tabular}}
\end{table}

\begin{figure*}[!tp]
\centering
\includegraphics[width=\textwidth]{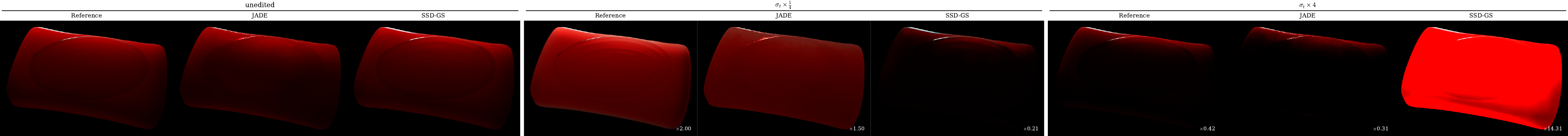}\\[2pt]
\includegraphics[width=\textwidth]{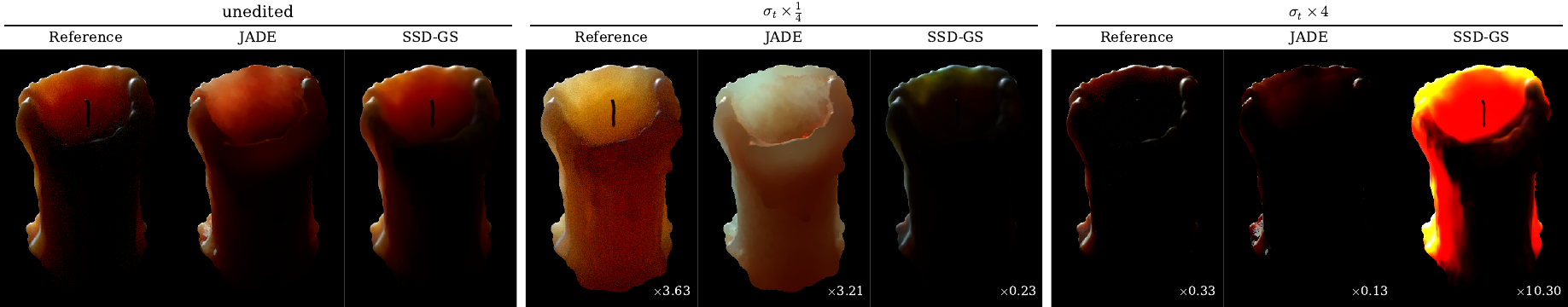}
\caption{\textbf{Editing the extinction}, soap~(top) and candle~(bottom) under a back light: unedited, $\boldsymbol{\sigma}_t\times\frac14$ and $\times4$, each as reference, \lucent~(physics only) and SSD-GS~(its $\sigma_s'$ and $\sigma_a$ scaled). Numbers: brightness change of the view against the unedited render of the same method; one display gain per object.}
\label{fig:editrows}
\end{figure*}

Under back light, \lucent changes in the direction of the reference at every multiplier on all four objects, and SSD-GS changes the opposite way on the soap, the candle and the dragon~(except the dragon at $\times8$) and not at all on the statue~(\cref{tab:editsweep,fig:editrows}).
On the statue the scatter term of the trained SSD-GS is zero, because its network's outputs sit at the upper bounds, so no multiplier changes its image.
The magnitudes of \lucent are not those of the reference: on the soap it barely brightens further below $\times\frac14$ and darkens more than the reference at $\times4$ and $\times8$, and on the statue it overshoots, the effect of the errors in the recovered extinction discussed in \cref{sec:supp:editgt}.
Under front light the reference barely changes on the candle and the statue, while \lucent darkens them at large multipliers.

\subsection{Edits of real captures}
\label{sec:supp:realedit}
\begin{figure*}[!tp]
\centering
\includegraphics[width=\textwidth]{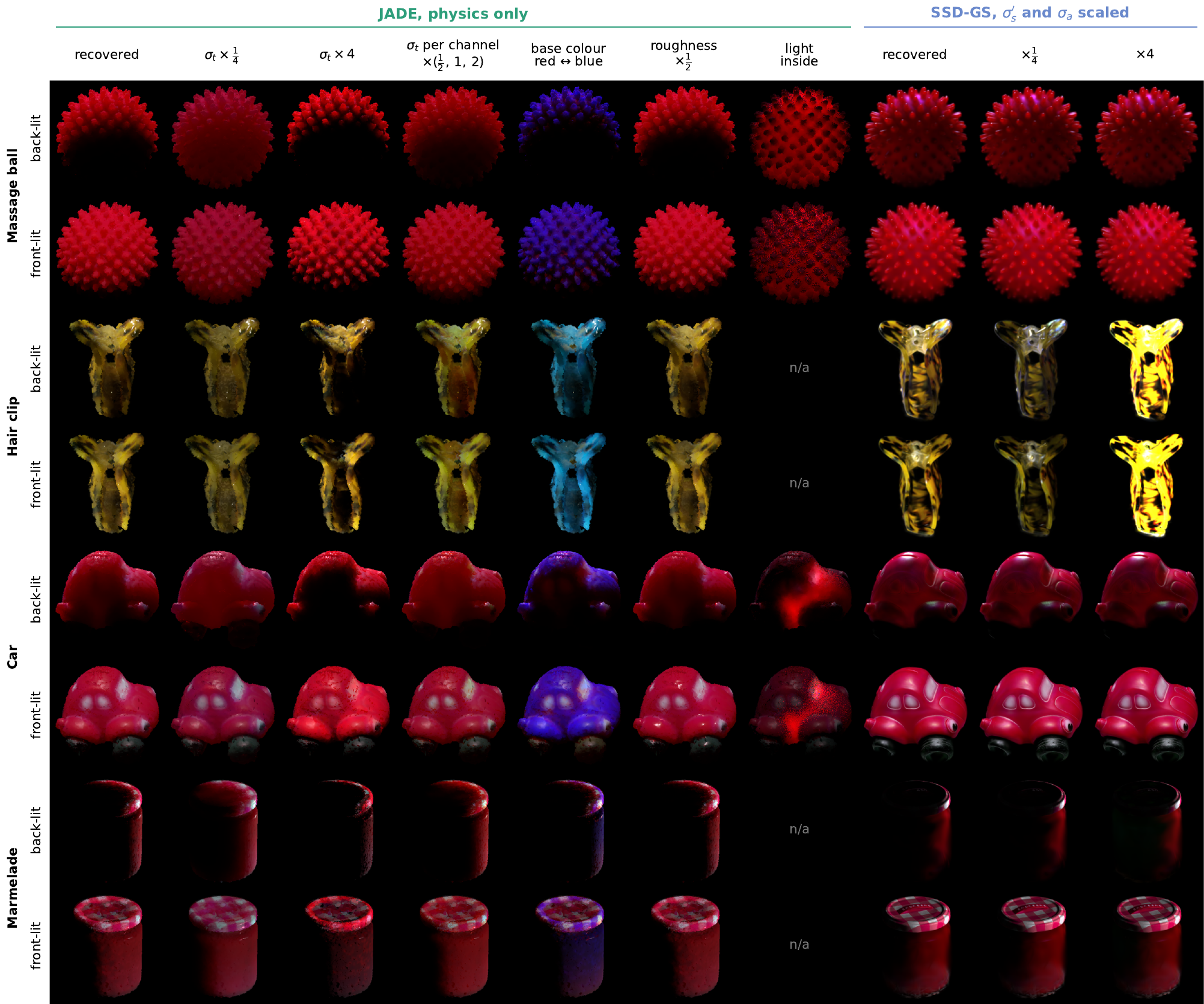}
\caption{\textbf{Edits of real captures}, held-out back and front light of one test camera, all at render time without retraining.
\lucent, physics only: extinction $\times\frac14$ and $\times4$, per channel $\times(\frac12,1,2)$, base color with red and blue swapped, roughness $\times\frac12$, and a light source inside the medium~(left out on the hair clip, a thin shell narrower than the light, and on the marmalade, where its light leaves mainly through the lid).
SSD-GS: its $\sigma_s'$ and $\sigma_a$ scaled as in \cref{sec:supp:ssdgsedit}.
One display gain per row for every tile.
The holes in the hair clip are real: it is a claw clip with open jaws, and the dataset's own mask has this gap in $15$ of its $158$ camera poses; the few specks next to it are small holes in our reconstruction.}
\label{fig:realedits}
\end{figure*}

Real captures have no reference for an edit, so we check two things: that edits look plausible~(\cref{fig:realedits}), and that the brightness of back-lit views moves the way an extinction change should, over all held-out views of all ten objects~(\cref{tab:direction}).
The views, the measure and the SSD-GS edit are those of \cref{sec:supp:ssdgsedit}, here over all $256$ held-out views of each object~($50$--$68$ back-lit); we leave out the back-lit views where the LED shines into the lens~(cosine below $-0.9$) and those of one dim LED.

\begin{table*}[!tp]\centering\small\setlength{\tabcolsep}{4pt}
\caption{\textbf{Direction of extinction edits on the real captures}: mean brightness change of the back-lit held-out views, edited against unedited render of the same method, and in brackets the share~(\%) of these views that move the expected way by more than $1\%$~(brighter below $\times1$, darker above).}
\label{tab:direction}
\begin{tabular}{llrcccccc}
\toprule
Object & Method & $n$ & $\times\tfrac{1}{8}$ & $\times\tfrac{1}{4}$ & $\times\tfrac{1}{2}$ & $\times2$ & $\times4$ & $\times8$ \\
\midrule
Car & \lucent & 54 & 1.11\,(57) & 1.20\,(72) & 1.16\,(91) & 0.88\,(89) & 0.79\,(93) & 0.66\,(100) \\
 & SSD-GS & & 1.00\,(0) & 1.00\,(0) & 1.00\,(0) & 1.00\,(0) & 1.00\,(0) & 1.00\,(0) \\
Hair clip & \lucent & 63 & 0.57\,(8) & 0.73\,(8) & 0.91\,(10) & 0.89\,(92) & 0.63\,(100) & 0.39\,(100) \\
 & SSD-GS & & 0.40\,(0) & 0.42\,(0) & 0.51\,(0) & 4.28\,(0) & 8.61\,(0) & 9.42\,(0) \\
Blue bottle & \lucent & 62 & 0.42\,(0) & 0.62\,(0) & 0.84\,(2) & 0.97\,(55) & 0.76\,(90) & 0.53\,(90) \\
 & SSD-GS & & 1.00\,(0) & 1.00\,(0) & 1.00\,(0) & 1.00\,(0) & 1.00\,(0) & 1.00\,(0) \\
Clear crystal & \lucent & 67 & 0.72\,(9) & 0.94\,(24) & 1.03\,(42) & 0.91\,(69) & 0.75\,(88) & 0.54\,(100) \\
 & SSD-GS & & 1.00\,(0) & 1.00\,(0) & 1.00\,(0) & 1.00\,(0) & 1.01\,(0) & 1.04\,(0) \\
Marmalade & \lucent & 53 & 4.89\,(79) & 2.13\,(85) & 1.25\,(91) & 0.82\,(100) & 0.66\,(100) & 0.51\,(100) \\
 & SSD-GS & & 1.00\,(0) & 1.00\,(0) & 1.00\,(0) & 1.00\,(0) & 1.02\,(0) & 1.08\,(0) \\
Liqueur & \lucent & 61 & 1.45\,(69) & 1.59\,(75) & 1.37\,(85) & 0.71\,(100) & 0.49\,(100) & 0.32\,(100) \\
 & SSD-GS & & 1.00\,(0) & 1.00\,(0) & 1.00\,(0) & 1.00\,(0) & 1.00\,(0) & 1.00\,(0) \\
Nail polish remover & \lucent & 58 & 0.34\,(0) & 0.51\,(0) & 0.76\,(0) & 1.05\,(26) & 0.89\,(74) & 0.70\,(84) \\
 & SSD-GS & & 0.96\,(0) & 0.96\,(0) & 0.97\,(0) & 1.10\,(0) & 1.22\,(0) & 1.09\,(0) \\
Massage ball & \lucent & 50 & 1.05\,(32) & 1.00\,(28) & 0.93\,(18) & 1.01\,(18) & 0.76\,(100) & 0.42\,(100) \\
 & SSD-GS & & 1.00\,(0) & 1.00\,(0) & 1.00\,(0) & 1.00\,(0) & 1.01\,(0) & 1.04\,(0) \\
Joyful yell & \lucent & 50 & 1.69\,(100) & 1.62\,(100) & 1.36\,(100) & 0.69\,(100) & 0.46\,(100) & 0.29\,(100) \\
 & SSD-GS & & 1.00\,(0) & 1.00\,(0) & 1.00\,(0) & 1.00\,(0) & 1.00\,(0) & 1.00\,(0) \\
Physalis & \lucent & 68 & 1.48\,(81) & 1.44\,(96) & 1.26\,(100) & 0.73\,(100) & 0.48\,(100) & 0.30\,(100) \\
 & SSD-GS & & 1.00\,(0) & 1.00\,(0) & 1.00\,(0) & 1.00\,(0) & 1.00\,(0) & 1.00\,(0) \\
\midrule
Mean of 10 & \lucent & & 1.37\,(44) & 1.18\,(49) & 1.09\,(54) & 0.87\,(75) & 0.67\,(95) & 0.47\,(97) \\
 & SSD-GS & & 0.94\,(0) & 0.94\,(0) & 0.95\,(0) & 1.34\,(0) & 1.79\,(0) & 1.87\,(0) \\
\bottomrule
\end{tabular}
\end{table*}

Raising the extinction darkens the back-lit views of \lucent on all ten objects at $\times4$ and $\times8$, and on eight at $\times2$~(not the nail polish remover, $1.05$, and the massage ball, $1.01$).
Lowering it brightens them on five~(car, marmalade, liqueur, joyful yell, physalis), gives a mixed response on two~(clear crystal, massage ball) and darkens them on three~(hair clip, blue bottle, nail polish remover).
The brighter-when-thinner rule holds for media that are optically thick; in an optically thin medium, less extinction also means less scattering towards the camera, so the image darkens.
The recovered media follow this rule.
We measure the optical thickness of each channel as its recovered $\sigma_t$ times twice the median distance of the medium's sites to the nearest empty site.
It is below $1$ in every channel of the hair clip~($0.55$--$0.70$), the nail polish remover~($0.32$--$0.55$) and the clear crystal~($0.66$--$0.85$), and in the green and blue channels of the blue bottle~($0.65$, $0.92$); it is above $2$ in every channel of the liqueur, the marmalade, the joyful yell and the physalis; the car and the massage ball, whose images are mostly red, have a red channel in between~($1.1$ and $1.6$) and respond weakly or mixed.
Within each object, too, the channel with the smallest recovered extinction brightens least or darkens most at $\times\frac14$ and $\times\frac18$~(e.g., blue bottle: red $1.83$, green $0.45$, blue $0.54$, at optical thickness $7.3$, $0.65$ and $0.92$).
SSD-GS moves the expected way in none of the $586$ back-lit views at any of the six multipliers: its image does not change at all on five objects, where its scatter term is inactive~(\cref{sec:supp:ssdgsscatter}), changes by at most $8\%$ on three more, and moves the opposite way on the hair clip~(up to $9\times$) and the nail polish remover~(up to $1.2\times$).

\subsection{Surface, glass and interior edits}
\label{sec:supp:editing}

\begin{figure*}[!tp]
  \centering
  \newlength{\figSuW}\setlength{\figSuW}{0.164\textwidth}
  \setlength{\tabcolsep}{0.5pt}\renewcommand{\arraystretch}{0.6}\scriptsize
  \begin{tabular}{@{}cccccc@{}}
  recovered & roughness $\times\frac12$ & roughness $\times2$ & base color R$\leftrightarrow$B & base color gray & base color $\times\frac12$ \\
  \includegraphics[width=\figSuW]{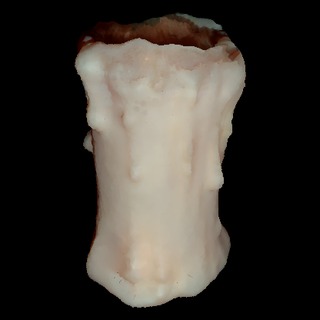} & \includegraphics[width=\figSuW]{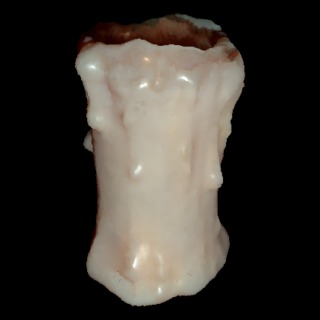} & \includegraphics[width=\figSuW]{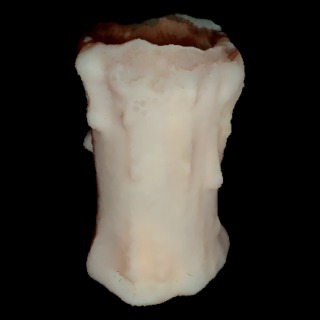} & \includegraphics[width=\figSuW]{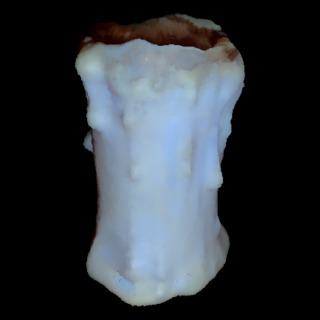} & \includegraphics[width=\figSuW]{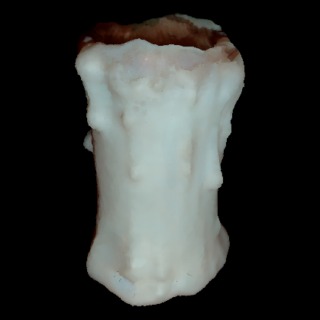} & \includegraphics[width=\figSuW]{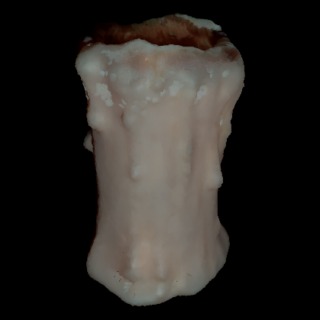} \\
  \end{tabular}
  \caption{\textbf{Surface edits} of the candle under its most front-lit light, physics only: per-corner GGX roughness and base color; the base color also sets the albedo of the medium under each wall.}
  \label{fig:editing_surface}
\end{figure*}

\begin{figure*}[!tp]
\centering
\includegraphics[width=\textwidth]{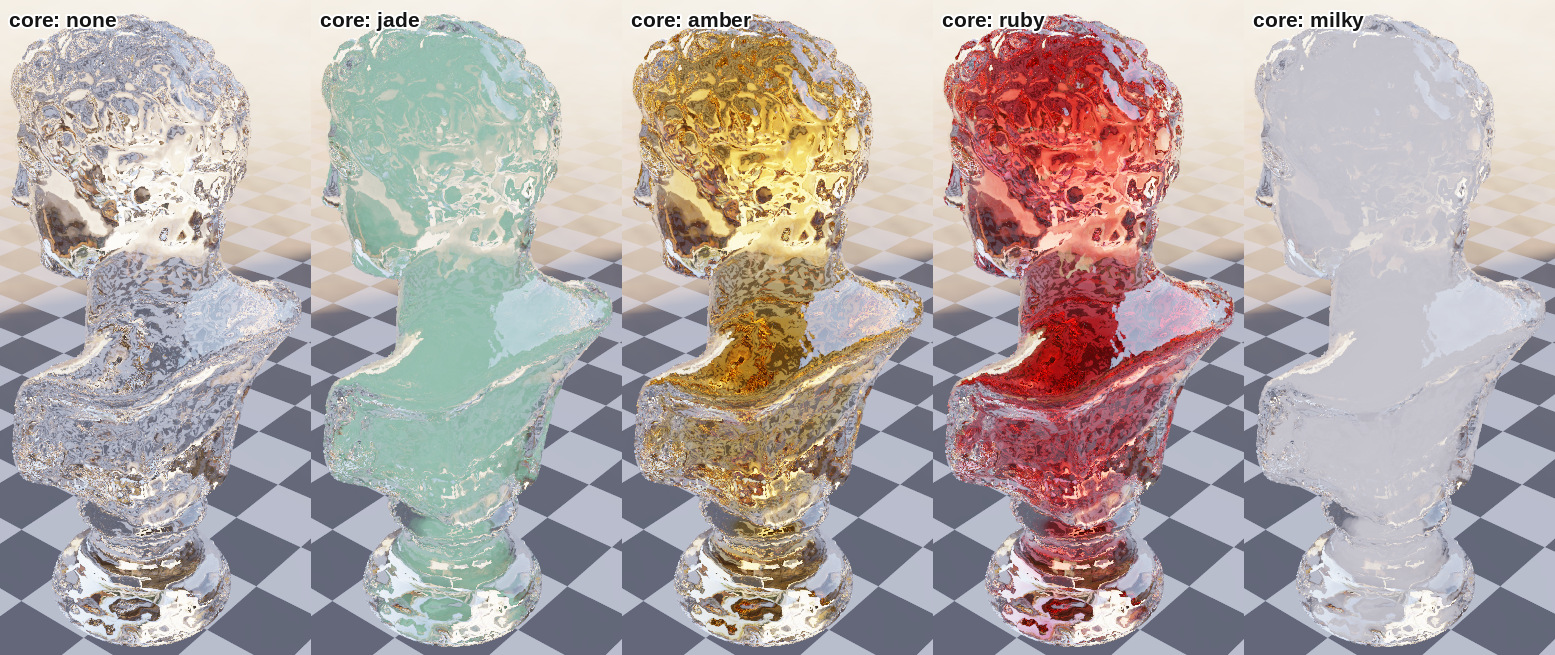}
\caption{\textbf{Glass with an editable inside.}
The reconstructed statue rendered as glass, with the inner medium edited per cell~(left to right: none, jade, amber, ruby, milky); the thin parts stay clear glass.
This is a render-time edit without retraining that uses only the recovered geometry: the walls become a smooth dielectric interface~($\eta=1.5$, up to $8$ internal bounces, $16$ rays per pixel) under an environment map with a checker floor, and the cells deeper than $0.03$ of the bounding-box diagonal hold the edited core.}
\label{fig:glass_core}
\end{figure*}

\begin{figure*}[p]
\centering
\includegraphics[width=\textwidth,height=0.85\textheight,keepaspectratio]{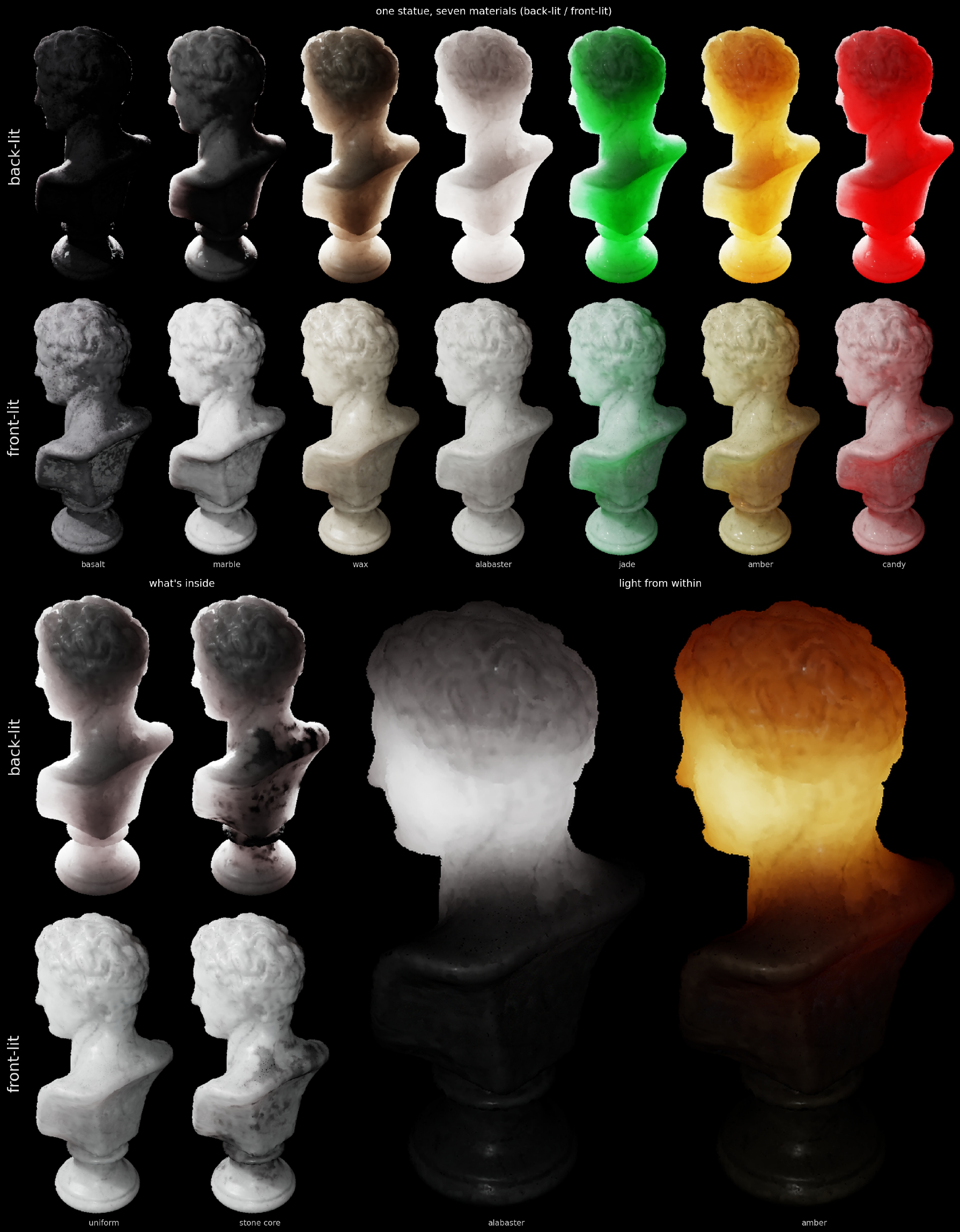}
\caption{\textbf{Editing the inside of a reconstructed statue.}
All edits are applied to the physics-only reconstruction at render time, without retraining.
\emph{Top:} seven materials from one reconstruction, under back lights~(top row) and front lights~(bottom row); the values are in \cref{tab:presets}.
Each changes the surface~(base color, roughness or both) and the medium's $\boldsymbol{\sigma}_t$: per channel for jade, amber and candy, by one factor for basalt, wax and alabaster; marble changes only the roughness.
\emph{Bottom left:} a stone core~($\boldsymbol{\sigma}_t\times16$, albedo $0.05$) fills the cells deeper than about a third of the largest depth below the surface and blocks the light passing through the statue.
\emph{Bottom right:} a light source placed inside the medium near the neck; the panel shows the diffusion term with a dim fill of the other terms, so the face and the thin parts near the light glow while the thick bust stays dark.}
\label{fig:editing_inside}
\end{figure*}

\Cref{fig:glass_core} renders the surface of the statue as glass around an edited medium, and \Cref{fig:editing_inside} edits its inside into seven materials and adds a stone core and a light inside the medium.
All material edits are multipliers on the recovered values~(\cref{tab:presets}).
The stone core multiplies $\boldsymbol{\sigma}_t$ and replaces the albedo in the cells whose distance to the nearest boundary cell exceeds a fraction of the largest such distance, with a smooth transition; the light traced to each cell then passes through the edited medium, so the core also shadows it.
The light inside replaces the source $q_i$ of the diffusion solve by an isotropic Gaussian blob around a point in the medium; it is not an extra term, and the outside light no longer feeds the diffusion.

\begin{table}[!tbp]\centering\small\setlength{\tabcolsep}{4pt}
\caption{\textbf{Material presets} of \cref{fig:editing_inside}, as multipliers on the recovered per-channel $\boldsymbol{\sigma}_t$ and base color~(RGB); roughness is set to the given value~(``--'': unchanged).}
\label{tab:presets}
\begin{tabular}{l ccc}
\toprule
Material & $\boldsymbol{\sigma}_t\times$ & base color$\times$ & roughness \\
\midrule
basalt & $(2,2,2)$ & $(0.4,0.4,0.42)$ & $1.0$ \\
marble & -- & -- & $0.05$ \\
wax & $(\frac14,\frac14,\frac14)$ & $(1,0.9,0.75)$ & -- \\
alabaster & $(\frac18,\frac18,\frac18)$ & $(1,0.97,0.92)$ & $1.0$ \\
jade & $(0.35,0.15,0.3)$ & $(0.6,0.95,0.7)$ & $0.05$ \\
amber & $(0.08,0.125,0.2)$ & $(1,0.78,0.45)$ & $0.08$ \\
candy & $(0.1,0.4,0.4)$ & $(1,0.5,0.5)$ & $0.05$ \\
\bottomrule
\end{tabular}
\end{table}

\section{Limitations and anticipated questions}
\label{sec:supp:faq}
\textit{Extends main \cref{sec:conclusion}.}
\paragraph{Are the views of main \cref{fig:qual} selected in our favor?}
Yes, and its caption says so; main \cref{tab:objects} scores $256$ held-out views per real object and $32$ per synthetic object.
\Cref{sec:supp:median} shows, for every object, the front-lit and the back-lit view at our median PSNR, next to both baselines.

\paragraph{Is the comparison with SSD-GS fair?}
Both baselines are retrained on the same images with their released settings~(\cref{sec:supp:calib}).
Our learned per-LED power and per-camera exposure use only the training images, and account for about $0.5$\,dB of our lead on the turbid real objects.
The pose and light refinement of SSD-GS is off, as in its released configuration for the synthetic scenes of this dataset.
Its release has no configuration for the real captures, and its script for other real data turns the refinement on.
Retrained with that script, which also trains longer~($100$k instead of $60$k iterations), SSD-GS changes by $-0.03$\,dB on average over the ten real captures~($33.92$ against $33.95$\,dB; \cref{tab:objects_full}), our mean lead goes from $1.73$ to $1.76$\,dB, and SSD-GS passes \lucent on the blue bottle.
These numbers follow the release's protocol, in which SSD-GS fits the pose and the LED position of each test view on the test image itself, which \lucent does not; with the corrections carried over from the training images instead it scores $33.99$\,dB, and with the test views uncorrected $33.87$\,dB.
On five of the ten real captures its scatter term is inactive in our runs, so there it has no subsurface term to edit~(\cref{sec:supp:ssdgsscatter}).
In the retrained runs the term becomes active on three of these five~(blue bottle, liqueur and joyful yell, $24$--$58\%$ of the back-lit image), carries $0.2\%$ on the car and stays inactive on the physalis.

\paragraph{Does editing work on real captures, where no reference exists?}
\Cref{sec:supp:realedit} edits four real captures and measures, over all held-out views of all ten objects, whether the back-lit brightness moves the way an extinction change should.
Raising the extinction by $4\times$ or more darkens the back-lit views of \lucent on all ten objects; lowering it brightens them on the optically thick ones and darkens them on the optically thin ones, as it should.
Scaling the scattering coefficients of SSD-GS moves the expected way in none of the views; on five objects it changes nothing, since its scatter term is inactive there.

\paragraph{Why do edits match the reference in direction but not in magnitude?}
The recovered extinction is too low in channels that are optically thick at the cell size, where the fit trades extinction for the learned diffusion gain~(\cref{sec:supp:editgt}).
The physics still predicts where the image changes, which no global change of brightness can~(19 of 24 edits, main \cref{tab:editpattern}).

\paragraph{Does the residual change the medium?}
No. It is fitted after the foam, the surface and the medium are fixed, and only rescales the rendered terms per pixel and adds a small correction~(main \cref{sec:method:residual}); every edit in this paper is rendered with the physics only.

\paragraph{Why do a few cells of the edited dragon stay green in main \cref{fig:teaser}?}\label{sec:supp:greencells}
As on every object, the dragon's medium albedo is tied to the surface base color: every cell inside the surface takes the albedo of its nearest surface cell~(main \cref{sec:method:transport}), so a color edit of the surface recolors the medium.
Cells outside the surface but within two cell lengths of it are not tied: they scatter with a per-cell albedo that is trained, the only albedo of the model that the base color does not set~(\cref{sec:supp:impl}).
On the dragon, such semi-transparent cells line the silhouette and keep their trained jade albedo; after a wax, amber or ruby edit they show as green on the thin spines and the tail.
Applying the same edit to these cells, with the channels the material absorbs made absorbing rather than scattering, keeps the look of each material and removes most of the green~($86\%$ fewer green pixels on the ruby edit).
The video uses these renders; for display only, it also recolors the few remaining green pixels to the surrounding color and fills the see-through gap between the back and the tail of the glass dragon.

\paragraph{How long does an edit take?}
One new diffusion solve per light; with the render of one view it takes $0.12$--$0.34$\,s on one A100~(\cref{tab:cost}).

\paragraph{Is the dragon's edit reference ground truth?}
No. The dataset ships no Blender file for the dragon, so its reference is a re-created scene fitted to the dataset images~(\cref{sec:supp:ssdgsedit}); main \cref{fig:editing} therefore shows the dragon's edit but leaves it out of the sweep against the reference.

\paragraph{Was the configuration tuned to the objects of the paper?}
The settings of the method are the same for every object; what differs is how the data are handled, listed here in full and in the object table of the code~(\texttt{configs/objects.py}).
The hair clip is trained on all its images, since the exclude list shipped with it removes every image; its masks cut more of the visual hull than on the other objects, so the visual-hull check of its surface export uses a threshold of $0.75$ instead of $0.30$, and its residual fit floors the per-image loss weights at $10\%$ of the physics error, since the default weighting diverged.
Five cameras of the joyful yell that the dataset mis-registers are left out of our training; every method is scored on the same views.
The synthetic soap and candle do not sit where the camera axes meet, so the image given to the normal prior is chosen about the object's center.
The statue uses $2$M sites instead of $262$k.
The paper's soap, candle and dragon models use an earlier surface extraction~(\cref{sec:supp:code}).
\Cref{sec:supp:newobjects} trains three further real captures once each with the released code at its defaults, except for the two data settings listed there; all three fall within the range of the main paper, and one, the dark green candle, is a failure case of the physics.

\paragraph{Where does \lucent fail?}
On glass, whose appearance is dominated by refraction, which a diffusion medium does not model~(\cref{fig:failures}); under back light near the optical axis, where camera glare dominates the missing light; in optically thin media, where lowering the extinction darkens instead of brightening, as physics predicts; and on media recovered as almost opaque, such as the dark green candle of \cref{sec:supp:newobjects}, where the residual carries much of the appearance.

\section{Code and reproducibility}
\label{sec:supp:code}
The code we release on the project page trains both stages and the object residual, relights, and applies every edit of this paper; its README gives the commands.
It uses one configuration of the method for all fourteen objects; the per-object data settings of \cref{sec:supp:faq} are in its object table.
The paper's soap, candle and dragon models were trained with an earlier surface extraction~(no hole filling and $20$ smoothing passes) that the released configuration replaces; \cref{tab:retrain} compares retrains from scratch with the released code to the paper's numbers.
On the car, trained with the released configuration both times, the retrain is within $0.2$\,dB of the paper; on the soap, the candle and the dragon, whose surface extraction changed, it is within $0.4$\,dB, except the dragon's physics-only render~($-0.9$\,dB).

\begin{table}[!tbp]\centering\small\setlength{\tabcolsep}{4pt}
\caption{\textbf{Retrained with the released code}, from scratch, against main \cref{tab:objects}: PSNR~(dB) of the physics only and with the object residual.}
\label{tab:retrain}
\begin{tabular}{l cc cc}
\toprule
 & \multicolumn{2}{c}{Physics only} & \multicolumn{2}{c}{+ residual} \\
\cmidrule(lr){2-3}\cmidrule(l){4-5}
 & paper & released & paper & released \\
\midrule
Car~(real) & 30.42 & 30.26 & 35.93 & 35.81 \\
Soap & 31.40 & 31.21 & 42.63 & 43.02 \\
Candle & 30.76 & 30.94 & 40.33 & 40.26 \\
Dragon & 31.31 & 30.40 & 35.09 & 34.65 \\
\bottomrule
\end{tabular}
\end{table}

\end{document}